\documentclass[aos,preprint]{imsart}
\usepackage{mathtools,amssymb,amsthm,booktabs,tabularx,array,float}
\usepackage[numbers]{natbib}
\usepackage{hyperref}
\hypersetup{hidelinks,hypertexnames=false,linktoc=all,bookmarksdepth=2,
  pdftitle={Pragmatic Deeply Debiased ML},
  pdfauthor={Victor Chernozhukov and Chengchun Shi},
  pdfsubject={Deep debiasing for linear regression functionals}}
\startlocaldefs
\allowdisplaybreaks[2]
\newcommand{\E}{\mathbb E}
\newcommand{\Var}{\operatorname{Var}}
\newcommand{\Cov}{\operatorname{Cov}}
\newcommand{\R}{\mathbb R}
\newcommand{\Pn}{\mathbb P_n}
\newcommand{\ip}[2]{\langle #1,#2\rangle}
\newcommand{\norm}[1]{\lVert #1\rVert}
\newcommand{\op}{\mathrm{op}}
\newcommand{\HS}{\mathrm{HS}}
\newcommand{\ad}{\rightsquigarrow}
\newcommand{\ind}{\mathbf 1}
\newcommand{\gh}{\widehat g}
\newcommand{\ah}{\widehat\alpha}
\newcommand{\Kh}{\widehat K}
\newcommand{\thh}{\widehat\theta}
\newcommand{\calT}{\mathcal T}
\theoremstyle{plain}
\newtheorem{theorem}{Theorem}
\newtheorem{proposition}{Proposition}
\newtheorem{corollary}[proposition]{Corollary}
\theoremstyle{definition}
\newtheorem{example}[proposition]{Example}
\newtheorem{remark}[proposition]{Remark}
\endlocaldefs

\begin{document}
\begin{frontmatter}
\pdfsubject{Deep debiasing for linear regression functionals}
\title{Pragmatic Deeply Debiased ML}
\runtitle{Pragmatic Deeply Debiased ML}
\begin{aug}
\author[mit]{\fnms{Victor}~\snm{Chernozhukov}}
\author[lse]{\fnms{Chengchun}~\snm{\href{https://scholar.google.com/citations?user=dDGy3N0AAAAJ\&hl=en}{Shi}}}
\runauthor{V. Chernozhukov and C. Shi}
\address[mit]{Department of Economics and Center for Statistics and Data Science,\protect\\
Massachusetts Institute of Technology}
\address[lse]{Department of Statistics, London School of Economics and Political Science}
\end{aug}
\thankstext[]{manuscript}{10 November 2026 (revised 24 November 2026). We developed this paper in collaboration with GPT Astra. We take full responsibility for its contents.}
\begin{abstract}
Debiased machine learning (DML) combines a fitted outcome regression with weighted residuals to remove first-order regularization bias when estimating parameters such as average treatment effects, yet a second-order bias remains whenever regression and weighting errors align. For linear regression functionals, we target this second-order bias by training a strong learner to predict first-stage residuals from many covariates, yielding a one-dimensional (or low-dimensional) working model whose bias contribution is estimated on an independent sample. To complement this targeted search, we also analyze general kernel-based corrections, including covariate partitions and smoothers. We prove that deeply debiased ML is algebraically identical to ordinary DML evaluated with a post-hoc regression update that adjusts each scored observation using kernel-weighted residuals from the remaining evaluation observations. In our empirical and simulation studies, strong initial DML learners often leave the second-order correction small relative to sampling uncertainty, in which case the correction serves as a diagnostic check alongside sensitivity bounds for any remaining misspecification bias.
\end{abstract}
\end{frontmatter}

\section{Introduction}\label{sec:intro}
Ordinary debiased machine learning (DML)~\cite{dml} removes the first-order regularization bias that arises when we plug a fitted outcome regression into a target parameter. For any linear functional of that regression, such as an average treatment effect, the bias left behind is second order. It equals the negative inner product of the outcome-regression error and the error in the estimated Riesz weights (for example, inverse-propensity weights). Because this remaining bias is a product of two errors, it is often small, yet whenever both learners miss the same covariate patterns, their errors align and the mixed bias can easily dominate sampling uncertainty. We study a single additional correction aimed directly at that alignment.

Suppose we are estimating an average treatment effect for households. Within an income group, the estimated treatment weights may fail to reproduce the group's true share of the population, while the fitted outcome regression systematically overpredicts or underpredicts outcomes in that same group. Neither true nuisance function is observed, yet both discrepancies can be estimated directly on an independent evaluation sample. Multiplying sample averages over the same observations would introduce own-observation correlation (a diagonal covariance term), so instead we pair the weight discrepancy from one observation with the outcome residual from a distinct observation. As a result, the bias correction takes the form of an average over all distinct pairs of observations, known as a U-statistic. A kernel chosen on training data specifies the groups (or, more generally, the directions) along which we apply this correction.

Choosing a basis for bias correction can be harder than choosing a learner for the initial regression. A prespecified dictionary forces the analyst to guess in advance which covariate patterns the fitted regression will miss, and projecting onto a large dictionary can add substantial variance. We learn the dictionary for bias correction from first-stage residuals. A flexible regression of these residuals on the covariates can produce a single feature that predicts where the initial fit errs. On an independent evaluation sample, we then pair outcome residuals with weighting discrepancies along that learned feature to estimate its contribution to bias in the target. If the learned feature captures the regression error up to scale, projecting onto that single direction removes the entire mixed bias while keeping the working space one-dimensional. Section~\ref{sec:learned-bias-rate} shows how a residual regression with a growing number of predictors can learn this direction accurately enough to make the remaining bias smaller than the sampling error.

Once we have learned where the regression errs, a natural alternative to correcting the target directly is to update the regression along the learned directions and plug it back into ordinary DML. Theorem~\ref{thm:regression-update} shows that the two procedures give exactly the same estimate when we update the regression for each observation using the residuals of the other evaluation observations and the same kernel. For a single learned direction, the update adds that direction to the initial regression with a slope estimated leaving out the observation being scored. Although each observation is held out of its own update, the updated scores share data. The pairwise representation lets us determine how much bias this post-hoc corrected DML estimator removes and what the update costs in sampling variance.

An exact projection onto a training-chosen working space eliminates all mixed bias inside that space, leaving only the inner product of the error components outside it (Proposition~\ref{prop:bias}). Neither true nuisance function needs to belong to the working space, since the space simply specifies which error directions are removed. The projection also preserves the usual worst-case product bound, although the realized bias of a single fit can increase when errors inside and outside the space happened to cancel (Example~\ref{ex:cancellation}).

The pairwise correction can alter both the first-order influence function and the sampling noise contributed by pairs. An exact Hoeffding decomposition separates these two components, yielding a closed-form, finite-sample unbiased estimator of the full conditional variance (Proposition~\ref{prop:vhat}) alongside explicit conditions under which the corrected estimator retains the ordinary DML asymptotic distribution and standard error (Proposition~\ref{prop:ordinary}).

Even after a second debiasing step, some misspecification and nuisance-learning bias may remain outside the working space. We adapt the two-factor sensitivity framework of Chernozhukov, Cinelli, Newey, Sharma, and Syrgkanis~\cite{sensitivity} to control this remaining misspecification bias, translating a stated allowance for residual nuisance error into bias-aware confidence intervals. When the initial DML learners are already strong, a further correction often moves the point estimate very little. In that case, the correction works as a diagnostic check along the learned directions, while the sensitivity bounds quantify how much misspecification bias in the remaining directions the conclusion can tolerate.

Our empirical exercises illustrate how the strength of the initial learner governs the value of a second debiasing step. In the 401(k) eligibility dataset, starting from linear DML, a single direction learned by a random forest raises the estimated effect of eligibility on capped net financial assets by \$1,849, and a direction learned with CausalPFN~\cite{causalpfn} raises it by \$2,808. Starting from random-forest DML on the same households, those two corrections move the estimate by only \$54 and $-$\$113. In the ACIC simulations, where the true treatment effect is known, a forest-learned correction lowers root mean squared error by 38--43\% after linear initial fits, by less than 2\% after gradient boosting, and by about 2\% after a CausalPFN outcome fit. In ACIC setting~29 with boosted nuisances, the ordinary and corrected 95\% intervals both cover the true effect in 84.5\% of draws. A common allowance for remaining bias raises coverage to 95.5\% for ordinary DML and 94.5\% for DDML at a 2\% sensitivity budget.

We also examine a setting where variable selection misses a diffuse confounding signal. Extending the high-dimensional linear design of Chetverikov, S\o rensen, and Tsyvinski~\cite{cst}, we spread a common confounding component across 400 weak controls out of 500. Post-Lasso refits the variables selected for each nuisance regression, yet many individually weak controls remain unselected and can still push the treatment coefficient in the same direction. Fitting a ridge regression to the treatment residuals combines all 500 controls into a single learned feature, and the resulting rank-one correction reduces root mean squared error by 7--22\% across the two tail strengths and two correlation levels we study. We also report the original geometric design, where post-Lasso refitting alone leaves little room for the Triple Lasso score to improve the estimate.

In these experiments, a single learned direction reduces bias and root mean squared error when the initial fits miss a systematic pattern. With strong initial fits, the estimated correction is often small relative to the standard error of the treatment estimate. We then use it to check the fit along the learned direction and report how much residual bias the conclusion can tolerate.

\subsection*{Relationship to the Literature}
\addcontentsline{toc}{subsection}{Relationship to the Literature}
Projection-based bias correction builds on the higher-order influence functions of Robins, Li, Tchetgen, and van der Vaart~\cite{robins} and Robins, Li, Mukherjee, Tchetgen Tchetgen, and van der Vaart~\cite{robins2017}, who project onto the span of prespecified basis functions and use pairs of observations to estimate the bias captured by that projection. Liu, Mukherjee, and Robins~\cite{liu-mukherjee-robins} likewise use estimates of projected bias to test whether a conventional confidence interval can attain its nominal coverage.

Related second-order corrections appear across several settings. D\'iaz, Carone, and van der Laan~\cite{diaz} develop a second-order estimator for a mean under missingness, and Shi, Wan, Chernozhukov, and Song~\cite{shi} apply successive debiasing steps to policy-value estimation in off-policy evaluation, naming the approach ``deep debiasing''. Bonhomme, Jochmans, Newey, and Weidner~\cite{bjnw} construct higher-order orthogonal moments for models with finite-dimensional nuisances, while Chetverikov, S\o rensen, and Tsyvinski~\cite{cst} introduce a second-order orthogonal score to correct regularization bias in high-dimensional linear regression.

Complementary post-hoc strategies address different failure modes of ordinary DML inference. Van der Laan, Luedtke, and Carone~\cite{vdl-calibration} apply isotonic calibration to the scalar outcome predictions and Riesz weights to achieve doubly robust inference when one nuisance converges to an incorrect limit while the other is sufficiently accurate. Zheng, Bonvini, and Guo~\cite{zheng-perturbed} instead construct Perturbed DML intervals by refitting Lasso nuisances under perturbations and taking a union of Wald intervals around estimates near the original DML fit. By contrast, our procedure targets correlated covariate-level errors across both nuisances: it centers a single interval after subtracting the estimated second-order mixed bias along learned covariate directions, and uses a two-factor sensitivity budget to account for any bias that remains.

We developed this paper independently and contemporaneously with Escanciano and Terschuur~\cite{escanciano-terschuur}. For affine moments and a finite dictionary, their Proposition~6 and our Proposition~\ref{prop:bias} give the same formula for the remaining mixed bias. Their analysis also covers nonlinear moments and characterizes the minimum variance cost of exact second-order orthogonality. As in the projection approach of Robins et al.~\cite{robins,robins2017}, their construction starts from a supplied dictionary $p(X)$ and uses training data to select or shrink directions within its span. Our focus is instead on learning the dictionary itself from first-stage residuals: a forest, ridge fit, or foundation model compresses many covariates into a single predictor of nuisance error, leaving one direction whose bias contribution we estimate on an independent sample. Alongside this targeted correction, we show how stated bounds on the two learning errors control the remaining misspecification bias outside the learned space.

The rest of the paper is organized as follows. Sections~\ref{sec:estimator} and~\ref{sec:kernels} develop the pairwise correction and describe practical working spaces. Sections~\ref{sec:inference} and~\ref{sec:sensitivity} cover variance estimation, asymptotic inference, and bias-aware sensitivity intervals. Section~\ref{sec:workflow} collects the implementation steps and rules for choosing an interval. Section~\ref{sec:empirical} compares the 401(k), ACIC, and Triple Lasso results. Section~\ref{sec:theory-connections} relates the correction to higher-order orthogonality and influence functions, and Section~\ref{sec:conclusion} discusses the implications for choosing and evaluating a correction. All proofs are collected in Appendix~\ref{app:proofs}.

\section{The pairwise correction and its bias geometry}\label{sec:estimator}
We work with an iid evaluation sample $W_1,\ldots,W_n$ of size~$n$ that is independent of a training sample with sigma-field~$\calT$. We take $n\ge 2$ for the pairwise estimator. All nuisance learners $\gh$ and $\ah$, together with the correction kernel~$\Kh$, are fitted solely on the training sample and held fixed during evaluation. Throughout Sections~\ref{sec:estimator}--\ref{sec:sensitivity}, all expectations, variances, and norms involving fitted objects are taken conditional on~$\calT$ unless stated otherwise. Section~\ref{sec:crossfitting} explains when these split-sample results extend to standard $K$-fold cross-fitting.

\subsection{Target functionals, motivating examples, and ordinary DML}\label{sec:motivating}
Let $W$ denote a full observation containing covariates $X \subset W$ and a scalar outcome~$Y$, and write $g_0(x):=\E[Y\mid X=x]$ for the true conditional mean function. Our target parameter is the expectation of a known linear functional of~$g_0$,
\begin{equation}\label{eq:target}
 g_0(x):=\E[Y\mid X=x],\qquad
 \theta_0:=\E[m(W,g_0)],
\end{equation}
where the map $h\mapsto m(W,h)$ is linear in~$h$. We assume that this functional is continuous in mean square, which guarantees the existence of a square-integrable Riesz representer $\alpha_0(X)$ satisfying
\begin{equation}\label{eq:riesz}
 \E[m(W,h)]=\E[\alpha_0(X)h(X)]
\end{equation}
for every function~$h$ in a linear domain containing the fitted objects below, under standard integrability conditions allowing expectations to pass through~$m$. We write $\ip{f}{h}:=\E[f(X)h(X)]$ for the $L^2(P_X)$ inner product and $\norm{f}_2^2:=\ip{f}{f}$ for the corresponding squared norm. Adding an integrable data-only term $r(W)$ to the score and its expectation $\E[r(W)]$ to the target leaves every bias formula below unchanged.

We use this notation for two familiar econometric targets.
\begin{itemize}\setlength{\itemsep}{3pt}\setlength{\parskip}{0pt}
 \item \textbf{Average treatment effect.} For a binary treatment~$D$ and baseline covariates~$Z$, set $W=(Y,D,Z)$, $X=(D,Z)$, and $e_0(Z):=\Pr(D=1\mid Z)$. The covariate-adjusted contrast is $m(W,g)=g(1,Z)-g(0,Z)$. Under overlap and square integrability, iterated expectations show that the inverse-propensity weight $\alpha_0(D,Z)=D/e_0(Z)-(1-D)/(1-e_0(Z))$ satisfies the Riesz identity~\eqref{eq:riesz}. Consistency and conditional exchangeability then give this contrast its causal interpretation as the average treatment effect.
 \item \textbf{Partially linear regression.} Suppose $Y=D\beta_0+g_0^{\rm pl}(X)+\epsilon$ and $D=\gamma_0(X)+\nu$, with $\E[\epsilon\mid D,X]=0$ and $\E[\nu\mid X]=0$. For any candidate value~$\beta$, define the residualized outcome $Z_\beta:=Y-\beta D$ and its conditional mean $g_\beta(X):=\E[Z_\beta\mid X]$. The Robinson orthogonal moment $\E[\{Z_\beta-g_\beta(X)\}\{D-\gamma_0(X)\}]$ vanishes at the true coefficient~$\beta_0$. Holding~$\beta$ fixed, expanding the product puts this moment into the linear-functional form \eqref{eq:target}--\eqref{eq:riesz} with $m(W,f)=-Df(X)$, Riesz representer $\alpha_0(X)=-\gamma_0(X)$, and observable addend $DZ_\beta$. Section~\ref{sec:triple-lasso} develops the high-dimensional linear specialization in detail.
\end{itemize}

Given training estimates $\gh$ and $\ah$, the ordinary DML score and estimator~\cite{dml} combine the plug-in term with a weighted outcome residual,
\begin{equation}\label{eq:db}
 s(W):=m(W,\gh)+\ah(X)\{Y-\gh(X)\},\qquad
 \thh_{\rm DB}:=\Pn s,
\end{equation}
where $\Pn f := n^{-1}\sum_{i=1}^n f(W_i)$ denotes the evaluation-sample average. Write $\delta_g:=\gh-g_0$ and $\delta_\alpha:=\ah-\alpha_0$ for the two nuisance estimation errors. Because $\E[Y-\gh(X)\mid X]=-\delta_g(X)$, the Riesz identity~\eqref{eq:riesz} makes the first-order plug-in error $\E[m(W,\delta_g)]=\ip{\alpha_0}{\delta_g}$ cancel with the first-order residual term $\E[\alpha_0(X)\{Y-\gh(X)\}]=-\ip{\alpha_0}{\delta_g}$. Only the product of the two errors survives, giving the familiar mixed-bias identity
\begin{equation}\label{eq:mixed}
 \E[s(W)]-\theta_0=-\ip{\delta_\alpha}{\delta_g}.
\end{equation}
In the ATE example, \eqref{eq:db} is the augmented inverse-probability weighted (AIPW) estimator, and \eqref{eq:mixed} shows that its bias equals minus the $L^2$ inner product of the IPW-weight error and the outcome-regression error. Ordinary DML relies on this product being smaller than the $n^{-1/2}$ sampling scale, whereas our correction directly estimates and removes the detectable portion of~\eqref{eq:mixed}.

\subsection{Turning observable moment violations into a pairwise correction}
Even though neither $g_0$ nor $\alpha_0$ is observed, the evaluation sample reveals projections of their errors along any training-chosen test function~$h(X)$. Applying the Riesz identity~\eqref{eq:riesz} and the conditional-mean identity $\E[Y-\gh(X)\mid X]=-\delta_g(X)$ gives two observable population diagnostics, a \emph{Riesz-equation (covariate-balance) violation} and a \emph{regression-residual predictability violation}:
\begin{align}\label{eq:violations}
 \mathcal R_\alpha(h)
 &:=\E[\ah(X)h(X)-m(W,h)]
   =\ip{\delta_\alpha}{h},\\
 \mathcal R_g(h)
 &:=\E[h(X)\{Y-\gh(X)\}]
   =-\ip{\delta_g}{h}.\nonumber
\end{align}
For the ATE, let $h(D,Z):=\ind\{D=1, Z\in C\}$ indicate that a unit received treatment and belongs to covariate cell~$C$. Then $\mathcal R_\alpha(h) = \E[(D/\widehat e(Z) - 1)\ind\{Z\in C\}]$ checks whether the fitted IPW weights recover the cell's population share, while $\mathcal R_g(h) = \E[\ind\{D=1,Z\in C\}(Y-\gh(1,Z))]$ checks whether the outcome regression leaves nonzero average residuals inside~$C$. Both moments can be estimated directly from the evaluation data. If we simply multiply their sample averages over the same observations, however, each observation~$i$ gets paired with itself, which introduces a diagonal covariance term that can be nonzero even when the population bias is zero. To estimate the product of the two population errors without this diagonal bias, we pair the balance violation from one observation with the outcome residual from a distinct observation.

To aggregate these violations across many directions at once, let $\Kh(x,x')$ be a training-chosen kernel whose columns lie in the domain of~$m$, and let $T$ denote its bounded integral operator,
\begin{equation}\label{eq:operator}
 (Tf)(x):=\E[\Kh(x,X)f(X)].
\end{equation}
The kernel~$\Kh$ does not need to be symmetric or positive definite. Symmetry is convenient when building stable projections, while the adjoint operator~$T^*$ enters the bias formula for a general kernel.

Using this kernel, we evaluate the Riesz discrepancy of observation~$W$ at covariate point~$x'$ as
\[
 A(W;x'):=\ah(X)\Kh(X,x')-m(W,\Kh(\cdot,x')),
\]
and pair it with the outcome residual $\varepsilon_g(W'):=Y'-\gh(X')$ from a separate observation~$W'$ to form the pairwise kernel $H(W,W'):=A(W;X')\varepsilon_g(W')$. Averaging over all ordered pairs of distinct observations yields the U-statistic bias correction and the \emph{Deeply Debiased ML (DDML)} estimator,
\begin{equation}\label{eq:pair}
 \widehat B_K:=\frac{1}{n(n-1)}\sum_{i\ne j}H(W_i,W_j),\qquad
 \thh_K:=\thh_{\rm DB}-\widehat B_K.
\end{equation}
Because the diagonal terms $i=j$ are excluded, the observations $W_i$ and $W_j$ are independent conditional on~$\calT$, making $\widehat B_K$ an unbiased estimator of the targeted component of the mixed bias.

\subsection{DDML as a post-hoc regression update}\label{sec:regression-update}
The same residuals that enter the correction in~\eqref{eq:pair} can update the outcome regression itself. To score observation~$i$, we add a kernel-weighted average of the other evaluation residuals to the initial fit. Define the \emph{leave-one-out regression update} by
\begin{equation}\label{eq:updated-regression}
 \widetilde g_{-i}(x):=\gh(x)+\frac{1}{n-1}\sum_{j\ne i}
       \Kh(x,X_j)\{Y_j-\gh(X_j)\}.
\end{equation}
The kernel carries the normalization chosen on training data. The updated regression is a function of~$x$, so in the ATE example we can evaluate it under both treatment and control for the same household.

\begin{theorem}[Exact equivalence of DDML and post-hoc corrected DML]\label{thm:regression-update}
Let $n\ge2$, and fix the training fits $\gh$, $\ah$, and $\Kh$. Suppose $m(w,\cdot)$ is linear on a domain containing $\gh$ and the kernel columns $\Kh(\cdot,X_j)$. For every evaluation sample, the DDML estimator in~\eqref{eq:pair} equals the average ordinary DML score evaluated with the updated regressions in~\eqref{eq:updated-regression} and the original Riesz weights,
\begin{equation}\label{eq:ddml-regression-score}
 \thh_K=\frac1n\sum_{i=1}^n
 \big[m(W_i,\widetilde g_{-i})
      +\ah(X_i)\{Y_i-\widetilde g_{-i}(X_i)\}\big].
\end{equation}
\end{theorem}

For a single learned direction~$f$ with training normalization $\widehat\nu_f>0$, the kernel $\Kh(x,x')=f(x)f(x')/\widehat\nu_f$ gives the update $\widetilde g_{-i}=\gh+\widehat c_{-i}f$, where
\[
 \widehat c_{-i}:=\frac{1}{(n-1)\widehat\nu_f}
    \sum_{j\ne i}f(X_j)\{Y_j-\gh(X_j)\}.
\]
Only one coefficient needs to be updated when we leave out observation~$i$. A coefficient fitted once on training data generally gives a different estimate, as the matched comparison in Section~\ref{sec:calibrated-comparison} illustrates. In~\eqref{eq:ddml-regression-score}, each updated regression is independent of its own evaluation observation conditional on training, but different scores use overlapping residuals. Sections~\ref{sec:bias-geometry} and~\ref{sec:inference} use the pairwise representation to derive the bias and variance of this post-hoc update.

\paragraph{A leave-one-out update of the Riesz weights}
DDML admits two symmetric interpretations: it can be read either as updating the outcome regression~$\gh$ while holding $\ah$ fixed (Theorem~\ref{thm:regression-update}), or as updating the Riesz weights~$\ah$ while holding $\gh$ fixed. Summing the same pairs first over the observations measuring weight imbalance gives this companion representation. Define the leave-one-out balance adjustment
\begin{align}\label{eq:updated-riesz}
 \widehat d_{-j}(x)
 &:=\frac{1}{n-1}\sum_{i\ne j}
 \{\ah(X_i)\Kh(X_i,x)-m(W_i,\Kh(\cdot,x))\},\\
 \thh_K
 &=\frac1n\sum_{j=1}^n
 \left[m(W_j,\gh)+\{\ah(X_j)-\widehat d_{-j}(X_j)\}
 \{Y_j-\gh(X_j)\}\right].\nonumber
\end{align}
Viewed this way, DDML leaves the outcome regression~$\gh$ untouched and updates the Riesz weight of each observation~$j$ by subtracting $\widehat d_{-j}(X_j)$. In population, $\E[\widehat d_{-j}(x)]=(T^*\delta_\alpha)(x)$, so an exact projection removes the projected component of~$\delta_\alpha$ from the weights. Because the leave-one-out weights share evaluation observations across different indices~$j$, the variance analysis in Section~\ref{sec:inference} accounts explicitly for their dependence.

\begin{remark}[Fast computation for finite-rank kernels]\label{rem:fast-compute}
Although evaluating a general kernel over all distinct pairs requires $O(n^2)$ operations, the finite-rank kernels used most often in practice are much faster to compute. Suppose $\Kh(x,x')=b(x)'Mb(x')$ for a feature vector $b(x)\in\R^k$, and set $Z_\alpha(W):=\ah(X)b(X)-m(W,b)$ and $Z_g(W):=b(X)\{Y-\gh(X)\}$. By summing over all pairs $(i,j)$ first and then subtracting the $n$ diagonal terms, the double sum factors into a single bilinear form of sample totals:
\[
 \widehat B_K
 =\frac{1}{n(n-1)}\left[
   \left(\sum_{i=1}^n Z_\alpha(W_i)\right)' M
   \left(\sum_{j=1}^n Z_g(W_j)\right)
   -\sum_{i=1}^n Z_\alpha(W_i)' M Z_g(W_i)
 \right].
\]
Once the features and the fitted matrix~$M$ are formed, the aggregate product costs $O(nk+k^2)$ operations. Evaluating the $n$ diagonal bilinear forms directly costs $O(nk^2)$ for a general dense~$M$, or $O(nk)$ when $M$ is diagonal. For the ATE cell kernel, each observation belongs to only a fixed number of arm-by-cell entries, so once cell memberships are assigned, computing the cell totals and subtracting the diagonal takes just $O(n+k)$ operations.
\end{remark}

\subsection{Exact conditional bias and working-model robustness}\label{sec:bias-geometry}
A central advantage of~$\thh_K$ is that its conditional bias obeys an exact finite-sample identity, without requiring $\Kh$ to approximate the identity across an infinite-dimensional function space.

\begin{proposition}[Exact remaining bias]\label{prop:bias}
Under \eqref{eq:target}--\eqref{eq:operator} and the stated integrability conditions,
\begin{equation}\label{eq:bias}
 \E[\widehat B_K]=-\ip{\delta_\alpha}{T\delta_g},\qquad
 b_K:=\E[\thh_K]-\theta_0
 =-\ip{\delta_\alpha}{(I-T)\delta_g}.
\end{equation}
Equivalently, $b_K=-\ip{(I-T^*)\delta_\alpha}{\delta_g}$.
\end{proposition}

Proposition~\ref{prop:bias} clarifies that bias removal is \emph{directional}. For bias to be negligible on the root-$n$ scale, it suffices that
\begin{equation}\label{eq:bias-small}
 \sqrt n\,\big|\ip{\delta_\alpha}{(I-T)\delta_g}\big|=o_p(1).
\end{equation}
If the operator~$T$ captures most of the outcome-regression error~$\delta_g$, in the sense that $\norm{(I-T)\delta_g}_2\le r_K\norm{\delta_g}_2$, then Cauchy--Schwarz gives $|b_K|\le r_K\norm{\delta_\alpha}_2\norm{\delta_g}_2$, strictly relaxing the standard DML product rate whenever $r_K=o_p(1)$. Requiring $\norm{I-T}_{\op}=o_p(1)$ uniformly across all functions in an infinite-dimensional $L^2$ space is impossible for any square-integrable kernel because its integral operator is compact. The bias condition~\eqref{eq:bias-small} concerns only the action of $I-T$ on the two fitted errors.

\paragraph{The working-model decomposition}
To see this geometry in familiar regression terms, let $\mathcal H_k$ be a training-chosen, $k$-dimensional \emph{working space}, such as indicators for income-by-treatment cells, leaves of a shallow regression tree, or a single learned residual feature. Let $\Pi$ denote the population $L^2(P_X)$ orthogonal projection onto~$\mathcal H_k$, which acts as the population analogue of a Frisch--Waugh--Lovell regression projection. Splitting both nuisance errors into their components inside and outside~$\mathcal H_k$,
\[
 \delta_\alpha^\perp:=(I-\Pi)\delta_\alpha,\qquad
 \delta_g^\perp:=(I-\Pi)\delta_g,
\]
orthogonality decomposes the ordinary DML bias into two pieces,
\begin{equation}\label{eq:working-bias-split}
 \E[\thh_{\rm DB}]-\theta_0
 =-\ip{\Pi\delta_\alpha}{\Pi\delta_g}
  -\ip{\delta_\alpha^\perp}{\delta_g^\perp}.
\end{equation}
When $T=\Pi$ is the exact projection onto~$\mathcal H_k$, DDML eliminates the first term completely and leaves only the inner product of the orthogonal remainders,
\begin{equation}\label{eq:projection-tail}
 b_\Pi=-\ip{\delta_\alpha^\perp}{\delta_g^\perp}.
\end{equation}
Estimating the Gram matrix or cell probabilities on training data produces an approximate projection $T=\Pi T\Pi$ whose range still lies in~$\mathcal H_k$. Its remaining bias splits into two interpretable parts, namely \emph{omitted-direction bias} outside~$\mathcal H_k$ and \emph{Gram-reproduction error} inside~$\mathcal H_k$:
\begin{equation}\label{eq:approx-projection}
 b_K=-\ip{\delta_\alpha^\perp}{\delta_g^\perp}
     -\ip{\Pi\delta_\alpha}{(I-T)\Pi\delta_g}.
\end{equation}
Neither $g_0$ nor $\alpha_0$ needs to belong to~$\mathcal H_k$, because the working space simply defines which directions of error we eliminate rather than imposing a parametric model on the data-generating process.

To make precise what DDML guarantees under global and residual error bounds, suppose the fitted nuisances satisfy global $L^2$ bounds $\norm{\delta_\alpha}_2\le\epsilon_\alpha$ and $\norm{\delta_g}_2\le\epsilon_g$, along with possibly tighter residual bounds $\norm{\delta_\alpha^\perp}_2\le\rho_\alpha\le\epsilon_\alpha$ and $\norm{\delta_g^\perp}_2\le\rho_g\le\epsilon_g$ outside~$\mathcal H_k$. Conditional on the training sample, define the error class
\begin{equation}\label{eq:working-class}
 \mathcal U_\Pi:=
 \left\{(a,v)\in L^2(P_X)^2:
 \begin{array}{ll}
 \norm{a}_2\le\epsilon_\alpha,&\norm{v}_2\le\epsilon_g,\\
 \norm{(I-\Pi)a}_2\le\rho_\alpha,&
 \norm{(I-\Pi)v}_2\le\rho_g
 \end{array}\right\}.
\end{equation}

\begin{proposition}[Sharp bias bounds over error balls]\label{prop:working-balls}
If the orthogonal complement of~$\mathcal H_k$ is nontrivial, then
\begin{equation}\label{eq:working-sup}
 \sup_{(a,v)\in\mathcal U_\Pi}
       |\ip{a}{(I-\Pi)v}|=\rho_\alpha\rho_g.
\end{equation}
In particular, under global bounds alone ($\rho_\alpha=\epsilon_\alpha$ and $\rho_g=\epsilon_g$), the supremum equals $\epsilon_\alpha\epsilon_g$. If $T=\Pi T\Pi$ and $\norm{\Pi-T}_{\op}\le\omega$, then for the actual nuisance errors,
\begin{equation}\label{eq:working-approx}
 |b_K|\le
 \norm{\delta_\alpha^\perp}_2\norm{\delta_g^\perp}_2
 +\omega\norm{\Pi\delta_\alpha}_2\norm{\Pi\delta_g}_2.
\end{equation}
\end{proposition}

The supremum in Proposition~\ref{prop:working-balls} shows an important boundary on what a projection can guarantee from global error norms alone. If we know only that $\norm{\delta_\alpha}_2\le\epsilon_\alpha$ and $\norm{\delta_g}_2\le\epsilon_g$, all of the nuisance error could in principle hide outside~$\mathcal H_k$, so the worst-case bias remains $\epsilon_\alpha\epsilon_g$. A tighter bound requires information about the errors left outside~$\mathcal H_k$. Section~\ref{sec:sensitivity} expresses their sizes relative to the outcome residuals and fitted weights. This deterministic calculation complements the structure-agnostic minimax lower bounds of Jin and Syrgkanis~\cite{jin}.

At the same time, any general kernel satisfies the operator-norm fallback
\begin{equation}\label{eq:fallback-bound}
 |b_K|\le(1+\norm{T}_{\op})\norm{\delta_\alpha}_2\norm{\delta_g}_2.
\end{equation}
A stable operator with $\norm{T}_{\op}=O_p(1)$ preserves the usual product-rate bound for bias, while valid inference also depends on the first-order and pairwise variance terms analyzed in Section~\ref{sec:inference}. Even an exact projection, however, can increase the absolute bias of a particular training fit when the two terms in~\eqref{eq:working-bias-split} happen to have opposite signs and partially cancel each other out.

\begin{example}[A correction can remove a cancellation]\label{ex:cancellation}
For orthonormal vectors $e_1$ and $e_2$, set $\delta_\alpha=t_n(e_1+e_2)$, $\delta_g=t_n(e_1-e_2)$, and $T=\Pi_{e_1}$. In ordinary DML, the two coordinates contribute equal biases of opposite signs, so their inner product cancels completely and gives $-\ip{\delta_\alpha}{\delta_g}=0$. Projecting onto~$e_1$ removes only the first contribution and leaves the second intact, producing a corrected bias of $b_K=t_n^2$. When $t_n=n^{-1/5}$, this corrected bias equals $n^{-2/5}$, which exceeds the $n^{-1/2}$ sampling scale, though the product of the two global error norms is $2n^{-2/5}$, so the usual DML product-rate condition also fails here.
\end{example}

\begin{remark}[Double robustness of consistency versus inference]
Proposition~\ref{prop:bias} immediately implies double robustness of the population moment, since $b_K=0$ whenever either $\delta_g=0$ or $\delta_\alpha=0$. Consistency follows if one nuisance is consistent, the other error stays bounded, the operator~$T$ is stable, and stochastic fluctuations vanish. Achieving doubly robust \emph{confidence intervals} when one nuisance is misspecified is harder because estimation error in the consistent nuisance and the first-order projection of~$\widehat B_K$ remain relevant~\cite{dukes}.
\end{remark}

\section{Choosing the correction kernel in practice}\label{sec:kernels}
Because the remaining bias equals $b_K=-\ip{\delta_\alpha}{(I-T)\delta_g}$, the choice of kernel~$\Kh$ governs both which error directions are removed and how much sampling variance is added. This section introduces practical kernel constructions, starting from the canonical finite-basis projection of Robins et al.~\cite{robins,robins2017} and moving to data-adaptive residual directions, covariate partitions, and smooth kernels (summarized in Table~\ref{tab:kernel-comparison}). Section~\ref{sec:learned-bias-rate} gives a bias rate for a learned direction. Section~\ref{sec:triple-lasso} shows how the finite-basis construction applies to high-dimensional linear regression with arbitrary penalized learners, extending the Triple Lasso of Chetverikov, S\o rensen, and Tsyvinski~\cite{cst}.

\subsection{Practical kernel constructions}\label{sec:kernel-constructions}

\paragraph{The canonical starting point: finite basis projections}
The classical projection approach, pioneered by Robins et al.~\cite{robins,robins2017} in the higher-order influence function literature, starts from a prespecified dictionary of basis functions $b(x)=(b_1(x),\ldots,b_k(x))'$, such as splines, polynomials, or interaction terms. Let $\Sigma:=\E[b(X)b(X)']$ be its population Gram matrix and let $M\approx\Sigma^{-1}$ be an inverse-Gram estimate computed on training covariates. Setting $\Kh(x,x')=b(x)'Mb(x')$ defines a finite-rank correction kernel whose range lies in $\mathcal H_k=\operatorname{span}\{b_1,\ldots,b_k\}$. When $M=\Sigma^{-1}$, the induced operator is the exact orthogonal projection~$\Pi$ onto~$\mathcal H_k$, with operator norm $\norm{\Pi}_{\op}=1$ and squared Hilbert--Schmidt norm $\norm{K_\Pi}_{\HS}^2=k$. As Section~\ref{sec:triple-lasso} shows, this construction gives the off-diagonal bias correction for high-dimensional regression with any first-stage penalized learner, generalizing Triple Lasso. In moderate or high dimensions, however, choosing a prespecified basis that captures where the initial ML learners err can be difficult, and enlarging~$k$ can increase both Gram estimation error and pairwise variance. This motivates the targeted and data-adaptive constructions below.

\paragraph{1. Targeted rank-one (or low-rank) residual directions}
Instead of projecting onto a wide prespecified dictionary, we can let the first-stage residuals themselves teach us which direction to correct. Conditional on a fixed first-stage fit~$\gh$, the population regression of the outcome residual $Y-\gh(X)$ on~$X$ equals $-\delta_g(X)=g_0(X)-\gh(X)$. Regressing these residuals on~$X$ supplies a direction along which to check the initial fit. The empirical examples use forests, ridge regression, and differences between CausalPFN~\cite{causalpfn} and the initial outcome predictions. The resulting predictor~$f(X)$ supplies a one-function dictionary with rank-one kernel
\begin{equation}\label{eq:rankone}
 K_f(x,x'):=\frac{f(x)f(x')}{\norm{f}_2^2}.
\end{equation}
With the population normalization and $\norm{f}_2>0$, this kernel induces the exact orthogonal projection~$\Pi_f$ onto the span of~$f$. Its squared Hilbert--Schmidt norm is~$1$, the smallest possible for a nonzero orthogonal projection. The remaining bias satisfies
\[
 |b_{\Pi_f}|=|\ip{(I-\Pi_f)\delta_\alpha}{(I-\Pi_f)\delta_g}|
 \le \norm{(I-\Pi_f)\delta_\alpha}_2\,\norm{(I-\Pi_f)\delta_g}_2.
\]
Consequently, the correction removes the entire mixed bias if either error is proportional to~$f$, and approximate proportionality controls the absolute remainder through this bound, though the fraction removed also depends on any cancellation in the original bias (Example~\ref{ex:cancellation}). Estimating $\norm{f}_2^2$ on training data (or on a separate unlabeled calibration split) produces an approximate projection while keeping the evaluation sample independent for inference. When the feature or its calibration second moment is zero, we use $K_f=0$ and recover ordinary DML. Section~\ref{sec:learned-bias-rate} quantifies the error from learning and normalizing the direction. With several learned directions, the same construction uses their Gram matrix.

Because the two-way split separates all training operations from evaluation, Propositions~\ref{prop:bias} and~\ref{prop:variance} condition on every step used to build the nuisances and kernel. To learn the error of a specific initial fit~$\gh$, one can freeze that fit, form residual labels on fresh training observations, and regress those labels on~$X$. When the residual labels instead come from out-of-fold predictions within training, each fold targets the error of its own fit, yet the exact bias identity still applies to the learned direction and the final refit used in evaluation. Letting $\bar g$ denote the average of the label-generating fits weighted by their held-out fold sizes, taking $c=1$ and applying the triangle inequality yields
\[
 \inf_c\norm{(g_0-\gh)-cf}_2
 \le \norm{f-(g_0-\bar g)}_2+\norm{\gh-\bar g}_2 .
\]
Here the first term measures how well the learned feature predicts the error of the fold-average fit, while transferring that rate to the final refit~$\gh$ also requires controlling the stability gap $\norm{\gh-\bar g}_2$.

\paragraph{2. Covariate partitions and residual-fitted regression trees}
A direct nonparametric alternative partitions the covariate space into $k$ disjoint cells~$\mathcal C$ of positive population probability (crossed with treatment arm in an ATE problem) using training data. These cells can come from domain-chosen quantiles, as in our 401(k) partition analysis, or from a single shallow regression tree fitted to standardized training residuals of~$Y$ and~$D$, as in our ACIC partition experiments. The population cell-projection kernel is
\begin{equation}\label{eq:forest}
 K_{\mathcal C}(x,x'):=\sum_{C\in\mathcal C}
 \frac{\ind\{x\in C\}\ind\{x'\in C\}}{P_X(C)}.
\end{equation}
Its integral operator averages functions within each cell, has operator norm~$1$, and has squared Hilbert--Schmidt norm $\norm{K_{\mathcal C}}_{\HS}^2:=\E[K_{\mathcal C}(X,X')^2]=k$. Replacing $P_X(C)$ with training cell frequencies~$\widehat p_C$ (stabilized by minimum leaf sizes and optional pseudo-counts) allows the correction to be computed directly from cell totals in $O(n+k)$ time (Remark~\ref{rem:fast-compute}). Because rare cells inflate $\widehat p_C^{-1}$ and introduce Gram-reproduction error, enforcing a minimum leaf size is important in practice. Averaging kernels across multiple trees also yields a positive contraction whenever population normalizations are used.

\paragraph{3. Local smoothers and RKHS ridge kernels}
When smooth nonlinear errors are expected without hard cell boundaries, we can smooth across neighboring covariate values. A local kernel smoother uses
\begin{equation}\label{eq:local-kernel}
 K_h(x,x'):=\frac{\kappa_h(x-x')}{\E[\kappa_h(x-X)]},
\end{equation}
with its denominator estimated on training covariates (cf.~\cite{diaz} for second-order targeted estimation). Alternatively, for a reproducing kernel Hilbert space (RKHS)~$\mathcal H$ with bounded kernel~$k$, feature map $k_x:=k(x,\cdot)$, covariance operator $C:=\E[k_X\otimes k_X]$, and ridge penalty $\lambda>0$, the equivalent kernel is
\begin{equation}\label{eq:rkhs-kernel}
 K_\lambda(x,x'):=\ip{k_x}{(C+\lambda I)^{-1}k_{x'}}_{\mathcal H}.
\end{equation}
If $T_0$ denotes the integral operator of~$k$ on $L^2(P_X)$ with eigenpairs $(\mu_\ell,e_\ell)$, then $T_\lambda$ shrinks each eigendirection smoothly according to
\begin{equation}\label{eq:rkhs}
 T_\lambda:=T_0(T_0+\lambda I)^{-1},\qquad
 T_\lambda e_\ell=\frac{\mu_\ell}{\mu_\ell+\lambda}e_\ell,
\end{equation}
with effective degrees of freedom $\norm{K_\lambda}_{\HS}^2=\sum_\ell\{\mu_\ell/(\mu_\ell+\lambda)\}^2$. More generally, for any self-adjoint square-integrable kernel with a complete orthonormal eigenbasis (including the null space), eigenpairs $(s_\ell,e_\ell)$ and error projections $a_\ell:=\ip{\delta_\alpha}{e_\ell}$ and $b_\ell:=\ip{\delta_g}{e_\ell}$, the spectral expansion
\begin{equation}\label{eq:spectral}
 b_K=-\sum_\ell(1-s_\ell)a_\ell b_\ell,\qquad
 \norm{\Kh}_{\HS}^2=\sum_\ell s_\ell^2
\end{equation}
displays the bias--variance trade-off across spectral directions. If both nuisance errors belong to~$\mathcal H$, spectral calculus gives the explicit bias bound
\begin{equation}\label{eq:rkhs-error-bound}
 |b_{K_\lambda}|=|\ip{\delta_\alpha}{(I-T_\lambda)\delta_g}|
 \le\lambda\norm{\delta_\alpha}_{\mathcal H}\norm{\delta_g}_{\mathcal H}.
\end{equation}
Reducing~$\lambda$ sharpens this bias bound at the cost of raising $\norm{K_\lambda}_{\HS}^2$ and estimation noise in~$\widehat C$. At fixed~$\lambda$, the remaining bias is still quadratic.

Finally, any kernel~$\Kh$ can also be damped toward ordinary DML by replacing~$\Kh$ with $\rho\Kh$ for a training-chosen factor $\rho\in[0,1]$, which scales the pairwise variance by~$\rho^2$ while leaving a fraction $1-\rho$ of the represented mixed bias (recovering ordinary DML at $\rho=0$).

\begin{table}[htbp]
\centering\small
\caption{Practical comparison of correction kernels for applied work}
\label{tab:kernel-comparison}
\begin{tabularx}{\textwidth}{@{}>{\raggedright\arraybackslash}p{0.19\textwidth}>{\raggedright\arraybackslash}p{0.24\textwidth}>{\raggedright\arraybackslash}p{0.18\textwidth}>{\raggedright\arraybackslash}p{0.15\textwidth}>{\raggedright\arraybackslash}X@{}}
\toprule
Kernel family & Directions corrected & Choice on training data & Evaluation cost & Example \\
\midrule
Basis projection / High-dim.\ regression & Span of $b(X)$ (splines, interactions, covariates) & Basis dimension $k$ or sparsity & $O(nk^2)$ for dense $M$ & Canonical HOIF~\cite{robins}, partially linear models. \\
Targeted rank-one \eqref{eq:rankone} & Single learned residual predictor $f(X)$ & Learner for $f$ (training only) & $O(n)$ & One learned residual direction with many covariates. \\
Cell partition \eqref{eq:forest} & Within-cell / leaf averages (crossed with $D$) & Leaf count $k$ and min leaf size & $O(n+k)$ for cell totals & ATE with quantile cells or residual-tree leaves. \\
Local smoother \eqref{eq:local-kernel} & Smooth local covariate neighborhoods & Bandwidth $h$ & $O(n^2)$ & Low-dimensional continuous $X$ with smooth nonlinear errors. \\
RKHS ridge \eqref{eq:rkhs-kernel} & Leading spectral modes of kernel $k$ & Ridge penalty $\lambda$ and kernel width & $O(n^2)$ kernel evaluations & Smooth nonparametric corrections without hard cell boundaries. \\
\bottomrule
\end{tabularx}
\end{table}

\subsection{Bias-rate conditions for a learned direction}\label{sec:learned-bias-rate}
As the initial regression improves, even the zero predictor achieves vanishing $L^2$ error when predicting the mean residual, so what matters for bias correction is how much of the initial regression error survives projection onto the learned feature. Given a training-chosen feature~$f$ with $\nu_f:=\E[f(X)^2]>0$ and a positive training estimate $\widehat\nu_f$, define
\[
 \Delta_f:=\inf_{a\in\R}\norm{\delta_g-af}_2
          =\norm{(I-\Pi_f)\delta_g}_2,
 \qquad \omega_f:=\left|\nu_f/\widehat\nu_f-1\right|.
\]
Here $\Delta_f$ measures how closely $f$ captures the shape of the regression error up to scale and sign, while $\omega_f$ measures the relative error in the inverse second moment used by the kernel.

\begin{corollary}[Remaining bias after rank-one correction]\label{cor:learned-bias-rate}
Under the conditions of Proposition~\ref{prop:bias}, the kernel $K_f(x,x')=f(x)f(x')/\widehat\nu_f$ satisfies
\begin{align}
 |b_{K_f}|&\le\norm{(I-\Pi_f)\delta_\alpha}_2\,\Delta_f
       +\omega_f\norm{\Pi_f\delta_\alpha}_2\norm{\Pi_f\delta_g}_2\notag\\
 &\le\norm{\delta_\alpha}_2\bigl(\Delta_f+\omega_f\norm{\delta_g}_2\bigr).
 \label{eq:learned-bias-rate}
\end{align}
If either upper bound is $o_p(n^{-1/2})$, the oracle expansion of Proposition~\ref{prop:ordinary} follows under its score and sampling-variation conditions.
\end{corollary}

While the first bound in~\eqref{eq:learned-bias-rate} can be small whenever $f$ captures either nuisance error, the second gives a direct criterion for a learner trained solely on outcome residuals. Freeze the initial nuisance fits and regress $Y-\gh(X)$ by least squares on a fixed dictionary of $p$ features $\xi_p(X)$ using $m$ fresh training observations, where the regression target is $v:=g_0-\gh=-\delta_g$. Let $a_p$ be a deterministic upper bound, holding with probability tending to one, on the approximation error $\inf_\beta\norm{v-\xi_p'\beta}_2$, where all $p$ coefficients may be nonzero. We then use $\ell$ further independent covariate observations to compute $\widehat\nu_f:=\ell^{-1}\sum_{j=1}^\ell f(X_j)^2$, reserve $n$ observations for evaluation, and let $m,\ell,n\to\infty$.

Suppose the features in $\xi_p(X)$ are uniformly bounded, their population Gram matrix has eigenvalues bounded above and away from zero, and $p=o(\sqrt m)$, with $\E[(Y-\gh(X))^2\mid\gh,\ah]=O_p(1)$. To calibrate a nonzero~$f$, suppose further that $\E[f(X)^4]/\nu_f^2=O_p(1)$ given the fitted feature (independent symmetric $\{-1,1\}$ features, for example, have identity Gram matrix and a fourth-moment ratio of at most~$3$ for every nonzero linear combination). If $\norm{\delta_j}_2=O_p(r_j)$ for $j\in\{g,\alpha\}$, then
\begin{equation}\label{eq:learned-series-rate}
 |b_{K_f}|=O_p\!\left(r_\alpha
       \left[a_p+\sqrt{p/m}+r_g/\sqrt\ell\right]\right).
\end{equation}
The first two terms in brackets capture the cost of approximating and fitting the residual regression, whereas the final term reflects calibrating its one-dimensional projection.

To illustrate the gain, take $m,\ell\asymp n$ and $r_g=r_\alpha=n^{-1/5}$. When both errors equal $n^{-1/5}u$ for a unit function~$u$, ordinary DML carries an absolute bias of $n^{-2/5}$. In the symmetric sign design, $u=p^{-1/2}\sum_{j=1}^p \xi_{pj}$ puts weight on every feature and gives $a_p=0$. More generally, whenever $a_p=o(n^{-3/10})$ and $p=o(n^{2/5})$, \eqref{eq:learned-series-rate} yields $|b_{K_f}|=o_p(n^{-1/2})$. In this way, $p$ can grow at the residual-learning stage while the evaluation sample estimates only a single bias contribution.

Ordinary DML can also attain negligible bias in this example by updating the regression on the training sample: for $\widetilde g:=\gh+f$, the regression error is $f-v$, so~\eqref{eq:mixed} bounds the absolute bias by $\norm{\delta_\alpha}_2\norm{f-v}_2=O_p\{r_\alpha(a_p+\sqrt{p/m})\}$. Whereas this training update depends on the fitted magnitude of~$f$, the exact rank-one projection depends only on its direction because $\Pi_f$ is invariant to nonzero rescaling, which matters when~$f$ is regularized or learned on an arbitrary scale. Section~\ref{sec:calibrated-comparison} compares training and evaluation updates along the same learned direction.

\subsection{Special case: High-dimensional regression}\label{sec:triple-lasso}
The finite-basis construction applies directly to high-dimensional partially linear regression with any first-stage linear learner. Consider the model
\begin{equation}\label{eq:tl-model}
 Y=D\beta_0+X'\vartheta_0+\epsilon,\qquad
 \E[\epsilon\mid D,X]=0,\qquad
 D=X'\gamma_0+\nu,\quad\E[X\nu]=0,
\end{equation}
and write $\varphi_0:=\vartheta_0+\beta_0\gamma_0$ for the reduced-form outcome coefficients. On the training sample, estimate $\gamma_0$ and $\varphi_0$ using any penalized or post-regularization linear learner, such as ridge regression~\cite{hoerl-kennard}, Lasso~\cite{tibshirani}, post-Lasso~\cite{belloni-chernozhukov}, or lava and post-lava~\cite{lava}, and form the first-stage residuals
\[
 v:=D-X'\widehat\gamma,\qquad
 u:=Y-X'\widehat\varphi,\qquad a_\beta:=u-\beta v.
\]
The ordinary DML (Robinson) score is the sample average $\Pn[a_\beta v]$. Given a training estimate $\widehat\Theta$ of the precision matrix $\Sigma^{-1}=\E[XX']^{-1}$, the off-diagonal U-statistic correction is
\begin{equation}\label{eq:tl-U}
 \widehat M_U(\beta):=\Pn[a_\beta v]
 -\frac1{n(n-1)}\sum_{i\ne j}a_{\beta,i}X_i'\widehat\Theta X_jv_j.
\end{equation}
To see that~\eqref{eq:tl-U} is an exact instance of~\eqref{eq:pair}, set $Z_\beta:=Y-\beta D$, restrict attention to the linear span of~$X$, define
\[
 \gh_\beta(x):=x'(\widehat\varphi-\beta\widehat\gamma),\qquad
 m(W,f):=-Df(X),\qquad \ah(x):=-x'\widehat\gamma,
\]
and add $DZ_\beta$ to the score and $\E[DZ_\beta]$ to the target. Under this substitution, the ordinary DML score simplifies to $DZ_\beta-D\gh_\beta(X)+\ah(X)\{Z_\beta-\gh_\beta(X)\}=v a_\beta$, and its Riesz discrepancy along any linear function~$f$ equals $vf(X)$. Choosing the kernel $\Kh(x,\widetilde x)=x^\top\widehat\Theta^\top\widetilde x$ and swapping the summation indices reproduces~\eqref{eq:tl-U}, where we retain the transpose because sparsified inverse-Gram estimates need not be symmetric. The linear projection suffices here because $\E[X\{Z_\beta-X'(\varphi_0-\beta\gamma_0)\}]=0$ and $\E[X(D-X'\gamma_0)]=0$. These are the only residual moments tested by the linear kernel.

Writing $e_\gamma:=\widehat\gamma-\gamma_0$ and $e_\varphi:=\widehat\varphi-\varphi_0-\beta_0e_\gamma$ for the two nuisance errors, evaluating the expected score at~$\beta_0$ gives
\begin{equation}\label{eq:tl-bias}
 \E[\widehat M_U(\beta_0)]
 =e_\varphi'(\Sigma-\Sigma\widehat\Theta\Sigma)e_\gamma,
 \qquad \Sigma:=\E[XX'],
\end{equation}
which is Proposition~\ref{prop:bias} in matrix form. When the precision matrix is exact ($\widehat\Theta=\Sigma^{-1}$), the matrix factor $\Sigma-\Sigma\widehat\Theta\Sigma$ vanishes and the corrected score has mean zero at~$\beta_0$, regardless of whether $(\widehat\gamma,\widehat\varphi)$ came from ridge, Lasso, post-Lasso, lava, or post-lava. This extends the Triple Lasso of Chetverikov, S\o rensen, and Tsyvinski~\cite{cst}, who focused specifically on first-stage Lasso fits.

\paragraph{U-statistic versus V-statistic implementation}
Chetverikov, S\o rensen, and Tsyvinski~\cite{cst} implement Triple Lasso as a product of sample averages (a V-statistic),
\begin{equation}\label{eq:tl-V}
 \widehat M_V(\beta)=\Pn[a_\beta v]
 -\Pn[a_\beta X']\widehat\Theta\Pn[Xv],
\end{equation}
which includes the diagonal terms $i=j$. For any pair array $F_{ij}$, the V-statistic and U-statistic averages are linked by the exact identity
\begin{equation}\label{eq:UV}
 \frac1{n^2}\sum_{i,j}F_{ij}
 =\frac{n-1}{n}\frac1{n(n-1)}\sum_{i\ne j}F_{ij}
   +\frac1{n^2}\sum_iF_{ii}.
\end{equation}
Excluding the diagonal in~\eqref{eq:tl-U} eliminates own-observation bias and lets us apply the unbiased variance formulas of Section~\ref{sec:inference} directly, whereas for~\eqref{eq:tl-V} the diagonal term must be tracked separately as the covariate dimension grows. For inference on the root $\widehat\beta$ of $\widehat M_U(\widehat\beta)=0$, the usual local linearization divides the score variance at~$\beta_0$ by the squared population Jacobian and requires that Jacobian to be nonzero. A mean-zero score need not have an unbiased root.

\section{Sampling variation, standard errors, and cross-fitting}\label{sec:inference}
Subtracting the pairwise correction~$\widehat B_K$ alters the conditional mean of the estimator and can also alter its sampling variance. This section decomposes that sampling variation into its two exact components, builds a finite-sample unbiased estimator of the full conditional variance, states when ordinary DML standard errors and $K$-fold cross-fitting remain valid, and warns against a double-counting pitfall with jackknife variance estimators.

\subsection{Two sources of sampling variation: the Hoeffding decomposition}\label{sec:hoeffding}
Because the off-diagonal sum in $\widehat B_K$ runs over all ordered pairs $i\ne j$, we can equivalently average the symmetrized pairwise kernel over unordered pairs $i<j$. Symmetrizing $H(w,w')$ from~\eqref{eq:pair} gives
\begin{equation}\label{eq:sym}
 h(w,w'):=\tfrac12\{H(w,w')+H(w',w)\},\qquad
 \mu_h:=\E[h(W,W')],
\end{equation}
so that $\widehat B_K=\mathbb{U}_n h$, where the \emph{U-average} $\mathbb{U}_n h:=\binom{n}{2}^{-1}\sum_{i<j}h(W_i,W_j)$ averages the symmetric kernel over all unordered pairs of evaluation observations. By Hoeffding's orthogonal decomposition~\cite{hoeffding}, any symmetric pairwise average splits into an individual-observation (linear) average plus a purely pairwise (degenerate) remainder. Define
\begin{align}\label{eq:hoeffding-components}
 L(w)&:=2\{\E[h(w,W')]-\mu_h\},\\
 d(w,w')&:=h(w,w')-\mu_h-\tfrac12\{L(w)+L(w')\},\nonumber
\end{align}
where $\E[L(W)]=0$ and $\E[d(W,W')\mid W]=0$. The linear term $L(w)$ measures how much observation~$w$ shifts the pairwise correction on average across all independent partners~$W'$, whereas $d(w,w')$ is the pure pairwise interaction left after removing those single-observation effects. Combining the linear term with the original DML score defines the \emph{effective influence function} of the corrected estimator,
\begin{equation}\label{eq:effective-if}
 \phi_K(w):=s(w)-\E[s(W)]-L(w).
\end{equation}

\begin{proposition}[Exact conditional expansion and variance]\label{prop:variance}
If $s\in L^2(P)$ and $h\in L^2(P\otimes P)$, then
\begin{align}\label{eq:exact-expansion}
 \thh_K-\theta_0&=b_K+\Pn\phi_K-\mathbb{U}_n d,\\
 \tau_n^2:=\Var(\thh_K\mid\calT)
 &=\frac{\sigma_{1,n}^2}{n}
   +\frac{2\sigma_{2,n}^2}{n(n-1)},\qquad
 \sigma_{1,n}^2:=\E[\phi_K^2],\quad
 \sigma_{2,n}^2:=\E[d^2].\label{eq:full-variance}
\end{align}
In particular, $\sigma_{1,n}^2=\Var(s)+\Var(L)-2\Cov(s,L)$.
\end{proposition}

The covariance term $-2\Cov(s,L)$ in Proposition~\ref{prop:variance} explains why simply adding the variance of the correction~$\widehat B_K$ to the ordinary DML variance gives the wrong answer. The first-order standard error falls exactly when $2\Cov(s,L)>\Var(L)$. The estimated first-order standard error falls in the 401(k) forest fits with two or four income cells (Table~\ref{tab:emp-var}). To see how the nuisance errors enter~$L$, write $f:=T\delta_g$ and $a:=T^*\delta_\alpha$. Integrating out the second observation in~$H$ yields
\begin{equation}\label{eq:L-explicit}
 L(W)=m(W,f)-\ah(X)f(X)
       +a(X)\{Y-\gh(X)\}
       +2\ip{\delta_\alpha}{T\delta_g}.
\end{equation}
Hence $L$ scales directly with the projected nuisance errors $(f,a)$ and the moments of the score and outcome residual. The continuity conditions following Proposition~\ref{prop:ordinary} below show how this formula bounds $\norm{L}_2$.

\subsection{When DDML attains the \texorpdfstring{root-$n$}{root-n} oracle limit}\label{sec:ddml-oracle}
Let $\psi_0(W):=m(W,g_0)+\alpha_0(X)\{Y-g_0(X)\}$ denote the oracle influence score and set $\phi_0:=\psi_0-\theta_0$. A central question for applied work is when the DDML estimator~$\thh_K$ is root-$n$ asymptotically normal with this oracle influence function~$\phi_0$, so that standard DML confidence intervals apply directly to~$\thh_K$ without adding a pairwise variance term. Proposition~\ref{prop:ordinary} covers two distinct cases at once. First, when ordinary DML fails because the global product $\norm{\delta_\alpha}_2\norm{\delta_g}_2$ is too large, DDML can still achieve the root-$n$ oracle limit as long as the kernel removes enough correlated error to make $b_K=o_p(n^{-1/2})$ while keeping $L$ and the pairwise term negligible. Second, when ordinary DML meets its usual product-rate condition, operator stability controls the remaining bias. The same oracle limit follows if the linear and pairwise fluctuations of the correction are negligible.

\begin{proposition}[Root-$n$ oracle inference for DDML]\label{prop:ordinary}
Suppose $0<\E[\phi_0^2]<\infty$ and
\begin{equation}\label{eq:ordinary-conditions}
 \norm{s-\psi_0}_{L^2(P)}=o_p(1),\qquad
 \norm{L}_{L^2(P)}=o_p(1),\qquad
 \E[d^2]=o_p(n),\qquad b_K=o_p(n^{-1/2}).
\end{equation}
Then the DDML estimator satisfies
\begin{equation}\label{eq:ordinary-limit}
 \sqrt n(\thh_K-\theta_0)
 =\frac1{\sqrt n}\sum_{i=1}^n\phi_0(W_i)+o_p(1)
 \ad N(0,\E[\phi_0^2]).
\end{equation}
The bias condition $b_K=o_p(n^{-1/2})$ holds either from directional bias removal~\eqref{eq:bias-small}, as illustrated by the residual-regression rate~\eqref{eq:learned-series-rate}, or from $\norm{T}_{\op}=O_p(1)$ combined with the standard DML product rate $\norm{\delta_\alpha}_2\norm{\delta_g}_2=o_p(n^{-1/2})$.
\end{proposition}

\paragraph{Allowable complexity of the DDML kernel}
The second moment of the pair kernel determines how large the working-space dimension~$k$ can be under~\eqref{eq:ordinary-conditions}. Suppose that with probability tending to one, $\norm{\ah}_\infty\le C$, $\E[\varepsilon_g^2\mid X]\le C$, and the functional is $L^2$-continuous,
\begin{equation}\label{eq:m-continuity}
 \E[m(W,f)^2]\le C\norm{f}_2^2,
\end{equation}
a condition satisfied by the ATE under uniform overlap, though it can fail for derivative functionals. If both nuisance learners are $L^2$-consistent and $\norm{T}_{\op}=O_p(1)$, then \eqref{eq:L-explicit} immediately implies $\norm{L}_2=o_p(1)$. Conditioning on~$X'$ bounds the pairwise second moment by the squared Hilbert--Schmidt norm of the kernel,
\begin{equation}\label{eq:hs}
 \E[d^2]\le\E[h^2]\le\E[H^2]\lesssim \norm{\Kh}_{\HS}^2:=\E[\Kh(X,X')^2].
\end{equation}
For a rank-$k$ orthogonal projection, $\norm{T}_{\op}=1$ and $\norm{\Kh}_{\HS}^2=k$. Because the degenerate term averages over $\binom{n}{2}\asymp n^2$ uncorrelated pairs, its conditional variance is $O_p(k/n^2)$ and its standard deviation is $O_p(\sqrt{k}/n)$. Choosing $k=o(n)$ makes DDML's pairwise noise asymptotically negligible relative to the $n^{-1/2}$ sampling scale. When $\sigma_{1,n}^2\le C_1$ and $\sigma_{2,n}^2\le C_2 k$, combining~\eqref{eq:full-variance} with Proposition~\ref{prop:working-balls} gives the conditional root mean squared error bound
\begin{equation}\label{eq:working-risk}
 \left\{\E[(\thh_\Pi-\theta_0)^2\mid\calT]\right\}^{1/2}
 \le
 \rho_\alpha\rho_g+
 \sqrt{\frac{C_1}{n}+\frac{2C_2k}{n(n-1)}}.
\end{equation}
Under these variance bounds and global error radii alone, $k\lesssim n$ gives a root mean squared error of at most order $\epsilon_\alpha\epsilon_g+n^{-1/2}$. Whenever $k=o(n)$ and the residual error product outside the working space satisfies $\rho_\alpha\rho_g=o(n^{-1/2})$, both the remaining bias and the pairwise noise vanish at the $n^{-1/2}$ scale, so the score and linear-term conditions in~\eqref{eq:ordinary-conditions} give the root-$n$ oracle limit even when ordinary DML fails. When $k$ is large enough (or nuisance errors are large enough) that $L$ or the pairwise term~$\mathbb{U}_n d$ is no longer negligible, the full variance estimator permits Gaussian inference under the variance-consistency and central limit conditions in Sections~\ref{sec:vhat} and~\ref{sec:general-clt}.

\subsection{Computing the full standard error in practice}\label{sec:vhat}
To estimate the first-order and pairwise variance components together, we combine the ordinary DML score and the correction in a single symmetric kernel,
\begin{equation}\label{eq:q}
 q(w,w'):=\tfrac12\{s(w)+s(w')\}-h(w,w'),\qquad \thh_K=\mathbb{U}_n q.
\end{equation}
Writing $\thh_K$ as a single U-statistic $\mathbb{U}_n q$ means its Hoeffding variance automatically incorporates $\Var(s)$, $\Var(L)$, $-2\Cov(s,L)$, and $\Var(\mathbb{U}_n d)$. Set $q_{ij}:=q(W_i,W_j)$ and define the three population moments
\begin{equation}\label{eq:ABC}
 A:=\E[q_{12}^2],\qquad B:=\E[q_{12}q_{13}],\qquad
 C:=\E[q_{12}q_{34}]=(\E q_{12})^2,
\end{equation}
which satisfy $\sigma_{1,n}^2=4(B-C)$ and $\sigma_{2,n}^2=A-2B+C$. The \emph{falling factorial} $(n)_r:=n(n-1)\cdots(n-r+1)$ counts the ordered $r$-tuples of distinct evaluation indices. For $n\ge 4$, averaging over the corresponding pairs, triples, and quadruples gives
\begin{equation}\label{eq:ABC-est}
 \widehat A:=\frac{\sum_{i\ne j}q_{ij}^2}{(n)_2},\qquad
 \widehat B:=\frac{\sum_{i,j,k\ {\rm distinct}}q_{ij}q_{ik}}{(n)_3},\qquad
 \widehat C:=\frac{\sum_{i,j,k,l\ {\rm distinct}}q_{ij}q_{kl}}{(n)_4}.
\end{equation}

\begin{proposition}[An unbiased estimator of the full variance]\label{prop:vhat}
Under the second-moment conditions of Proposition~\ref{prop:variance},
\begin{equation}\label{eq:vhat}
 \widehat V
 :=\underbrace{\frac4n(\widehat B-\widehat C)}_{\widehat V_1\text{ (first-order)}}
  +\underbrace{\frac{2}{n(n-1)}(\widehat A-2\widehat B+\widehat C)}_{\widehat V_2\text{ (pairwise)}}
\end{equation}
satisfies $\E[\widehat V\mid\calT]=\tau_n^2$.
\end{proposition}

This construction specializes unbiased U-statistic variance estimation~\cite{wanglindsay}. Evaluating~\eqref{eq:ABC-est} never requires looping over triples or quadruples, because row sums and sums of squares suffice. Define
\[
 r_i:=\sum_{j\ne i}q_{ij},\qquad
 S:=\sum_i r_i,\qquad Q:=\sum_{i\ne j}q_{ij}^2,
 \qquad R:=\sum_i r_i^2.
\]
Separating equal from distinct indices reduces~\eqref{eq:ABC-est} to $O(n^2)$ operations,
\begin{equation}\label{eq:fast-var}
 \widehat A=\frac{Q}{(n)_2},\qquad
 \widehat B=\frac{R-Q}{(n)_3},\qquad
 \widehat C=\frac{S^2-4R+2Q}{(n)_4}.
\end{equation}
Centering each off-diagonal entry~$q_{ij}$ by subtracting~$\mathbb{U}_n q$ leaves~\eqref{eq:vhat} unchanged and improves numerical stability. For a rank-$k$ kernel, computing all row sums from the feature totals takes $O(nk)$ operations with diagonal $M$ and $O(nk^2)$ with a general dense $M$ (Remark~\ref{rem:fast-compute}). When $\widehat V>0$, the ratio $\widehat V_2/\widehat V$ estimates the pairwise share.

Asymptotic validity of studentized inference requires ratio consistency $\widehat V/\tau_n^2\to_p 1$, for which it suffices that
\begin{equation}\label{eq:v-consistency}
 |\widehat A-A|=o_p(n^2\tau_n^2),\qquad
 |\widehat B-B|+|\widehat C-C|=o_p(n\tau_n^2).
\end{equation}
Because $\widehat V$ is an unbiased linear combination of moment estimates rather than a sum of squares, it can occasionally turn nonpositive in finite samples when kernel complexity is high relative to~$n$ (this occurs in 2 of 1,200 ACIC draws in Appendix~\ref{app:acic}). A nonpositive $\widehat V$ is a warning to inspect kernel complexity and rare cells rather than blindly truncating to zero.

\paragraph{Row-projection shortcut and a double-counting pitfall}
When pairwise noise is negligible ($\E[d^2]/n=o_p(1)$ with $\sigma_{1,n}^2$ bounded away from zero), the empirical row projections (the U-statistic jackknife) give a nonnegative estimator of the first-order variance,
\begin{equation}\label{eq:row-var}
 \widehat\phi_i:=2\left\{\frac1{n-1}\sum_{j\ne i}q_{ij}-\mathbb{U}_n q\right\},
 \qquad \widehat V_{\rm lin}:=\frac1{n^2}\sum_i\widehat\phi_i^2.
\end{equation}
Under uniform integrability of~$\phi_K^2$ and negligible pairwise noise, $\widehat V_{\rm lin}$ consistently estimates the effective first-order variance, capturing both $s$ and~$L$. Adding a separate estimate of pairwise variance on top of~$\widehat V_{\rm lin}$, however, double counts pairwise variation because the empirical rows~$\widehat\phi_i$ already absorb it. In a purely degenerate case ($q=d_q$ with mean zero), direct algebra shows that
\begin{equation}\label{eq:jackknife-inflation}
 \E[\widehat V_{\rm lin}\mid\calT]
 =\frac{4(n-2)}{n^2(n-1)}\E[d_q^2],\qquad
 \Var(\mathbb{U}_n q\mid\calT)=\frac{2}{n(n-1)}\E[d_q^2],
\end{equation}
so $\widehat V_{\rm lin}$ already overestimates the degenerate variance by a factor approaching~$2$. Whenever pairwise noise is non-negligible, one should use the exact unbiased decomposition~\eqref{eq:vhat} instead.

\subsection{General inference regimes}\label{sec:general-clt}
When the pairwise term contributes appreciably to~$\tau_n^2$, valid Gaussian inference requires a joint central limit theorem covering both the linear and degenerate parts. To express $\thh_K-\E[\thh_K\mid\calT]$ as a sum of martingale differences, let $\mathcal F_i:=\sigma(\calT,W_1,\ldots,W_i)$ and define
\begin{equation}\label{eq:martingale}
 \Delta_i:=\frac{\phi_K(W_i)}n
       -\frac{2}{n(n-1)}\sum_{j<i}d(W_i,W_j).
\end{equation}
Conditional on~$\mathcal F_{i-1}$, both $\phi_K(W_i)$ and each degenerate term $d(W_i,W_j)$ for $j<i$ have mean zero, so $\E[\Delta_i\mid\mathcal F_{i-1}]=0$ and $\sum_{i=1}^n\Delta_i=\thh_K-\E[\thh_K\mid\calT]$.

\begin{proposition}[Inference allowing additional variance]\label{prop:general-clt}
Suppose $\tau_n>0$ with probability tending to one, and
\begin{align}\label{eq:martingale-conditions}
 \tau_n^{-2}\sum_i\E[\Delta_i^2\mid\mathcal F_{i-1}]&\to_p1,\\
 \tau_n^{-2}\sum_i\E[\Delta_i^2\ind\{|\Delta_i|>\epsilon\tau_n\}
                         \mid\mathcal F_{i-1}]&\to_p0
 \quad\text{for every }\epsilon>0.\nonumber
\end{align}
If $\widehat V/\tau_n^2\to_p1$, then the centered estimator satisfies
\begin{equation}\label{eq:studentized-centered}
 \frac{\thh_K-\E[\thh_K\mid\calT]}{\sqrt{\widehat V}}\ad N(0,1).
\end{equation}
If, in addition, $b_K=o_p(\tau_n)$, then
\begin{equation}\label{eq:studentized}
 \frac{\thh_K-\theta_0}{\sqrt{\widehat V}}\ad N(0,1).
\end{equation}
When $n\tau_n^2$ is bounded above and away from zero in probability, the resulting confidence intervals have root-$n$ width, and otherwise their width scales with~$\tau_n$.
\end{proposition}

In practice, the conditional variance condition~\eqref{eq:martingale-conditions} holds when the linear component~$\Pn\phi_K$ dominates or when the degenerate pairwise component~$\mathbb{U}_n d$ has growing effective rank (as in Hall~\cite{hall}), whereas a fixed-rank degenerate component has a non-Gaussian weighted chi-squared limit if it dominates the linear part. Because $\E[\thh_K\mid\calT]=\theta_0+b_K$, the law of total variance separates variation across evaluation samples from variation in the conditional bias across training samples,
\begin{equation}\label{eq:training-var}
 \Var(\thh_K)=\E[\tau_n^2]+\Var(b_K).
\end{equation}
Because $\widehat V$ estimates the conditional evaluation variance~$\tau_n^2$, the rate condition $b_K=o_p(\tau_n)$ makes random training fluctuations in~$b_K$ negligible for the studentized limit. When $b_K$ is non-negligible relative to~$\tau_n$, bias-aware intervals from Section~\ref{sec:sensitivity} are needed instead.

\subsection{Cross-fitting}\label{sec:crossfitting}
A correction can shift the bias by more than a standard error while still leaving the standard error unchanged to first order. To see why, consider one evaluation fold $I_k$ of size $n_k$ with fold average $\mathbb P_k$, conditioning on the data used to train its nuisances and kernel. The decomposition in~\eqref{eq:hoeffding-components} splits the estimated correction into three parts,
\[
 \widehat B_k
 =\underbrace{\mu_k}_{\text{mean correction}}
  +\underbrace{\mathbb P_k L_k}_{\text{individual-observation fluctuation}}
  +\underbrace{\mathbb{U}_{n_k} d_k}_{\text{pairwise fluctuation}}.
\]
Subtracting $\widehat B_k$ from ordinary DML shifts its conditional mean by $-\mu_k$ (enough to remove a bias of order $n_k^{-2/5}$, as in Section~\ref{sec:learned-bias-rate}), while only the remaining two terms fluctuate across evaluation samples for fixed training fits.

Suppose Proposition~\ref{prop:ordinary} holds within each fold. Because the centered function~$L_k$ has $L^2$ norm $o_p(1)$, its fold average is $o_p(n_k^{-1/2})$, and the degenerate pairwise remainder is likewise $o_p(n_k^{-1/2})$ by~\eqref{eq:full-variance} and $\E[d_k^2]=o_p(n_k)$ (with the expectation taken over an independent evaluation pair given training). Combining these bounds with $L^2$ convergence of the fitted score to the oracle score and a remaining bias of $o_p(n_k^{-1/2})$ yields the fold expansion
\[
 \widehat\theta_k-\theta_0
 =\frac{1}{n_k}\sum_{i\in I_k}\phi_0(W_i)+o_p(n_k^{-1/2}).
\]
Although each fold uses its own fitted nuisances and correction, its leading term averages the same centered oracle DML score~$\phi_0$. Weighting the fold estimates by their evaluation sizes therefore gives every observation equal weight despite overlapping training samples, and shows directly which scores supply the standard error.

\begin{corollary}[Cross-fitted DDML with the ordinary DML standard error]\label{cor:crossfit-ordinary}
Partition $N$ iid observations into a fixed number $K$ of folds independently of their values, with $n_k:=|I_k|\asymp N$ for every~$k$. For each fold, fit the nuisances and correction kernel using only data outside $I_k$. Suppose the conditions of Proposition~\ref{prop:ordinary} hold fold by fold with evaluation sample size $n_k$. The cross-fitted estimator $\widehat\theta_{\rm CF}:=\sum_{k=1}^K(n_k/N)\widehat\theta_k$ then satisfies
\begin{equation}\label{eq:crossfit-ordinary}
 \sqrt N(\widehat\theta_{\rm CF}-\theta_0)
 =\frac1{\sqrt N}\sum_{i=1}^N\phi_0(W_i)+o_p(1).
\end{equation}
For $i\in I_k$, let $\widehat s_i:=s_{-k}(W_i)$ be the ordinary DML score~\eqref{eq:db} evaluated with the nuisances trained outside $I_k$. Its sample variance $\widehat\sigma^2:=(N-1)^{-1}\sum_{i=1}^N(\widehat s_i-\overline s)^2$, where $\overline s:=N^{-1}\sum_{i=1}^N\widehat s_i$, satisfies
\[
 \widehat\sigma^2\to_p\E[\phi_0^2],\qquad
 \frac{\sqrt N(\widehat\theta_{\rm CF}-\theta_0)}{\widehat\sigma}
 \ad N(0,1).
\]
\end{corollary}

In this regime, the ordinary out-of-fold scores supply the standard error $\widehat\sigma/\sqrt N$ for an interval centered at the corrected estimate. The proof in Appendix~\ref{app:proofs} uses the independence of each evaluation fold from its own training fit and the common oracle score across folds.

When the correction changes the first-order influence function or contributes appreciable pairwise variation, estimating the variance of the pooled cross-fitted estimator remains an open question here. Proposition~\ref{prop:vhat} gives the full conditional variance estimator for a single independent training and evaluation split. With overlapping training samples, a pooled variance calculation must also account for any leading covariances across folds.

\section{Bias-aware inference and sensitivity analysis for misspecification}\label{sec:sensitivity}
With independent training and evaluation samples, the correction estimates how much the two nuisance errors align along the chosen directions. To ask how much error the correction may have missed, we adapt the two-factor sensitivity analysis of Chernozhukov, Cinelli, Newey, Sharma, and Syrgkanis~\cite{sensitivity}. The analyst specifies plausible sizes for the two remaining errors. Their product determines how far to widen the interval for the observed-data target.

\subsection{Controlling misspecification bias via short-versus-long sensitivity}\label{sec:ovb-connection}
Chernozhukov, Cinelli, Newey, Sharma, and Syrgkanis~\cite{sensitivity} express the difference between a ``long'' target and a ``short'' regression target as an inner product of omitted outcome and Riesz components. Although that framework is often introduced for unobserved confounding, the exact same Hilbert-space geometry governs functional misspecification and remaining learning bias in DDML.

To see this connection in its simplest form, hold the exact population projection~$\Pi$ onto the training-chosen working space~$\mathcal H_k$ fixed, and define the \emph{short target} associated with~$\mathcal H_k$ by
\begin{equation}\label{eq:short-target}
 g_{\mathcal H}:=\Pi g_0,\qquad
 \alpha_{\mathcal H}:=\Pi\alpha_0,\qquad
 \theta_{\mathcal H}:=\E[m(W,g_{\mathcal H})].
\end{equation}
Here $g_{\mathcal H}$ is the best $L^2(P_X)$ approximation of the true regression~$g_0$ inside~$\mathcal H_k$, and $\alpha_{\mathcal H}$ is the Riesz representer of the target functional restricted to~$\mathcal H_k$, meaning that $\E[m(W,h)]=\ip{\alpha_{\mathcal H}}{h}$ for every $h\in\mathcal H_k$.

\begin{proposition}[The short target and the corrected center]\label{prop:short-target}
Under the Riesz identity and square-integrability conditions,
\begin{equation}\label{eq:short-long}
 \theta_0-\theta_{\mathcal H}
 =\ip{g_0-g_{\mathcal H}}{\alpha_0-\alpha_{\mathcal H}}.
\end{equation}
If both fitted nuisances $\gh$ and $\ah$ belong to~$\mathcal H_k$, then the exact projection correction satisfies
\begin{equation}\label{eq:short-center}
 \E[\thh_\Pi\mid\calT]=\theta_{\mathcal H},
\end{equation}
without requiring either fitted nuisance to equal its population projection.
\end{proposition}

When both fitted nuisances lie inside~$\mathcal H_k$, Proposition~\ref{prop:short-target} shows that the exact projection correction removes \emph{all} estimation bias within~$\mathcal H_k$ and centers $\thh_\Pi$ at the short target~$\theta_{\mathcal H}$, leaving only the functional-misspecification gap~\eqref{eq:short-long}. Normalizing this gap by the short-model scales $\sigma_{\mathcal H}^2:=\E(Y-g_{\mathcal H})^2>0$ and $\nu_{\mathcal H}^2:=\norm{\alpha_{\mathcal H}}_2^2>0$, define
\begin{equation}\label{eq:short-sensitivity-scales}
 C_{Y,\mathcal H}:=
       \frac{\norm{g_0-g_{\mathcal H}}_2}{\sigma_{\mathcal H}},\qquad
 C_{\alpha,\mathcal H}:=
       \frac{\norm{\alpha_0-\alpha_{\mathcal H}}_2}{\nu_{\mathcal H}},\qquad
 S_{\mathcal H}:=\sigma_{\mathcal H}\nu_{\mathcal H},
\end{equation}
and let $\lambda_{\mathcal H}\in[-1,1]$ be the $L^2$ cosine between $g_0-g_{\mathcal H}$ and $\alpha_0-\alpha_{\mathcal H}$ (set to zero if either remainder vanishes). The misspecification gap then factors as
\begin{equation}\label{eq:short-sensitivity-normalized}
 \theta_0-\theta_{\mathcal H}
 =\lambda_{\mathcal H}
       C_{Y,\mathcal H}C_{\alpha,\mathcal H}S_{\mathcal H},
\end{equation}
which is the working-space analogue of Theorem~2 in~\cite{sensitivity}. Here $C_{Y,\mathcal H}^2\le 1$ automatically because $\E(Y-g_{\mathcal H})^2=\E(Y-g_0)^2+\norm{g_0-g_{\mathcal H}}_2^2$, while $C_{\alpha,\mathcal H}$ equals the root mean square of the omitted part of the Riesz representer divided by the root mean square of its projection. For example, if $e\perp\mathcal H_k$ is a unit vector with $g_0=g_{\mathcal H}+ue$ and $\alpha_0=\alpha_{\mathcal H}+ve$, then $\theta_0-\theta_{\mathcal H}=uv$. Improving coefficient estimates inside~$\mathcal H_k$ cannot remove this gap, whereas adding~$e$ to~$\mathcal H_k$ removes it, with the change in variance given by~\eqref{eq:full-variance}.

\paragraph{Extension to arbitrary ML fits}
A fitted forest, boosted regression, or CausalPFN prediction can lie outside the working space~$\mathcal H_k$. In this case the same norm and alignment identities apply to the \emph{residual estimation errors} left outside~$\mathcal H_k$. Define
\begin{equation}\label{eq:residual-sensitivity}
 r_g:=(I-\Pi)\delta_g,\qquad
 r_\alpha:=(I-\Pi)\delta_\alpha,\qquad
 \overline\theta_\Pi:=\E[\thh_\Pi\mid\calT],
\end{equation}
so that Proposition~\ref{prop:bias} gives
\begin{equation}\label{eq:residual-target-gap}
 \theta_0-\overline\theta_\Pi=\ip{r_\alpha}{r_g}.
\end{equation}
To see why this behaves just like a short-versus-long misspecification gap, define the \emph{population-refined nuisances}
\begin{equation}\label{eq:population-refinement}
 g^\dagger:=\gh-\Pi\delta_g=g_0+r_g,\qquad
 \alpha^\dagger:=\ah-\Pi\delta_\alpha=\alpha_0+r_\alpha.
\end{equation}
These functions project $g_0$ and $\alpha_0$ onto the affine spaces $\gh+\mathcal H_k$ and $\ah+\mathcal H_k$, and evaluating the ordinary DML score at $(g^\dagger,\alpha^\dagger)$ yields the exact corrected center:
\begin{equation}\label{eq:refined-score-center}
 \E\!\left[m(W,g^\dagger)
       +\alpha^\dagger(X)\{Y-g^\dagger(X)\}\right]
 =\theta_0-\ip{r_\alpha}{r_g}
 =\overline\theta_\Pi.
\end{equation}
Writing $R_\alpha:=\norm{r_\alpha}_2^2$ and $R_g:=\norm{r_g}_2^2$ for the two residual mean squares, together with their $L^2$ cosine (equal to their correlation when $1\in\mathcal H_k$ and both residual variances are positive)
\begin{equation}\label{eq:alignment}
 \lambda_\perp:=\frac{\ip{r_\alpha}{r_g}}{\sqrt{R_\alpha R_g}}
 \quad(\text{or }0\text{ if }R_\alpha R_g=0),
\end{equation}
we can rewrite~\eqref{eq:residual-target-gap} as
\begin{equation}\label{eq:alignment-gap}
 \theta_0-\overline\theta_\Pi
 =\lambda_\perp\sqrt{R_\alpha R_g}.
\end{equation}
\paragraph{A two-factor budget for the fitted correction}
To apply the same sensitivity analysis to an estimated projection or a regularized kernel, we also account for the error that its normalization leaves inside the working space. Suppose the fitted operator~$T$ is bounded and self-adjoint, as in all of our empirical comparisons. Write $A:=I-T$ and define
\begin{equation}\label{eq:q-errors}
 R_{j,T}:=\ip{\delta_j}{|A|\delta_j},\qquad j\in\{g,\alpha\},
\end{equation}
where $|A|$ takes the absolute value of each spectral value of~$A$. Because $A=|A|^{1/2}\operatorname{sgn}(A)|A|^{1/2}$ and the sign operator has norm at most one, Cauchy--Schwarz in~\eqref{eq:bias} yields $|b_K|\le\sqrt{R_{g,T}R_{\alpha,T}}$. For an exact projection, $R_{j,\Pi}=\|(I-\Pi)\delta_j\|_2^2$, whereas for a rank-one correction $T=c\Pi_f$ the squared error along~$f$ enters with weight $|1-c|$ and the squared error orthogonal to~$f$ enters with weight~$1$. Ordinary DML corresponds to $T=0$ and uses the full squared error of each fit.

To express the outcome-error budget in the units of a particular regression fit, we use its training residual mean square $s_Y^2:=\mathbb P_{\rm tr}[(Y-\gh(X))^2]$, where $\mathbb P_{\rm tr}$ denotes the training average. For the weighting error, the natural reference is the population second moment $\E[\alpha_0(X)^2]$, which is common to every method targeting the same functional. Following the orthogonal score of Chernozhukov et al.~\cite{sensitivity}, any candidate representer~$a$ satisfies
\[
 \E[2m(W,a)-a(X)^2]=\E[\alpha_0(X)^2]-\norm{a-\alpha_0}_2^2
\]
by the Riesz identity~\eqref{eq:riesz}, so minimizing the expected Riesz loss $\E[a(X)^2-2m(W,a)]$ selects the candidate closest to~$\alpha_0$ in mean square.

We estimate one shared reference scale per training split by dividing the training sample into folds $J_1,\ldots,J_V$. For each fold~$v$, we select a reference learner by cross-validated Riesz loss on the complement of~$J_v$ and refit it there to obtain~$\widehat a_{-v}$, so that every observation in~$J_v$ is scored by a learner selected and fitted without it. With $n_{\rm tr}$ training observations, the resulting debiased second-moment estimate is
\begin{equation}\label{eq:shared-riesz-scale}
 \widehat\nu_\alpha:=\frac{1}{n_{\rm tr}}\sum_{v=1}^V\sum_{i\in J_v}
 \big\{2m(W_i,\widehat a_{-v})-\widehat a_{-v}(X_i)^2\big\}.
\end{equation}
Whenever $\widehat\nu_\alpha>0$, we freeze $s_\alpha:=\sqrt{\widehat\nu_\alpha}$ across all methods on that split: each method still uses its own weights in the estimating equation, yet sharing $s_\alpha$ puts every weighting-error budget on a common scale while allowing $s_Y$ to vary with the outcome fit.

For a common sensitivity parameter $0\le q<1$, consider the budget
\begin{equation}\label{eq:q-budget}
 R_{g,T}\le q\,s_Y^2,\qquad
 R_{\alpha,T}\le \frac{q}{1-q}\,s_\alpha^2
 \quad\Longrightarrow\quad
 |b_K|\le B_q:=\frac{q}{\sqrt{1-q}}\,s_Ys_\alpha .
\end{equation}
The first inequality budgets outcome error as a fraction of the training residual mean square, while the second uses the odds $q/(1-q)$ from the two-factor Riesz calibration of~\cite{sensitivity}, and taking the geometric mean yields the bias bound~$B_q$ for either sign of alignment. Separate budgets $(q_Y,q_\alpha)$ give $s_Ys_\alpha\sqrt{q_Yq_\alpha/(1-q_\alpha)}$, and we set $q_Y=q_\alpha=q$ to report each empirical comparison on a single scale. In these training-determined units, $q=1\%$ permits a root mean squared outcome error of up to one tenth of~$s_Y$ in the geometry of~\eqref{eq:q-errors}. Whether the unknown nuisance errors actually satisfy~\eqref{eq:q-budget} remains a sensitivity assumption, and a learner that nearly interpolates the training outcomes produces a small~$s_Y$ and therefore a more demanding outcome-error assumption at a given~$q$.

\begin{remark}[Extension to nonsymmetric kernels]\label{rem:nonsymmetric-budget}
For a bounded nonsymmetric kernel operator, the same bound $|b_K|\le\sqrt{R_{g,T}R_{\alpha,T}}$ follows from the polar decomposition $A=U|A|$: because $A=|A^*|^{1/2}U|A|^{1/2}$ with $\|U\|_{\op}\le 1$, it suffices to set $R_{g,T}:=\ip{\delta_g}{|A|\delta_g}$ for the outcome error and $R_{\alpha,T}:=\ip{\delta_\alpha}{|A^*|\delta_\alpha}$ for the weighting error.
\end{remark}

\paragraph{Benchmarking residual error with nested working spaces}
To gauge the plausibility of a proposed budget, we can compare two nested, training-chosen working spaces with projections $\Pi_1$ and $\Pi_2$. Setting $D:=\Pi_2-\Pi_1$ decomposes the remaining bias of the coarser space as
\begin{equation}\label{eq:nested-bias}
 \theta_0-\overline\theta_{\Pi_1}
 =\ip{D\delta_\alpha}{D\delta_g}
  +\theta_0-\overline\theta_{\Pi_2}.
\end{equation}
The difference between the two corrected estimates measures the bias contribution of the added directions. Its size gives the analyst a reference for judging $B_q$. Inferring an upper bound on what lies outside the richer space would require an extrapolation assumption.

\begin{remark}[Distinguishing misspecification bias from unobserved confounding]\label{rem:two-biases}
The sensitivity bounds in this section control remaining misspecification and learning bias for the observed-data functional $\theta_0=\theta_{\rm obs}$. If causal identification is also in doubt, allowing unobserved confounders introduces a separate gap between the causal parameter~$\theta_{\rm causal}$ and~$\theta_{\rm obs}$,
\begin{equation}\label{eq:two-biases}
 \theta_{\rm causal}-\overline\theta_\Pi
 =\underbrace{\theta_{\rm causal}-\theta_{\rm obs}}_{
       \text{omitted-confounder discrepancy}}
  +\underbrace{\theta_{\rm obs}-\overline\theta_\Pi}_{
       \text{remaining learning bias}}.
\end{equation}
Stronger learners or richer working spaces can shrink only the second term. If an analyst separately bounds $|\theta_{\rm causal}-\theta_{\rm obs}|\le B_{\rm conf}$, widening a valid bias-aware interval by $B_{\rm conf}$ on each side displays both sources of uncertainty separately.
\end{remark}

\subsection{Bias-aware confidence intervals}\label{sec:bias-aware-ci}
Choose $q$ before evaluation and suppose the two inequalities in~\eqref{eq:q-budget} hold with probability tending to one over training samples. Let $\widehat{\rm se}_K$ be a consistent estimator of the conditional standard error in relative terms. Under the centered normal approximation~\eqref{eq:studentized-centered} of Proposition~\ref{prop:general-clt},
\begin{equation}\label{eq:centered-clt}
 \frac{\thh_K-\E[\thh_K\mid\calT]}{\widehat{\rm se}_K}\ad N(0,1).
\end{equation}
For nominal coverage $1-\gamma$, write $z:=z_{1-\gamma/2}$ for the standard normal quantile. The bias-aware interval is
\begin{equation}\label{eq:bias-aware-interval}
 \mathcal I_n(q):=
 [\thh_K-z\widehat{\rm se}_K-B_q,\
   \thh_K+z\widehat{\rm se}_K+B_q].
\end{equation}
Whenever the centered Wald interval contains $\E[\thh_K\mid\calT]$ and $|b_K|\le B_q$, the triangle inequality gives $\theta_0\in\mathcal I_n(q)$. Hence $\liminf_{n\to\infty}\Pr\{\theta_0\in\mathcal I_n(q)\}\ge1-\gamma$. For this statement, the budgets are in units of the realized training scales. Conditioning on $\calT$ fixes the selected reference fits and $s_\alpha$, so estimating the common scale adds no evaluation-sample uncertainty to $B_q$. Sampling uncertainty enters through the center alone. If the scientific question instead specifies a bound in terms of an unknown population scale, inference on its endpoints must also account for estimation of that scale and its covariance with the center, as in Theorem~4 of~\cite{sensitivity}.

For a positive estimated effect, it is useful to report the budget at which the lower endpoint reaches zero. Suppose that the $q=0$ Wald lower endpoint, $L_0:=\thh_K-z\widehat{\rm se}_K$, and the scale product $s_Ys_\alpha$ are both positive. Setting $B_q=L_0$ and solving for $q$ gives
\begin{equation}\label{eq:sensitivity-frontier}
 t:=\frac{L_0}{s_Ys_\alpha},\qquad
 q^*:=\frac{2t}{\sqrt{t^2+4}+t}.
\end{equation}
The interval supports a positive effect for $0\le q<q^*$. If $L_0\le0$, it already includes zero for a positive point estimate at $q=0$.

\section{Implementation and choice of interval}\label{sec:workflow}
Applying DDML in an empirical study begins with choosing the target functional and the working space. Once the nuisances and correction kernel are frozen on training data, the evaluation sample performs two jobs at once: it estimates the bias correction and it reveals how much the correction changes sampling variation relative to ordinary DML.

\subsection{Implementation}
\begin{enumerate}\setlength{\itemsep}{3pt}\setlength{\parskip}{0pt}
 \item Choose the target functional and partition the sample into training and evaluation splits. For an ATE, specify the outcome, the treatment contrast, and any propensity clipping in advance, consulting Section~\ref{sec:crossfitting} before pooling cross-fitted estimates.
 \item On training data, fit the outcome regression and the Riesz representer, learn the direction, and estimate its second moment (or Gram matrix). Freeze all three fits and the normalization before evaluation. An internal split can supply fresh residual labels or calibration covariates. Section~\ref{sec:empirical} gives the allocation used in each experiment.
 \item On the evaluation sample, compute the ordinary DML estimate~$\thh_{\rm DB}$, the off-diagonal U-statistic correction~$\widehat B_K$ in~\eqref{eq:pair}, and the corrected estimate~$\thh_K$. Reporting both point estimates side by side makes the shift in the center easy to see.
 \item Compute the original DML standard error, the corrected first-order variance estimate~$\widehat V_1$, and the full unbiased variance estimate $\widehat V=\widehat V_1+\widehat V_2$ in~\eqref{eq:vhat}. When both variances are positive, report the percentage change in the full standard error relative to ordinary DML along with the pairwise share $\widehat V_2/\widehat V$, and flag any nonpositive variance estimates as a sign to inspect kernel complexity or rare cells.
 \item Use the training scales and selected budgets $q$ in~\eqref{eq:q-budget} to express the allowance $B_q$ in units of the effect. Nested working spaces in~\eqref{eq:nested-bias} provide a benchmark for its magnitude.
 \item Report the confidence interval supported by the sampling and bias assumptions in Table~\ref{tab:decision-guide}. When residual learning error enters as a sensitivity parameter, report interval width alongside coverage in simulations and report interval endpoints across error budgets in applications, adding any separate confounding allowance via Remark~\ref{rem:two-biases} if causal identification is also under study.
\end{enumerate}

\subsection{Choosing an interval}
A small pairwise variance share does not guarantee that the correction leaves the first-order influence function unchanged. In the 401(k) analysis with 16 income cells, updating the first-order influence function raises the random-forest standard error from \$987 to \$1,015. Including the pairwise variance raises it to \$1,018. The full standard error rises by 3.1\% overall, even though the pairwise term accounts for only 0.55\% of the variance (Table~\ref{tab:emp-var}). While these diagnostics describe the behavior of the fitted estimator, the theoretical conditions collected in Table~\ref{tab:decision-guide} determine which confidence interval carries a formal coverage guarantee.

\begin{table}[htbp]
\centering\small
\caption{Choosing an interval for the original target}\label{tab:decision-guide}
\begin{tabularx}{\textwidth}{@{}p{.32\textwidth}X@{}}
\toprule
Conditions & Standard error and interval\\
\midrule
Ordinary influence function and negligible bias
& Under \eqref{eq:ordinary-conditions}, use the ordinary DML standard error and Wald interval. The conditions include score convergence, negligible $L$, negligible pairwise noise, and $b_K=o_p(n^{-1/2})$. For fixed-fold cross-fitting, apply Corollary~\ref{cor:crossfit-ordinary}.\\[3pt]
Changed influence function or appreciable pairwise noise
& With independent evaluation, a centered CLT, a ratio-consistent full variance estimate, and $b_K=o_p(\tau_n)$, use the Wald interval with $\sqrt{\widehat V}$. If only the first-order term matters, a consistent row-based standard error suffices. See Section~\ref{sec:crossfitting} for pooling restrictions.\\[3pt]
Residual learning errors satisfy the two-factor budget
& Under a centered CLT and consistent studentization, add the training allowance $B_q$ to each half-width as in~\eqref{eq:bias-aware-interval}. Coverage for the observed-data target holds under the specified budget.\\

\bottomrule
\end{tabularx}
\end{table}

\section{Empirical application and Monte Carlo evidence}\label{sec:empirical}
In the 401(k) application we compare how the estimated effect and its confidence interval change across learners on the same households, whereas the ACIC simulations provide a known treatment effect so we can measure both empirical coverage and the interval width needed to allow for residual bias. Both exercises rely on independent training and evaluation splits, while the high-dimensional linear simulations compare five-fold cross-fitted estimates on matched draws. Table~\ref{tab:sample-splits} summarizes the internal sample-splitting allocations used across all four setups.

\begin{table}[H]
\centering\small
\caption{Sample-splitting allocations across the empirical and simulation studies}\label{tab:sample-splits}
\begin{tabularx}{\textwidth}{@{}>{\raggedright\arraybackslash}p{0.26\textwidth}>{\raggedright\arraybackslash}p{0.16\textwidth}>{\raggedright\arraybackslash}p{0.16\textwidth}>{\raggedright\arraybackslash}X@{}}
\toprule
Experiment & Total / eval & Baseline fit & Direction learning and norm calibration \\
\midrule
401(k) main (Sec.~\ref{sec:401k}) & 9,915 / 4,958 & All 4,957 train & Forest: 2-fold residuals on first 3,717. PFN context: first 1,800. Norm: last 1,240. \\
ACIC linear or boosted (Sec.~\ref{sec:acic}) & 4,800 / 2,400 & All 2,400 train & Forest: 2-fold residuals on first 1,800. PFN context: first 1,800. Norm: last 600. \\
ACIC PFN baseline (Sec.~\ref{sec:acic}) & 4,800 / 2,400 & PFN 1,800, prop.\ 2,400 & Forest direction from residuals on next 400, norm calibrated on final 200. \\
Matched updates (Sec.~\ref{sec:calibrated-comparison}) & 401(k) 4,958, ACIC 2,400 & 401(k) 3,717, ACIC 1,800 & Calibration split for slope and norm: 1,240 (401(k)) or 600 (ACIC), with 1,800 PFN context. \\
High-dim.\ linear (Sec.~\ref{sec:lasso-results}--\ref{sec:lasso-geometric}) & 1,000 (5-fold) & 800 train / 200 eval & Residual learner on same 800 train obs., norm on 400 independent unlabeled covariates. \\
\bottomrule
\end{tabularx}
\end{table}

\subsection{401(k) eligibility and household financial assets}\label{sec:401k}
We estimate the effect of 401(k) eligibility on household net financial assets using 9,915 households from the 1991 Survey of Income and Program Participation~\cite{dml,doublemldata}. Throughout, we cap net financial assets at $-$\$50,000 and \$250,000 and work with the standard nine household covariates, splitting the data into 4,957 training households and 4,958 evaluation households. Interpreting the adjusted contrast causally requires conditional exchangeability, consistency, and overlap. We compare two baseline DML specifications: a simple parametric specification using linear outcome regressions within each eligibility arm and a logistic propensity model, and a flexible specification using random forests for both nuisances.

Following the allocation in Table~\ref{tab:sample-splits}, we fit the baseline nuisances on all 4,957 training households. To learn the forest direction, we divide the first 3,717 training households in half, fit the baseline outside each half to form out-of-fold residuals, and train arm-specific forests on those pooled residuals. For the CausalPFN direction, we fit the arm means on the first 1,800 training households and subtract the full-training baseline predictions, and both directions estimate their second moment on the remaining 1,240 training households (which also enter the final baseline fit). We then evaluate once on the held-out 4,958 households using the full-training nuisances, and include a third comparison that substitutes the same CausalPFN arm means directly into ordinary DML with the original propensity.

\begin{table}[H]
\centering\small
\caption{Effect of 401(k) eligibility on capped assets, with nominal 95\% intervals}\label{tab:rank401k}
\begin{tabular}{@{}l*{6}{@{\hspace{.7em}}r}@{}}
\toprule
Estimator & Estimate & SE & 95\% interval & Change & SE of change & SE red. (\%) \\
\midrule
\multicolumn{7}{l}{\textit{Linear outcome, logistic propensity}}\\
Ordinary DML & 5,914 & 1,090 & [3,778, 8,050] & --- & --- & --- \\
DML, direct PFN outcome & 8,680 & 1,046 & [6,630, 10,730] & +2,766 & 438 & 4.0 \\
DDML, PFN direction & 8,722 & 1,020 & [6,722, 10,722] & +2,808 & 561 & 6.4 \\
DDML, forest direction & 7,763 & 1,049 & [5,708, 9,818] & +1,849 & 321 & 3.8 \\
\addlinespace
\multicolumn{7}{l}{\textit{Forest outcome and propensity}}\\
Ordinary DML & 7,559 & 987 & [5,625, 9,493] & --- & --- & --- \\
DML, direct PFN outcome & 7,446 & 983 & [5,518, 9,373] & -113 & 232 & 0.3 \\
DDML, PFN direction & 7,446 & 983 & [5,520, 9,372] & -113 & 224 & 0.4 \\
DDML, forest direction & 7,613 & 998 & [5,657, 9,569] & +54 & 54 & -1.2 \\
\bottomrule
\end{tabular}
\par\vspace{4pt}\begin{minipage}{\textwidth}\footnotesize
All entries except percentages are in 1991 dollars. Within each panel all estimators use the same propensity and 4,958 evaluation households. Changes and SE reductions are relative to the panel's original DML fit. Wald intervals use the ordinary score SE for DML and the full conditional SE for DDML, assuming negligible remaining bias.
\end{minipage}
\end{table}

With the logistic propensity held fixed, ordinary linear DML yields \$5,914, whereas direct CausalPFN raises the estimate to \$8,680 and the rank-one PFN correction gives \$8,722 with nearly the same interval (Table~\ref{tab:rank401k}), while the forest direction raises the linear estimate by \$1,849. Under random-forest initial fits, by contrast, the largest shift is only \$113 against an ordinary standard error of \$987. Although every centered interval excludes zero, the true effect is unknown here, so these shifts alone cannot establish bias reduction.

The forest propensity achieved the lower validation Riesz loss in all three training folds, giving a shared reference scale $s_\alpha=2.443$. At $q=1\%$, the linear outcome fit gives an allowance of \$740 on each side of the interval. Applying the PFN direction raises the lower endpoint from \$3,038 to \$5,983 (Table~\ref{tab:sipp-q}). At $q=2\%$, the corrected interval ranges from \$5,235 to \$12,209. The lower endpoint reaches zero at a budget of 5.00\% for ordinary linear DML, 8.73\% for the PFN direction, and 9.08\% for direct PFN outcome substitution. With forest initial fits, the thresholds for ordinary DML and the two learned directions range from 7.81\% to 8.00\%. Direct PFN outcome substitution has the same allowance in both panels because both its outcome fit and its Riesz reference scale are shared.

\begin{table}[H]
\centering\small
\caption{Allowances for remaining bias and intervals for the 401(k) effect}\label{tab:sipp-q}
\begin{tabular}{@{}l*{5}{@{\hspace{.75em}}r}@{}}
\toprule
 & \multicolumn{2}{c}{$q=1\%$} & \multicolumn{2}{c}{$q=2\%$} & \\
\cmidrule(lr){2-3}\cmidrule(lr){4-5}
Estimator & $B_q$ & 95\% interval & $B_q$ & 95\% interval & $q^*$ (\%) \\
\midrule
\multicolumn{6}{l}{\textit{Linear outcome, logistic propensity}}\\
Ordinary DML & 740 & [3,038, 8,790] & 1,487 & [2,291, 9,537] & 5.00 \\
DML, direct PFN outcome & 700 & [5,931, 11,429] & 1,406 & [5,224, 12,136] & 9.08 \\
DDML, PFN direction & 740 & [5,983, 11,462] & 1,487 & [5,235, 12,209] & 8.73 \\
DDML, forest direction & 740 & [4,968, 10,558] & 1,487 & [4,221, 11,305] & 7.46 \\
\addlinespace
\multicolumn{6}{l}{\textit{Forest outcome and propensity}}\\
Ordinary DML & 682 & [4,943, 10,175] & 1,370 & [4,255, 10,863] & 7.96 \\
DML, direct PFN outcome & 700 & [4,819, 10,073] & 1,406 & [4,112, 10,779] & 7.62 \\
DDML, PFN direction & 682 & [4,839, 10,054] & 1,370 & [4,150, 10,743] & 7.81 \\
DDML, forest direction & 682 & [4,975, 10,251] & 1,370 & [4,286, 10,940] & 8.00 \\
\bottomrule
\end{tabular}
\par\vspace{4pt}\begin{minipage}{\textwidth}\footnotesize
The allowances $B_q$ and interval endpoints are in 1991 dollars. The same $q$ applies to both nuisance-error budgets in~\eqref{eq:q-budget}. All rows share the debiased Riesz scale and use the residual scale of their own outcome fit and the SE in Table~\ref{tab:rank401k}. The interval excludes zero for $q<q^*$ (thresholds rounded).
\end{minipage}
\end{table}

\subsection{ACIC simulations with strong initial fits}\label{sec:acic}
In ACIC settings 27 and 29, each of four fixed response surfaces specifies the conditional outcome means and treatment probabilities across 4,802 covariate records, held fixed across repetitions. Each repetition draws independent training and evaluation samples of 2,400 observations each, using the internal training splits summarized in Table~\ref{tab:sample-splits}. For linear or boosted baselines, the forest and PFN directions follow the same protocol as in the 401(k) application, whereas when CausalPFN supplies the initial outcome regression we freeze its fit on 1,800 observations, learn a forest direction from residuals on the next 400, and calibrate its norm on the final 200 while fitting the propensity on all 2,400 training observations. To form the shared sensitivity scale in~\eqref{eq:shared-riesz-scale}, we split training into three folds and select the propensity learner on each two-fold complement by two-fold Riesz loss (comparing logistic regression with boosting in ACIC and with a forest in the 401(k) data). The evaluation sample enters only the point and variance estimates, and every rank-one DDML interval uses the full variance estimate~\eqref{eq:vhat}, which delivers nominal Wald coverage under a centered Gaussian approximation and negligible remaining bias.

\begin{table}[H]
\centering\small
\caption{RMSE and standard error before and after a rank-one forest correction on ACIC}\label{tab:rankacic}
\begin{tabular}{@{}llrrrrr@{}}
\toprule
Setting & Outcome / propensity & Draws & DML & DDML & RMSE red. (\%) & SE red. (\%) \\
\midrule
27 & CausalPFN / Boosting & 20 & 0.02976 & 0.02921 & 1.9 & 0.7 \\
 & CausalPFN / Logistic & 20 & 0.03075 & 0.03023 & 1.7 & 0.8 \\
 & Boosting / Boosting & 200 & 0.02222 & 0.02187 & 1.6 & 1.3 \\
 & Linear / Logistic & 200 & 0.06681 & 0.03781 & 43.4 & 22.4 \\
\addlinespace
29 & CausalPFN / Boosting & 20 & 0.02050 & 0.01997 & 2.6 & 0.8 \\
 & CausalPFN / Logistic & 20 & 0.02682 & 0.02629 & 2.0 & 0.3 \\
 & Boosting / Boosting & 200 & 0.02832 & 0.02790 & 1.5 & 1.8 \\
 & Linear / Logistic & 200 & 0.09756 & 0.06086 & 37.6 & 22.9 \\
\bottomrule
\end{tabular}
\par\vspace{4pt}\begin{minipage}{\textwidth}\footnotesize
The DML and DDML columns report RMSE in marginal outcome standard deviations. Both percentage reductions use ordinary DML as the baseline. For SE we compute the percentage reduction in the estimated standard error on each paired draw, then average across draws. Each setting uses four fixed response surfaces and five draws per surface for CausalPFN or fifty for the other learners.
\end{minipage}
\end{table}

With linear/logistic nuisances, the forest direction reduces RMSE by 43.4\% in setting~27 and 37.6\% in setting~29 (Table~\ref{tab:rankacic}). With boosted nuisances, both reductions are below 2\%. Table~\ref{tab:rankacic-ci} asks whether the intervals cover the true effect. Within each pair, DML and DDML have the same initial outcome fit and share the training scales $s_Y$ and $s_\alpha$. At each budget $q$ in~\eqref{eq:q-budget}, we therefore add the same $B_q$ to each method's own Wald half-width. For ordinary DML the budget applies to the full nuisance errors ($T=0$), and for DDML it applies to the errors weighted by $|I-T|$. Widths and allowances are in units of the marginal outcome standard deviation on each surface.

\begin{table}[H]
\centering\small
\caption{Coverage and mean interval width for ordinary DML and rank-one DDML on ACIC}\label{tab:rankacic-ci}
\begin{tabular}{@{}lrl*{6}{@{\hspace{.75em}}r}@{}}
\toprule
 & & & \multicolumn{2}{c}{$q=0$} & \multicolumn{2}{c}{$q=1\%$} & \multicolumn{2}{c}{$q=2\%$}\\
\cmidrule(lr){4-5}\cmidrule(lr){6-7}\cmidrule(lr){8-9}
Initial fits / direction & Draws & Method & Cov. & Width & Cov. & Width & Cov. & Width \\
\midrule
\multicolumn{9}{l}{\textit{Setting 27}}\\
Linear, logistic / forest & 200 & DML & 74.0 & 0.143 & 79.0 & 0.166 & 83.5 & 0.189 \\
 &  & DDML & 88.0 & 0.111 & 93.5 & 0.133 & 96.0 & 0.156 \\
Boosted, boosted / forest & 200 & DML & 93.0 & 0.081 & 96.0 & 0.088 & 99.0 & 0.096 \\
 &  & DDML & 94.0 & 0.080 & 96.5 & 0.087 & 99.0 & 0.095 \\
Linear, logistic / PFN & 20 & DML & 85.0 & 0.137 & 85.0 & 0.159 & 90.0 & 0.182 \\
 &  & DDML & 80.0 & 0.088 & 90.0 & 0.111 & 95.0 & 0.134 \\
PFN, logistic / forest & 20 & DML & 85.0 & 0.084 & 85.0 & 0.092 & 90.0 & 0.099 \\
 &  & DDML & 85.0 & 0.084 & 85.0 & 0.091 & 90.0 & 0.099 \\
PFN, boosted / forest & 20 & DML & 90.0 & 0.082 & 90.0 & 0.090 & 90.0 & 0.097 \\
 &  & DDML & 90.0 & 0.082 & 90.0 & 0.089 & 90.0 & 0.097 \\
\addlinespace
\multicolumn{9}{l}{\textit{Setting 29}}\\
Linear, logistic / forest & 200 & DML & 52.5 & 0.152 & 63.5 & 0.175 & 71.0 & 0.198 \\
 &  & DDML & 67.0 & 0.118 & 74.0 & 0.141 & 86.5 & 0.164 \\
Boosted, boosted / forest & 200 & DML & 84.5 & 0.083 & 91.5 & 0.094 & 95.5 & 0.104 \\
 &  & DDML & 84.5 & 0.082 & 91.0 & 0.092 & 94.5 & 0.103 \\
Linear, logistic / PFN & 20 & DML & 65.0 & 0.145 & 70.0 & 0.167 & 70.0 & 0.189 \\
 &  & DDML & 95.0 & 0.095 & 100.0 & 0.117 & 100.0 & 0.139 \\
PFN, logistic / forest & 20 & DML & 95.0 & 0.083 & 95.0 & 0.090 & 100.0 & 0.097 \\
 &  & DDML & 95.0 & 0.082 & 100.0 & 0.090 & 100.0 & 0.097 \\
PFN, boosted / forest & 20 & DML & 95.0 & 0.080 & 95.0 & 0.088 & 95.0 & 0.095 \\
 &  & DDML & 95.0 & 0.079 & 95.0 & 0.087 & 95.0 & 0.094 \\
\bottomrule
\end{tabular}
\par\vspace{4pt}\begin{minipage}{\textwidth}\footnotesize
Cov.\ is the percentage of paired draws whose interval contains the true effect. Width is the mean full interval width in marginal outcome SDs. On each draw both methods add the same $B_q$ to their own Wald half-width. The direction in the first column is used by DDML. An unavailable interval counts as noncoverage and is omitted from mean width. Table~\ref{tab:acic-q} reports how often $B_q$ covers DDML's absolute conditional bias.
\end{minipage}
\end{table}

Under linear/logistic fits, the forest correction raises coverage from 74\% to 88\% in setting~27 and from 52.5\% to 67\% in setting~29, while shortening the intervals in both. At $q=2\%$, ordinary DML covers in 83.5\% of draws in setting~27 and 71.0\% in setting~29. With the same allowance, DDML reaches 96.0\% and 86.5\%, respectively, with shorter mean intervals (0.156 versus 0.189 in setting~27 and 0.164 versus 0.198 in setting~29). The PFN pilot also permits a matched comparison on its 20 draws. In setting~29 with the logistic propensity, the PFN direction applied to the linear fit and direct PFN outcome substitution both cover in 19 of 20 draws, with mean widths 0.095 and 0.083. With only five draws per surface, these pilot comparisons are descriptive.

With boosted initial fits in setting~29, both unadjusted intervals cover in 84.5\% of draws. At $q=2\%$, ordinary DML covers in 95.5\% with mean width 0.104, and DDML covers in 94.5\% with mean width 0.103. Here the coverage gain comes from widening the intervals, and ordinary DML benefits at least as much on these draws. For DDML, subtracting each fit's true conditional bias from its estimation error gives a standard deviation of 0.0216, close to the root mean square estimated standard error of 0.0217. The unadjusted DDML intervals cover their conditional means in 94\% of draws, although the 2\% allowance bounds the true absolute conditional bias in only 40.5\% of fits. Near-nominal target coverage can thus coexist with frequent violations of the assumed bias bound. Appendix~\ref{app:acic} reports this diagnostic for every rank-one fit and for the 32-leaf correction.

\subsection{Updating the regression on training or evaluation data}\label{sec:calibrated-comparison}
Theorem~\ref{thm:regression-update} identifies the leave-one-out regression update that reproduces DDML, which invites comparison with a simpler alternative that fits a single slope before evaluation. Using calibration observations held apart from both regression fits, we regress $Y-\gh(X)$ on $f(X):=\widehat g_{\rm PFN}(X)-\gh(X)$ without an intercept and evaluate ordinary DML at $\gh+\widehat c f$ with the original weights (setting $\widehat c=1$ recovers direct PFN substitution). DDML uses the same calibration estimate of $\E[f(X)^2]$ and updates the slope on evaluation data leaving out each scored observation.

For a matched comparison, we reserve the calibration split before fitting either baseline nuisance (Table~\ref{tab:sample-splits}), so every panel compares the same learned direction and weighting fit on identical evaluation observations. The training-fitted alternatives use the ordinary conditional score standard error, whereas DDML uses the full conditional standard error of Section~\ref{sec:vhat}.

\begin{table}[H]
\centering\small
\caption{Matched regression updates for the 401(k) effect}\label{tab:matched-sipp}
\begin{tabular}{@{}lrrrr@{}}
\toprule
Initial fit / update & Estimate & SE & 95\% Wald interval & Interval, $q=2\%$ \\
\midrule
\multicolumn{5}{l}{\textit{Linear / logistic}}\\
Initial DML & 5,876 & 1,072 & [3,774, 7,977] & [2,286, 9,465] \\
Direct PFN & 8,607 & 1,022 & [6,603, 10,611] & [5,197, 12,017] \\
Training slope & 7,534 & 1,022 & [5,531, 9,537] & [4,111, 10,957] \\
DDML & 8,617 & 993 & [6,672, 10,563] & [5,184, 12,051] \\
\multicolumn{5}{l}{\textit{Forest / forest}}\\
Initial DML & 7,389 & 980 & [5,468, 9,310] & [4,059, 10,718] \\
Direct PFN & 7,697 & 966 & [5,804, 9,590] & [4,397, 10,996] \\
Training slope & 7,649 & 965 & [5,758, 9,540] & [4,358, 10,940] \\
DDML & 7,702 & 965 & [5,810, 9,593] & [4,402, 11,002] \\
\bottomrule
\end{tabular}
\par\vspace{4pt}\begin{minipage}{\textwidth}\footnotesize
All amounts are 1991 dollars. The slope and direction norm use the same 1,240 calibration households. The $q=2\%$ intervals use the shared Riesz scale and the training residual scale of each outcome fit.
\end{minipage}
\end{table}

With linear 401(k) fits, the estimated calibration slope is 0.61. The resulting estimate lies between initial DML and direct PFN substitution, while DDML nearly reproduces direct substitution. With forest fits the slope is 0.84 and the training-slope and DDML estimates differ by only \$53 (Table~\ref{tab:matched-sipp}). All four methods in each panel give a positive lower endpoint even at the 2\% budget.

\begin{table}[H]
\centering\small
\caption{Matched regression updates on ACIC}\label{tab:matched-acic}
\begin{tabular}{@{}l*{6}{@{\hspace{.72em}}r}@{}}
\toprule
Update & RMSE & Mean $|b|$ & Cov. & Width & Cov., $q=2\%$ & Width, $q=2\%$ \\
\midrule
\multicolumn{7}{l}{\textit{Setting 27, linear / logistic}}\\
Initial DML & 0.0609 & 0.0428 & 75 & 0.150 & 95 & 0.196 \\
Direct PFN & 0.0312 & 0.0113 & 90 & 0.088 & 90 & 0.103 \\
Training slope & 0.0314 & 0.0103 & 85 & 0.088 & 90 & 0.103 \\
DDML & 0.0320 & 0.0077 & 80 & 0.090 & 100 & 0.137 \\
\multicolumn{7}{l}{\textit{Setting 27, boosted / boosted}}\\
Initial DML & 0.0291 & 0.0070 & 90 & 0.084 & 90 & 0.102 \\
Direct PFN & 0.0337 & 0.0051 & 85 & 0.085 & 90 & 0.100 \\
Training slope & 0.0298 & 0.0056 & 85 & 0.079 & 90 & 0.093 \\
DDML & 0.0299 & 0.0054 & 85 & 0.079 & 90 & 0.096 \\
\multicolumn{7}{l}{\textit{Setting 29, linear / logistic}}\\
Initial DML & 0.1118 & 0.0774 & 60 & 0.178 & 65 & 0.223 \\
Direct PFN & 0.0317 & 0.0200 & 90 & 0.092 & 100 & 0.106 \\
Training slope & 0.0305 & 0.0187 & 100 & 0.092 & 100 & 0.106 \\
DDML & 0.0254 & 0.0127 & 90 & 0.089 & 100 & 0.134 \\
\multicolumn{7}{l}{\textit{Setting 29, boosted / boosted}}\\
Initial DML & 0.0264 & 0.0153 & 90 & 0.092 & 100 & 0.113 \\
Direct PFN & 0.0217 & 0.0115 & 90 & 0.077 & 95 & 0.092 \\
Training slope & 0.0223 & 0.0118 & 90 & 0.077 & 95 & 0.092 \\
DDML & 0.0199 & 0.0101 & 90 & 0.074 & 100 & 0.096 \\
\bottomrule
\end{tabular}
\par\vspace{4pt}\begin{minipage}{\textwidth}\footnotesize
Each setting has 20 repetitions, five per fixed response surface. RMSE, mean absolute conditional bias, and interval width are in marginal outcome standard deviations. Coverage is the percentage of nominal 95\% intervals containing the true effect. The first coverage and width columns use zero bias allowance.
\end{minipage}
\end{table}

In ACIC setting~27 the training-slope update and DDML achieve similar RMSE under either baseline, while in setting~29 DDML attains lower RMSE under both: with boosted fits, RMSE drops from 0.0223 to 0.0199 while both unadjusted intervals cover in 18 of 20 repetitions (Table~\ref{tab:matched-acic}). Rather than establishing a universal ranking across these small 20-draw panels, the comparison isolates the effect of estimating the slope on training versus evaluation data once the direction is fixed.

\subsection{Learning a correction for a dense tail of weak confounders}\label{sec:lasso-results}
Sparse variable-selection methods such as post-Lasso can miss many individually weak controls that push treatment and outcome in the same direction. To measure how much bias a dense tail of weak confounders creates, and how much a single learned direction can remove, we extend the Gaussian linear design of Chetverikov, S\o rensen, and Tsyvinski~\cite{cst}. Each draw consists of $n=1{,}000$ observations and 500 standardized Gaussian controls with Toeplitz covariance $\E[X_jX_k]=\rho^{|j-k|}$. The leading sparse component $S(X):=\sum_{j=1}^{500}2^{1-j}X_j$ puts large weights on the first few controls, while $U(X)$ equals the sum of the last 400 controls divided by its population standard deviation so that the dense tail has unit variance at every correlation level. We generate
\[
 D=S(X)+aU(X)+\nu,\qquad
 Y=D+S(X)+aU(X)+\varepsilon,
\]
with independent standard normal errors $\nu$ and $\varepsilon$ independent of~$X$, so the true treatment coefficient is~1. Even when $\rho=0$ and $a=0.50$, each individual tail control carries a coefficient of only 0.025 in each nuisance equation, yet together the 400 tail controls contribute a common signal $aU(X)$ with standard deviation 0.50. We evaluate tail strengths $a\in\{0.25,0.50\}$ and correlations $\rho\in\{0,0.8\}$.

Across five cross-fitting folds, we fit separate Lasso regressions for the treatment and the reduced-form outcome and refit the selected variables by least squares. We then compare Post-Lasso DML, Triple Post-Lasso (the selected-row Triple score evaluated with those same refits), and our off-diagonal rank-one score~\eqref{eq:tl-U}. Because the Riesz representer in this model is minus the treatment regression, predicting the treatment residuals on training data learns the direction of a Riesz update up to sign. We implement this step first with a ridge regression over all 500 controls, and second with CausalPFN (supplying artificial binary arm labels independent of the training data and averaging the two arm predictions while keeping the actual treatment continuous in the estimating equation). For each learned direction, we estimate its squared norm using 400 independent unlabeled calibration covariates per fold. In each fold we fit the residual learner on the same 800 observations used for post-Lasso and evaluate on the other 200. We solve the corrected moment on the held-out fold and average the five roots. Table~\ref{tab:ranklasso} measures their dispersion across independent draws. Section~\ref{sec:crossfitting} gives the conditions for pooling fold estimates for inference.

\begin{table}[H]
\centering\small
\caption{RMSE when 400 weak controls enter both nuisance functions}\label{tab:ranklasso}
\begin{tabular}{@{}lrrrr@{}}
\toprule
 & \multicolumn{2}{c}{Independent controls} & \multicolumn{2}{c}{Correlated controls ($\rho=0.8$)}\\
\cmidrule(lr){2-3}\cmidrule(l){4-5}
Tail strength $a$ & $0.25$ & $0.50$ & $0.25$ & $0.50$ \\
\midrule
\multicolumn{5}{l}{\textit{A. Ridge comparison on 400 paired draws}}\\
Post-Lasso DML & 0.0663 & 0.1945 & 0.0619 & 0.0748 \\
Triple Post-Lasso & 0.0663 & 0.1961 & 0.0639 & 0.0881 \\
DDML, ridge direction & 0.0617 & 0.1568 & 0.0482 & 0.0637 \\
RMSE reduction (\%) & 6.8 & 19.4 & 22.1 & 14.8 \\
Sampling SD reduction (\%) & 0.5 & 1.9 & 0.2 & 2.8 \\
\addlinespace
\multicolumn{5}{l}{\textit{B. CausalPFN comparison on the first 12 draws}}\\
Post-Lasso DML & 0.0607 & 0.1991 & 0.0595 & 0.0743 \\
Triple Post-Lasso & 0.0569 & 0.1964 & 0.0555 & 0.0750 \\
DDML, ridge direction & 0.0550 & 0.1597 & 0.0407 & 0.0582 \\
DDML, CausalPFN direction & 0.0595 & 0.1865 & 0.0403 & 0.0549 \\
RMSE reduction (\%) & 1.9 & 6.3 & 32.3 & 26.1 \\
Sampling SD reduction (\%) & -3.4 & -7.3 & -1.6 & -2.9 \\
\bottomrule
\end{tabular}
\par\vspace{4pt}\begin{minipage}{\textwidth}\footnotesize
The true treatment coefficient is 1. All methods within a panel share draws, five folds, and post-Lasso nuisance fits. Each percentage compares the featured direction with Post-Lasso DML. Sampling SD is the sample standard deviation of the treatment estimates across draws. A negative reduction indicates greater dispersion.
\end{minipage}
\end{table}

Across 400 paired draws, correcting along the single ridge-learned direction reduces RMSE by 6.8--22.1\% across all four dense-tail designs while slightly lowering sampling dispersion (Table~\ref{tab:ranklasso}, panel A). With independent controls and $a=0.50$, the treatment Lasso selects only 0.17 of the 400 tail variables on average, leaving a bias of 0.192 in Post-Lasso DML. Because the ridge direction achieves a mean squared held-out correlation of 0.242 with the true tail, projecting onto that single direction cuts the bias to 0.153 and lowers RMSE from 0.1945 to 0.1568, whereas Triple Post-Lasso gives an RMSE of 0.1961.

Panel B compares the ridge and CausalPFN directions on the first 12 draws. With independent controls and $a=0.50$, ridge delivers the larger RMSE reduction, whereas with correlated controls ($\rho=0.8$) CausalPFN lowers RMSE slightly further, from 0.0743 to 0.0549 compared with 0.0582 for ridge. At $a=0.25$, the CausalPFN direction reduces RMSE by 1.9\% under independent controls and by 32.3\% under correlated controls. Across all four designs on these 12 draws, the CausalPFN correction increases sampling standard deviation by 1.6--7.3\% relative to Post-Lasso DML, confirming that its RMSE improvements come from lower bias.

\subsection{The geometric design and the value of post-Lasso refitting}\label{sec:lasso-geometric}
When the confounding signal is approximately sparse rather than diffuse, much of the shrinkage bias of Lasso can already be removed by refitting the selected variables. Setting $a=0$ recovers the original geometric design of Chetverikov et al.~\cite{cst}, where all 500 coefficients are nonzero and decay geometrically as $2^{1-j}$. Table~\ref{tab:geometric-lasso} crosses the choice of nuisance fit (penalized Lasso versus post-Lasso refitting) with the choice of score (Double versus Triple) on the same 400 paired draws. The comparison averages the five fold roots and uses the standard deviation of their pooled estimated influence values divided by $\sqrt{1{,}000}$ for the reported Wald intervals. Coverage here is measured against the known coefficient over repeated draws.

\begin{table}[H]
\centering\small
\caption{Refitting and the Triple Lasso score with geometrically decaying coefficients}\label{tab:geometric-lasso}
\begin{tabular}{@{}lrrrrrr@{}}
\toprule
 & \multicolumn{3}{c}{Independent controls} & \multicolumn{3}{c}{Correlated controls ($\rho=0.8$)}\\
\cmidrule(lr){2-4}\cmidrule(l){5-7}
Nuisance fits and score & Bias & RMSE & Cov. (\%) & Bias & RMSE & Cov. (\%) \\
\midrule
Post-Lasso, Double & 0.00184 & 0.0313 & 96.75 & -0.00019 & 0.0315 & 96.25 \\
Post-Lasso, Triple & 0.00172 & 0.0315 & 96.25 & 0.00017 & 0.0323 & 95.00 \\
Lasso, Double & 0.04570 & 0.0542 & 67.50 & 0.01674 & 0.0348 & 93.00 \\
Lasso, Triple & 0.01496 & 0.0353 & 93.00 & -0.00680 & 0.0336 & 95.75 \\
\bottomrule
\end{tabular}
\par\vspace{4pt}\begin{minipage}{\textwidth}\footnotesize
The added tail has strength $a=0$. The 500 geometric coefficients remain in both structural nuisances. All estimators use the same 400 paired draws and five folds. Coverage refers to nominal 95\% intervals.
\end{minipage}
\end{table}

Under independent controls, applying Triple Lasso to penalized Lasso fits cuts bias from 0.04570 to 0.01496 and raises nominal 95\% coverage by 25.5 percentage points (from 67.50\% to 93.00\%), at the cost of a 4.3\% average increase in estimated standard error (and 8.1\% under correlated controls). Switching to post-Lasso nuisance fits with the Double score, however, already brings bias down to 0.00184 and coverage to 96.75\%, so moving to the Triple score changes coverage by only $-0.5$ percentage points and increases the standard error by 0.1\% (1.8\% under correlated controls). At $\rho=0.8$, both post-Lasso estimators have absolute bias below 0.0002. In the geometric design, post-Lasso refitting alone captures essentially all of the practical gain from removing shrinkage bias, whereas Table~\ref{tab:ranklasso} shows that learning a direction beyond the selected controls still yields substantial reductions when a dense tail of weak confounders is present.

There is a simple algebraic reason why a score correction on selected controls closely tracks a least-squares refit. On a common sample, let $z$ collect a shared set of selected controls with invertible sample Gram matrix $G_n=\Pn[zz']$, and let $\Pi_n$ denote sample least-squares projection onto~$z$. For the residuals $v$ and $a_\beta=u-\beta v$ from Section~\ref{sec:triple-lasso}, the normal equations imply
\[
 \Pn[a_\beta v]-\Pn[a_\beta z']G_n^{-1}\Pn[zv]
 =\Pn[(a_\beta-\Pi_n a_\beta)(v-\Pi_n v)].
\]
Whenever both initial linear fits lie in the span of~$z$ and the residualized treatment has positive sample variance, solving the right-hand estimating equation produces the exact treatment coefficient obtained by refitting both nuisances on~$z$. In practice, the Triple Lasso score uses selected rows of a training-fold inverse-Gram estimate together with residual moments on the held-out fold, while post-Lasso enforces the normal equations separately on each regression's own training support, which is why the two estimators can still differ slightly on a given draw.

\section{Connections to higher-order orthogonality and influence functions}\label{sec:theory-connections}
Proposition~\ref{prop:bias} gives an exact formula for the bias left by any pairwise kernel correction. To situate that formula within the broader econometric theory of second-order debiasing, we first separate four distinct notions of ``higher order'' that are often conflated in practice, then show how working-space projection achieves exact second-order Neyman orthogonality inside~$\mathcal H_k$ while leaving quadratic sensitivity outside it.

\subsection{Separating four notions of ``higher order''}
The literature on second-order debiasing uses ``order'' in four different senses, summarized in Table~\ref{tab:order-terms}. For a moment function~$\psi$ evaluated on $\ell$ independent copies of the data, let $\Psi(\theta,\eta):=\E[\psi(W_1,\ldots,W_\ell;\theta,\eta)]$ denote its expectation under the true data-generating law. The moment is \emph{Neyman orthogonal of order~$q$} if it is centered at the truth, $\Psi(\theta_0,\eta_0)=0$, and its directional derivatives with respect to the nuisance~$\eta$ vanish through order~$q$,
\begin{equation}\label{eq:orthogonality-definition}
 D_\eta^j\Psi(\theta_0,\eta_0)[h_1,\ldots,h_j]=0,
 \qquad j=1,\ldots,q,
\end{equation}
for all admissible nuisance directions. When the $(q+1)$-st derivative is locally bounded, Taylor expansion leaves a bias remainder of order $\norm{\eta-\eta_0}^{q+1}$, which is quadratic for $q=1$ (ordinary DML) and cubic for $q=2$.

\begin{table}[H]
\centering\small
\caption{Distinguishing four notions of ``higher order'' in debiased estimation}
\label{tab:order-terms}
\begin{tabularx}{\textwidth}{@{}p{0.30\textwidth}X@{}}
\toprule
Term & Property being described \\
\midrule
Neyman orthogonality of order $q$ & Derivatives through order $q$ vanish in a specified nuisance space at a fixed data distribution. \\
Higher-order influence function & A kernel represents higher pathwise derivatives of a target as the data distribution varies. \\
U-statistic of order $\ell$ & The estimator kernel takes $\ell$ distinct observations as arguments. \\
Higher-order bias reduction & The remaining bias is strictly smaller along the actual realized nuisance estimation errors. \\
\bottomrule
\end{tabularx}
\end{table}

Using an order-2 pairwise U-statistic establishes neither second-order orthogonality nor faster bias rates by itself. In the moment-condition framework of Bonhomme, Jochmans, Newey, and Weidner~\cite{bjnw} (Section~2.3), the nuisance vector explicitly includes an inverse Jacobian, and orthogonality covers mixed derivatives jointly in the original nuisance and this inverse using independent copies (their eq.~(9) takes the form of a U-statistic). By contrast, the likelihood construction of Bonhomme, Jochmans, and Weidner~\cite{bjw-likelihood} (Section~5) uses a specified conditional density~$f$ to project onto generalized scores $f^{-1}\partial_\eta^j f$ ($j=1,\ldots,q$) from single model observations such as panels or networks, with \emph{full orthogonality} demanding zero expected moment globally across nuisance values.

\paragraph{Finite-dimensional linear contrast illustration}
A simple linear-regression example shows how DDML reproduces second-order Neyman orthogonality when the model is finite-dimensional. Consider $g_0(x)=b(x)'\beta_0$ with invertible Gram matrix $\Sigma:=\E[b(X)b(X)']$, linear contrast $\theta_0=c'\beta_0$, and functional $m(W,b'\beta):=c'\beta$. Given a symmetric training estimate $M\approx\Sigma^{-1}$, choosing
\begin{equation}\label{eq:finite-choice}
 \gh=b'\widehat\beta,\qquad \ah=b'Mc,\qquad
 \Kh(x,x')=b(x)'Mb(x')
\end{equation}
in~\eqref{eq:pair} produces the corrected estimator
\begin{equation}\label{eq:finite-estimator}
 \thh_K=c'\widehat\beta+\frac1n\sum_j
 c'\{2M-M\widehat\Sigma_{-j}M\}b(X_j)
       \{Y_j-b(X_j)'\widehat\beta\},
\end{equation}
where $\widehat\Sigma_{-j}:=(n-1)^{-1}\sum_{i\ne j}b(X_i)b(X_i)'$. Comparing the ordinary DML estimator with~$\thh_K$, their conditional biases are
\begin{equation}\label{eq:finite-bias}
 \E[\thh_{\rm DB}]-\theta_0=c'(I-M\Sigma)(\widehat\beta-\beta_0),
 \qquad
 \E[\thh_K]-\theta_0=c'(I-M\Sigma)^2(\widehat\beta-\beta_0).
\end{equation}
Because $I-M\Sigma=-(M-\Sigma^{-1})\Sigma$, the corrected bias is cubic jointly in $(\widehat\beta-\beta_0,M-\Sigma^{-1})$, achieving exact second-order Neyman orthogonality. Setting $r(W,\beta)=b(X)\{Y-b(X)'\beta\}$ and $\Lambda=-M$ in eq.~(35) of~\cite{bjnw} yields the score
\[
 c'\widehat\beta+2c'Mb(X_2)\varepsilon_g(W_2)
 -c'Mb(X_3)b(X_3)'Mb(X_2)\varepsilon_g(W_2),
\]
whose average over distinct observations reproduces~\eqref{eq:finite-estimator} exactly. Only two copies are needed in this example because the linear contrast vector~$c$ is data-independent, whereas random target derivatives or nonlinear functionals require additional structure.

\subsection{Second-order orthogonality inside a working space versus bias outside}
Returning to the general linear functional~\eqref{eq:target}, fix a working space $\mathcal H_k=\operatorname{span}\{b_1,\ldots,b_k\}$ with positive-definite Gram matrix $\Sigma=\E[b(X)b(X)']$ and kernel $K_M(x,x'):=b(x)'Mb(x')$. Let $\mathcal M_M(g,\alpha)$ denote the population mean of the DDML estimator~\eqref{eq:pair} evaluated at candidate functions $(g,\alpha)$, and define the projected error moments $u_g:=\E[b(X)\delta_g(X)]$ and $u_\alpha:=\E[b(X)\delta_\alpha(X)]$.

\begin{proposition}[Working-space orthogonality and approximation bias]\label{prop:working-orthogonality}
Under the conditions of Proposition~\ref{prop:bias} and the preceding basis assumptions, let $\Pi$ be the population orthogonal projection onto~$\mathcal H_k$. For arbitrary square-integrable candidate errors,
\begin{equation}\label{eq:orthogonality-bias-decomposition}
 \mathcal M_M(g,\alpha)-\theta_0
 =-\ip{(I-\Pi)\delta_\alpha}{(I-\Pi)\delta_g}
   +u_\alpha'(M-\Sigma^{-1})u_g.
\end{equation}
Define $\beta_{\mathcal H},\gamma_{\mathcal H}\in\R^k$ via $g_{\mathcal H}=b'\beta_{\mathcal H}$ and $\alpha_{\mathcal H}=b'\gamma_{\mathcal H}$ in~\eqref{eq:short-target}. If the candidates lie in~$\mathcal H_k$, then
\begin{equation}\label{eq:orthogonality-cubic}
 \mathcal M_M(b'\beta,b'\gamma)-\theta_{\mathcal H}
 = (\gamma-\gamma_{\mathcal H})'\Sigma
   (M-\Sigma^{-1})\Sigma(\beta-\beta_{\mathcal H}).
\end{equation}
Consequently, $\mathcal M_M(b'\beta,b'\gamma)-\theta_{\mathcal H}$ is second-order Neyman orthogonal jointly in $(\beta,\gamma,M)$ at $(\beta_{\mathcal H},\gamma_{\mathcal H},\Sigma^{-1})$. If $g_0,\alpha_0\in\mathcal H_k$, this property holds for~$\theta_0$ itself.
\end{proposition}

At fixed~$k$, the cubic remainder in~\eqref{eq:orthogonality-cubic} is bounded by $\norm{\Sigma}_{\op}^2\norm{\gamma-\gamma_{\mathcal H}}\norm{M-\Sigma^{-1}}_{\op}\norm{\beta-\beta_{\mathcal H}}$, while as $k$ grows, operator norms, convergence rates, and sampling variance must still be controlled explicitly. When $\mathcal H_k$ is misspecified, exact second-order orthogonality for the short target~$\theta_{\mathcal H}$ coexists with the approximation bias $\theta_0-\theta_{\mathcal H}$. For unrestricted ML fits, \eqref{eq:orthogonality-bias-decomposition} preserves the residual inner product outside~$\mathcal H_k$, which is the exact term controlled by the sensitivity bounds of Section~\ref{sec:sensitivity}.

This decomposition also explains why no fixed square-integrable kernel can achieve exact second-order orthogonality over \emph{all} of $L^2(P_X)$. Holding~$K$ fixed, Proposition~\ref{prop:bias} gives the mixed second derivative
\begin{equation}\label{eq:orthogonality-mixed-derivative}
 D_\alpha D_g\mathcal M_K(g_0,\alpha_0)[a,h]
 =-\ip{a}{(I-T)h}.
\end{equation}
For this derivative to vanish for all $a,h\in L^2(P_X)$, the operator~$T$ would have to equal the identity~$I$, which is impossible for a compact integral operator on an infinite-dimensional $L^2(P_X)$ space. An exact working projection achieves second-order orthogonality \emph{within}~$\mathcal H_k$ while controlling remaining quadratic sensitivity \emph{outside}~$\mathcal H_k$.

\subsection{Higher-order influence functions and repeated corrections}
Our pairwise estimator connects closely to higher-order influence functions (HOIFs)~\cite{robins,robins2017}, which likewise estimate products of nuisance errors using independent observation copies, and to stabilized HOIFs~\cite{stabilized}, which address instability in estimated Gram inverses. In the Markov off-policy setting of Shi et al.~\cite{shi}, Bellman residuals and conditional visitation weights play the operator roles of regression residuals and covariate kernels, though temporal dependence requires its own asymptotic analysis.

To apply the correction~$r$ times, associate each independent observation~$W_j$ with the update operator $(\mathcal D_j g)(x):=g(x)+\Kh(x,X_j)\{Y_j-g(X_j)\}$, whose expectation satisfies $\E[\mathcal D_j g]-g_0=(I-T)(g-g_0)$. Composing $r$ updates over distinct observations, evaluating the ordinary DML score on one more distinct observation, and averaging over all distinct $(r+1)$-tuples yields
\begin{equation}\label{eq:iterate}
 \E[\thh^{(r)}]-\theta_0
 =-\ip{\delta_\alpha}{(I-T)^r\delta_g},\qquad 0\le r\le n-1,
\end{equation}
provided the iterated scores are integrable. Here $r=0$ recovers ordinary DML and $r=1$ gives the leave-one-out regression update of Theorem~\ref{thm:regression-update}. For an exact projection~$\Pi$, however, idempotence gives $(I-\Pi)^r=I-\Pi$ for every $r\ge 1$. Repeating a fixed projection cannot remove omitted-direction bias outside~$\mathcal H_k$, while it adds higher-order Hoeffding terms that are not covered by our $r=1$ variance formula~\eqref{eq:vhat}.

\section{Conclusion}\label{sec:conclusion}
Deeply debiased machine learning offers a pragmatic middle ground between ordinary DML and full higher-order influence function theory. Because no square-integrable kernel can achieve exact second-order orthogonality over all of $L^2(P_X)$, any practical second-order correction must choose which error directions to target. Learning a low-dimensional working space directly from first-stage residuals lets a flexible learner compress high-dimensional confounding patterns into one or a few directions whose mixed-bias contribution can be estimated on an independent sample. Theorem~\ref{thm:regression-update} shows that this pairwise U-statistic correction is algebraically identical to ordinary DML evaluated with a leave-one-out post-hoc regression (or Riesz-weight) update, while the Hoeffding decomposition provides exact finite-sample variance accounting and clear conditions under which ordinary DML standard errors and cross-fitting remain valid.

Across our empirical and simulation studies, the practical role of a second debiasing step depends on the strength of the initial learners. When the initial fits miss systematic nonlinearities or a diffuse tail of weak confounders, a single learned direction substantially lowers bias and root mean squared error. When the initial learners are already strong, the estimated correction is typically small relative to sampling uncertainty. In that regime, DDML acts primarily as a diagnostic check along the searched directions, while the two-factor sensitivity framework of Section~\ref{sec:sensitivity} quantifies how much remaining misspecification bias (both outside the learned space and along it from normalization error) the substantive conclusion can tolerate.

Two main directions stand out for future work. First, our residual learners currently target either the outcome error or the Riesz error on its own, whereas the mixed bias depends on their alignment $\ip{\delta_\alpha}{\delta_g}$. Designing regularized training objectives that search directly for covariate directions where both observable moment violations align, without inflating variance, could sharpen bias removal further. Second, while a single pairwise correction already exhausts the gain from a low-rank projection for static linear functionals, extending targeted residual-direction learning and working-space sensitivity bounds to nonlinear moment conditions and dynamic sequential settings such as Markov off-policy evaluation (connecting to the multi-step deep debiasing of Shi et al.~\cite{shi}), alongside cross-fitted variance estimation beyond the oracle-variance regime, remains an important agenda.

\clearpage
\begin{appendix}
\section{Notation}\label{app:notation}
Table~\ref{tab:notation} collects the main symbols used across the paper. Conditional on the training sample~$\calT$, all expectations, variances, inner products, and $L^2(P_X)$ norms are taken over the independent evaluation distribution~$P$ unless stated otherwise.

\begin{table}[htbp]
\centering\small
\caption{Summary of main notation}
\label{tab:notation}
\begin{tabularx}{\textwidth}{@{}p{0.23\textwidth}X@{}}
\toprule
Symbol & Meaning \\
\midrule
$W, X, Y$ & Observation $W$ with covariates $X\subset W$ and scalar outcome $Y$ \\
$g_0(x),\ \alpha_0(X),\ \theta_0$ & True regression $\E[Y\mid X=x]$, Riesz representer, and target $\E[m(W,g_0)]$ \\
$\calT,\ \Pn$ & Training-sample sigma-field and evaluation-sample average $n^{-1}\sum_{i=1}^n$ \\
$\mathbb{U}_n h$ & U-average $\binom{n}{2}^{-1}\sum_{i<j}h(W_i,W_j)$ over unordered pairs for a symmetric kernel $h$ \\
$(n)_r$ & Number of ordered $r$-tuples of distinct indices, $n(n-1)\cdots(n-r+1)$. In particular, $(n)_3=n(n-1)(n-2)$. \\
$\gh,\ \ah,\ \delta_g,\ \delta_\alpha$ & Training fits and nuisance errors $\delta_g:=\gh-g_0$, $\delta_\alpha:=\ah-\alpha_0$ \\
$\ip{f}{h},\ \norm{f}_2$ & Inner product $\E[f(X)h(X)]$ and $L^2(P_X)$ norm $\ip{f}{f}^{1/2}$ \\
$s(W),\ \thh_{\rm DB}$ & Ordinary DML score $m(W,\gh)+\ah(X)\{Y-\gh(X)\}$ and estimator $\Pn s$ \\
$\Kh(x,x'),\ T,\ T^*$ & Training-chosen kernel, integral operator $(Tf)(x):=\E[\Kh(x,X)f(X)]$, and adjoint \\
$\mathcal H_k,\ \Pi,\ \delta_j^\perp$ & $k$-dimensional working space, $L^2(P_X)$ projection onto $\mathcal H_k$, and $(I-\Pi)\delta_j$ \\
$A(W;x'),\ \varepsilon_g(W),\ H$ & Riesz discrepancy, outcome residual $Y-\gh(X)$, and pair kernel $A(W;X')\varepsilon_g(W')$ \\
$\widehat B_K,\ \thh_K,\ b_K$ & U-statistic correction, DDML estimator $\thh_{\rm DB}-\widehat B_K$, and bias $\E[\thh_K]-\theta_0$ \\
$h,\ L,\ d,\ \phi_K$ & Symmetrized kernel, Hoeffding linear and degenerate parts, and effective IF $s-\E[s]-L$ \\
$q(w,w'),\ \tau_n^2$ & Combined pair kernel $\tfrac12\{s(w)+s(w')\}-h(w,w')$ and conditional variance $\Var(\thh_K\mid\calT)$ \\
$\widehat V,\ \widehat V_1,\ \widehat V_2,\ \widehat V_{\rm lin}$ & Unbiased full variance estimator ($\widehat V_1+\widehat V_2$) and jackknife linear variance \\
$g_{\mathcal H},\ \alpha_{\mathcal H},\ \theta_{\mathcal H}$ & Short-model projections $\Pi g_0$, $\Pi\alpha_0$, and short target $\E[m(W,g_{\mathcal H})]$ \\
$r_g,\ r_\alpha,\ \lambda_\perp$ & Residual errors $(I-\Pi)\delta_g$, $(I-\Pi)\delta_\alpha$, and their $L^2$ cosine \\
$\epsilon_j,\ \rho_j,\ B_q$ & Global error radii, residual error radii, and two-factor $q$-budget bias allowance \\
$R_{j,T},\ s_Y,\ s_\alpha$ & Weighted mean squared nuisance error, training outcome residual RMS, and shared debiased Riesz scale \\
\bottomrule
\end{tabularx}
\end{table}

\section{Proofs}\label{app:proofs}
Throughout this appendix, all expectations, variances, covariances, and $L^2(P_X)$ norms involving training-fitted objects ($\gh$, $\ah$, $\Kh$, $T$, $\Pi$, and $\mathcal H_k$) are taken conditional on the training sigma-field~$\calT$ unless stated otherwise.

\begin{proof}[Proof of Theorem~\ref{thm:regression-update}]
Fix the training fits $(\gh,\ah,\Kh)$ and the evaluation sample $W_1,\ldots,W_n$. For each $i\in\{1,\ldots,n\}$, write the leave-one-out regression increment in~\eqref{eq:updated-regression} as
\[
 u_i(x):=\widetilde g_{-i}(x)-\gh(x)
 =\frac{1}{n-1}\sum_{j\ne i}\Kh(x,X_j)\varepsilon_g(W_j),
\]
where $\varepsilon_g(W_j)=Y_j-\gh(X_j)$. Because $\widetilde g_{-i}=\gh+u_i$ and the map $g\mapsto m(W_i,g)$ is linear on a domain containing $\gh$ and every kernel column $\Kh(\cdot,X_j)$, substituting $\widetilde g_{-i}$ into the ordinary DML score~\eqref{eq:db} for observation~$i$ gives
\begin{align*}
 m(W_i,\widetilde g_{-i})+\ah(X_i)\{Y_i-\widetilde g_{-i}(X_i)\}
 &=m(W_i,\gh)+\ah(X_i)\{Y_i-\gh(X_i)\}\\
 &\quad+m(W_i,u_i)-\ah(X_i)u_i(X_i)\\
 &=s(W_i)-\big\{\ah(X_i)u_i(X_i)-m(W_i,u_i)\big\}.
\end{align*}
Expanding the bracketed adjustment using the definition of~$u_i$ and linearity of $m(W_i,\cdot)$ yields
\begin{align*}
 &\ah(X_i)u_i(X_i)-m(W_i,u_i)\\
 &\quad =\frac{1}{n-1}\sum_{j\ne i}
   \big\{\ah(X_i)\Kh(X_i,X_j)-m(W_i,\Kh(\cdot,X_j))\big\}\varepsilon_g(W_j)\\
 &=\frac{1}{n-1}\sum_{j\ne i}A(W_i;X_j)\varepsilon_g(W_j)
  =\frac{1}{n-1}\sum_{j\ne i}H(W_i,W_j),
\end{align*}
by the definitions of $A(W_i;x')$ and $H(W_i,W_j)$ preceding~\eqref{eq:pair}. Averaging over $i=1,\ldots,n$ gives
\[
 \frac{1}{n}\sum_{i=1}^n
 \big[m(W_i,\widetilde g_{-i})+\ah(X_i)\{Y_i-\widetilde g_{-i}(X_i)\}\big]
 =\Pn s-\frac{1}{n(n-1)}\sum_{i\ne j}H(W_i,W_j)
 =\thh_{\rm DB}-\widehat B_K,
\]
which equals $\thh_K$ and establishes~\eqref{eq:ddml-regression-score}.
\end{proof}

\begin{proof}[Proof of Proposition~\ref{prop:bias}]
Let $W$ and $W'$ be independent draws from the evaluation distribution~$P$ conditional on~$\calT$, with covariate components $X$ and~$X'$. Because $\widehat B_K$ is an average over all $n(n-1)$ ordered pairs $(i,j)$ with $i\ne j$ and each pair $(W_i,W_j)$ has the same joint distribution as $(W,W')$, linearity of expectation gives $\E[\widehat B_K]=\E[H(W,W')]$.

We evaluate $\E[H(W,W')]$ in two steps, integrating out the residual observation~$W'$ first. By the definition of the conditional mean $g_0(x')=\E[Y'\mid X'=x']$, the outcome residual $\varepsilon_g(W')=Y'-\gh(X')$ satisfies
\[
 \E[\varepsilon_g(W')\mid W, X']
 =\E[Y'-\gh(X')\mid X']
 =g_0(X')-\gh(X')
 =-\delta_g(X'),
\]
where the first equality uses independence of $W'$ from~$W$ conditional on~$\calT$. Multiplying by $\Kh(x,X')$ and taking expectations over~$X'$ gives, for every fixed~$x$,
\[
 \E[\Kh(x,X')\varepsilon_g(W')\mid W]
 =-\E[\Kh(x,X')\delta_g(X')]
 =-(T\delta_g)(x).
\]
Now substitute this identity into $H(W,W')=A(W;X')\varepsilon_g(W')$. Using the definition of $A(W;x')$ preceding~\eqref{eq:pair} together with linearity of~$m(W,\cdot)$ and the integrability condition allowing expectation over~$W'$ to pass inside~$m$, we obtain
\begin{align*}
 \E[H(W,W')\mid W]
 &=\E[\{\ah(X)\Kh(X,X')-m(W,\Kh(\cdot,X'))\}\varepsilon_g(W')\mid W]\\
 &=-\ah(X)(T\delta_g)(X)+m(W,T\delta_g).
\end{align*}
Taking the outer expectation over~$W$ and applying the Riesz representation identity $\E[m(W,T\delta_g)]=\ip{\alpha_0}{T\delta_g}$ from~\eqref{eq:riesz} yields
\[
 \E[\widehat B_K]
 =\E[H(W,W')]
 =-\E[\ah(X)(T\delta_g)(X)]+\E[\alpha_0(X)(T\delta_g)(X)]
 =-\ip{\delta_\alpha}{T\delta_g}.
\]
Finally, combining this expression with the ordinary DML bias identity $\E[\thh_{\rm DB}]-\theta_0=-\ip{\delta_\alpha}{\delta_g}$ in~\eqref{eq:mixed} gives
\[
 b_K
 =\E[\thh_{\rm DB}-\widehat B_K]-\theta_0
 =-\ip{\delta_\alpha}{\delta_g}+\ip{\delta_\alpha}{T\delta_g}
 =-\ip{\delta_\alpha}{(I-T)\delta_g}.
\]
The equivalent form $b_K=-\ip{(I-T^*)\delta_\alpha}{\delta_g}$ follows directly from the definition of the $L^2(P_X)$ adjoint~$T^*$.
\end{proof}

\begin{proof}[Proof of Proposition~\ref{prop:working-balls}]
Because $\Pi$ is the $L^2(P_X)$ orthogonal projection onto~$\mathcal H_k$, the complementary operator $I-\Pi$ is also self-adjoint and idempotent, meaning $(I-\Pi)^*=I-\Pi$ and $(I-\Pi)^2=I-\Pi$. For any pair $(a,v)\in\mathcal U_\Pi$, these two properties together with the Cauchy--Schwarz inequality give
\begin{align*}
 |\ip{a}{(I-\Pi)v}|
 &=|\ip{(I-\Pi)a}{(I-\Pi)v}|\\
 &\le \norm{(I-\Pi)a}_2\norm{(I-\Pi)v}_2
 \le \rho_\alpha\rho_g.
\end{align*}
To show that this upper bound is sharp, suppose the orthogonal complement $\mathcal H_k^\perp$ is nontrivial, so we can pick a unit vector $e\in\mathcal H_k^\perp$ with $\norm{e}_2=1$ and $(I-\Pi)e=e$. Setting $a:=\rho_\alpha e$ and $v:=\rho_g e$, we have $\norm{a}_2=\norm{(I-\Pi)a}_2=\rho_\alpha\le\epsilon_\alpha$ and $\norm{v}_2=\norm{(I-\Pi)v}_2=\rho_g\le\epsilon_g$, so $(a,v)\in\mathcal U_\Pi$. In addition, $|\ip{a}{(I-\Pi)v}|=\rho_\alpha\rho_g\ip{e}{e}=\rho_\alpha\rho_g$, which establishes~\eqref{eq:working-sup}. Setting $\rho_\alpha=\epsilon_\alpha$ and $\rho_g=\epsilon_g$ gives the global-ball supremum $\epsilon_\alpha\epsilon_g$.

For the approximate projection bound~\eqref{eq:working-approx}, suppose $T=\Pi T\Pi$ and $\norm{\Pi-T}_{\op}\le\omega$. Because $\Pi^2=\Pi$, we have $(I-T)\Pi\delta_g=(\Pi-T)\Pi\delta_g$. Applying the triangle inequality and Cauchy--Schwarz to the exact decomposition~\eqref{eq:approx-projection} yields
\begin{align*}
 |b_K|
 &\le |\ip{\delta_\alpha^\perp}{\delta_g^\perp}|
      +|\ip{\Pi\delta_\alpha}{(\Pi-T)\Pi\delta_g}|\\
 &\le \norm{\delta_\alpha^\perp}_2\norm{\delta_g^\perp}_2
      +\norm{\Pi\delta_\alpha}_2\norm{(\Pi-T)\Pi\delta_g}_2\\
 &\le \norm{\delta_\alpha^\perp}_2\norm{\delta_g^\perp}_2
      +\omega\norm{\Pi\delta_\alpha}_2\norm{\Pi\delta_g}_2,
\end{align*}
which completes the proof.
\end{proof}

\begin{proof}[Proof of Corollary~\ref{cor:learned-bias-rate}]
Conditional on the training sample~$\calT$, the integral operator~$T$ induced by $K_f(x,x')=f(x)f(x')/\widehat\nu_f$ acts on any $u\in L^2(P_X)$ as
\[
 (Tu)(x)
 =\E\!\left[\frac{f(x)f(X)}{\widehat\nu_f}u(X)\right]
 =\frac{\nu_f}{\widehat\nu_f}\cdot\frac{\ip{f}{u}}{\nu_f}f(x)
 =\frac{\nu_f}{\widehat\nu_f}(\Pi_f u)(x),
\]
where $\Pi_f$ is the $L^2(P_X)$ orthogonal projection onto the one-dimensional working space $\mathcal H_1:=\operatorname{span}\{f\}$. Thus $T=\Pi_f T\Pi_f$ and
\[
 \norm{\Pi_f-T}_{\op}
 =\left|1-\frac{\nu_f}{\widehat\nu_f}\right|\norm{\Pi_f}_{\op}
 =\omega_f.
\]
Applying the approximate-projection bound~\eqref{eq:working-approx} from Proposition~\ref{prop:working-balls} with $\Pi=\Pi_f$ and $\omega=\omega_f$, and substituting $\norm{(I-\Pi_f)\delta_g}_2=\Delta_f$, gives the first inequality in~\eqref{eq:learned-bias-rate}:
\[
 |b_{K_f}|
 \le \norm{(I-\Pi_f)\delta_\alpha}_2\Delta_f
     +\omega_f\norm{\Pi_f\delta_\alpha}_2\norm{\Pi_f\delta_g}_2.
\]
Because both $\Pi_f$ and $I-\Pi_f$ are orthogonal projections, they are contractions ($\norm{\Pi_f}_{\op}\le 1$ and $\norm{I-\Pi_f}_{\op}\le 1$), so $\norm{(I-\Pi_f)\delta_\alpha}_2\le\norm{\delta_\alpha}_2$, $\norm{\Pi_f\delta_\alpha}_2\le\norm{\delta_\alpha}_2$, and $\norm{\Pi_f\delta_g}_2\le\norm{\delta_g}_2$. Substituting these bounds yields the second inequality in~\eqref{eq:learned-bias-rate}. If either upper bound is $o_p(n^{-1/2})$, then $b_{K_f}=o_p(n^{-1/2})$, which verifies the bias condition in~\eqref{eq:ordinary-conditions} so Proposition~\ref{prop:ordinary} applies.
\end{proof}

\begin{proof}[Proof of the residual-regression rate~\eqref{eq:learned-series-rate}]
We establish the rate in four steps, first conditioning on the initial nuisance fits $(\gh,\ah)$ to bound the shape error~$\Delta_f$, and then conditioning on the learned feature~$f$ to bound the calibration error~$\omega_f$.

\emph{Step 1: Population projection and residual orthogonality.}
Condition on $(\gh,\ah)$ and write $Z:=Y-\gh(X)$, $\xi:=\xi_p(X)$, and $G:=\E[\xi\xi']$. Let $\xi'\beta_p$ be the population $L^2(P_X)$ least-squares projection of $v=g_0-\gh=-\delta_g$ onto $\operatorname{span}\{\xi_p\}$, and set $e:=Z-\xi'\beta_p$. By definition of~$v$, we have $\E[Z\mid X]=v(X)$, so the population normal equations give
\[
 \E[\xi e]=\E[\xi(v-\xi'\beta_p)]=0,
\]
and by orthogonality of $v-\xi'\beta_p$ and $\xi'\beta_p$ (together with $\E[(Z-v)\xi'\beta_p]=0$),
\[
 \E[e^2]=\E[(Z-v)^2]+\E[(v-\xi'\beta_p)^2]
 \le \E[(Z-v)^2]+\E[v^2]
 =\E[Z^2]=O_p(1).
\]

\emph{Step 2: Gram matrix concentration and estimation error for $\Delta_f$.}
Let $\widehat G:=m^{-1}\sum_{i=1}^m\xi_i\xi_i'$ be the sample Gram matrix on the $m$ fresh residual-learning observations, where $\xi_i:=\xi_p(X_i)$. Because the entries of $\xi$ are uniformly bounded, each entry of $\widehat G-G$ is an average of $m$ iid mean-zero bounded random variables with variance $O(m^{-1})$. Summing over all $p^2$ entries gives $\E[\norm{\widehat G-G}_{\mathrm F}^2]=O(p^2/m)$, so $\norm{\widehat G-G}_{\op}\le\norm{\widehat G-G}_{\mathrm F}=O_p(p/\sqrt m)=o_p(1)$ under $p=o(\sqrt m)$. Since the smallest eigenvalue $\lambda_{\min}(G)$ is bounded away from zero, Weyl's inequality implies that $\lambda_{\min}(\widehat G)\ge \tfrac12\lambda_{\min}(G)>0$ with probability tending to one.

On that event, the least-squares estimator $\widehat\beta$ satisfies the sample normal equations $\widehat G(\widehat\beta-\beta_p)=m^{-1}\sum_{i=1}^m \xi_i e_i$, where $e_i:=Z_i-\xi_i'\beta_p$. Conditional on $(\gh,\ah)$, the summands $\xi_i e_i$ are iid with mean zero, so
\[
 \E\!\left[\left\|\frac1m\sum_{i=1}^m \xi_i e_i\right\|^2\;\middle|\;\gh,\ah\right]
 =\frac{\E[\|\xi\|^2 e^2\mid\gh,\ah]}{m}
 \lesssim \frac{p\,\E[e^2\mid\gh,\ah]}{m}
 =O_p(p/m),
\]
where $\|\cdot\|$ denotes the Euclidean norm and we used uniform boundedness of the coordinates of~$\xi$ to bound $\|\xi\|^2\lesssim p$. Combining this with the lower bound on $\lambda_{\min}(\widehat G)$ gives $\|\widehat\beta-\beta_p\|=O_p(\sqrt{p/m})$. Because the largest eigenvalue of~$G$ is bounded above, the prediction error of $f=\xi'\widehat\beta$ obeys
\[
 \norm{\xi'(\widehat\beta-\beta_p)}_2
 =\left\{(\widehat\beta-\beta_p)'G(\widehat\beta-\beta_p)\right\}^{1/2}
 \lesssim \|\widehat\beta-\beta_p\|
 =O_p(\sqrt{p/m}).
\]
Taking $a=-1$ in the definition of $\Delta_f$ and applying the triangle inequality with $v=-\delta_g$ therefore yields
\[
 \Delta_f
 \le \norm{\delta_g+f}_2
 =\norm{v-f}_2
 \le \norm{v-\xi'\beta_p}_2+\norm{\xi'(\widehat\beta-\beta_p)}_2
 \le a_p+O_p(\sqrt{p/m}).
\]

\emph{Step 3: Calibration ratio $\omega_f$.}
Now condition on both $(\gh,\ah)$ and the fitted feature~$f$. Whenever $\nu_f=\E[f(X)^2]>0$, the calibration estimator $\widehat\nu_f=\ell^{-1}\sum_{j=1}^\ell f(X_j)^2$ is an average of $\ell$ iid draws with mean~$\nu_f$, so the ratio $\widehat\nu_f/\nu_f$ has conditional mean~$1$ and conditional variance
\[
 \Var\!\left(\frac{\widehat\nu_f}{\nu_f}\;\middle|\;\gh,\ah,f\right)
 =\frac{\E[f(X)^4]/\nu_f^2-1}{\ell}
 =O_p(\ell^{-1}).
\]
By conditional Chebyshev's inequality, $\widehat\nu_f/\nu_f=1+O_p(\ell^{-1/2})$. In particular, $\Pr(\widehat\nu_f=0\mid\gh,\ah,f)\to_p 0$ on the event $\nu_f>0$, and taking reciprocals gives $\omega_f=|\nu_f/\widehat\nu_f-1|=O_p(\ell^{-1/2})$. (For independent symmetric $\{-1,1\}$ features, expanding $(\xi'\beta)^4$ and dropping odd powers leaves $\E[(\xi'\beta)^4]=\sum_j\beta_j^4+6\sum_{j<k}\beta_j^2\beta_k^2=3\|\beta\|^4-2\sum_j\beta_j^4\le 3\nu_f^2$, verifying the fourth-moment condition.)

\emph{Step 4: Combining the bounds.}
On the event $\nu_f>0$, substituting $\Delta_f=O_p(a_p+\sqrt{p/m})$, $\omega_f=O_p(\ell^{-1/2})$, and $\norm{\delta_j}_2=O_p(r_j)$ into the second bound of~\eqref{eq:learned-bias-rate} directly yields~\eqref{eq:learned-series-rate}. Finally, on the event $f=0$ almost surely under~$P_X$, the zero-kernel convention ($K_f=0$) recovers ordinary DML with $|b_{K_f}|=|\ip{\delta_\alpha}{\delta_g}|\le\norm{\delta_\alpha}_2\norm{v}_2$, and because $f=0$ a.s., $\norm{v}_2=\norm{v-f}_2\le a_p+O_p(\sqrt{p/m})$, so the same rate~\eqref{eq:learned-series-rate} still holds.
\end{proof}

\begin{proof}[Proof of Proposition~\ref{prop:variance}]
We first verify the Hoeffding expansion $\mathbb{U}_n h=\mu_h+\Pn L+\mathbb{U}_n d$. From the definition of~$d$ in~\eqref{eq:hoeffding-components}, each unordered pair $(i,j)$ with $1\le i<j\le n$ satisfies $h(W_i,W_j)=\mu_h+\tfrac12\{L(W_i)+L(W_j)\}+d(W_i,W_j)$. Summing over all $\binom{n}{2}=n(n-1)/2$ unordered pairs, each index $i\in\{1,\ldots,n\}$ appears in exactly $n-1$ pairs, so
\[
 \binom{n}{2}^{-1}\sum_{1\le i<j\le n}\frac{L(W_i)+L(W_j)}{2}
 =\frac{2}{n(n-1)}\cdot\frac{n-1}{2}\sum_{i=1}^n L(W_i)
 =\Pn L.
\]
Hence $\mathbb{U}_n h=\mu_h+\Pn L+\mathbb{U}_n d$. Subtracting this expansion from $\Pn s$ and adding and subtracting $\E[s(W)]$ gives
\[
 \thh_K-\theta_0
 =\{\E[s(W)]-\mu_h-\theta_0\}
  +\Pn\{s-\E[s(W)]-L\}
  -\mathbb{U}_n d
 =b_K+\Pn\phi_K-\mathbb{U}_n d,
\]
where the second equality uses Proposition~\ref{prop:bias} ($\E[s(W)]-\mu_h-\theta_0=b_K$) and the definition of~$\phi_K$ in~\eqref{eq:effective-if}. This proves~\eqref{eq:exact-expansion}.

Next we compute the conditional variance $\tau_n^2=\Var(\Pn\phi_K-\mathbb{U}_n d\mid\calT)$. Note that $\E[\phi_K(W)]=0$ and that $d$ is symmetric and degenerate, meaning $\E[d(W,W')\mid W]=0$ and $\E[d(W,W')\mid W']=0$. We check two orthogonality relations. First, for any index~$r$ and any pair $i<j$, either $r\notin\{i,j\}$ (so $W_r$ is independent of $(W_i,W_j)$ and $\E[\phi_K(W_r)d(W_i,W_j)]=\E[\phi_K(W_r)]\E[d(W_i,W_j)]=0$), or $r\in\{i,j\}$ (say $r=i$, in which case conditioning on $W_i$ gives $\E[\phi_K(W_i)\E[d(W_i,W_j)\mid W_i]]=0$). Hence $\Cov(\Pn\phi_K,\mathbb{U}_n d\mid\calT)=0$. Second, for any two distinct unordered pairs $\{i,j\}\ne\{k,l\}$, either the pairs are disjoint (giving zero covariance by independence) or they share exactly one index (say $i=k$ with $j\ne l$, in which case conditioning on $(W_i,W_j)$ gives $\E[d(W_i,W_j)\E[d(W_i,W_l)\mid W_i]]=0$). Consequently, all $\binom{n}{2}$ terms in $\mathbb{U}_n d$ are pairwise uncorrelated with common variance $\sigma_{2,n}^2=\E[d^2]$, which yields
\[
 \Var(\mathbb{U}_n d\mid\calT)
 =\binom{n}{2}^{-2}\cdot\binom{n}{2}\sigma_{2,n}^2
 =\frac{2\sigma_{2,n}^2}{n(n-1)}.
\]
Since the $n$ terms of $\Pn\phi_K$ are iid conditional on~$\calT$ with mean zero and variance $\sigma_{1,n}^2=\E[\phi_K^2]$, we obtain $\tau_n^2=\sigma_{1,n}^2/n+2\sigma_{2,n}^2/\{n(n-1)\}$, proving~\eqref{eq:full-variance}. Finally, $\sigma_{1,n}^2=\Var(s-L\mid\calT)=\Var(s)+\Var(L)-2\Cov(s,L)$.
\end{proof}

\begin{proof}[Proof of Proposition~\ref{prop:ordinary}]
Starting from the exact expansion~\eqref{eq:exact-expansion} in Proposition~\ref{prop:variance}, add and subtract the centered oracle average $\Pn\phi_0$ (noting $\E[\psi_0]=\theta_0$) and multiply by $\sqrt{n}$ to write
\[
 \sqrt{n}(\thh_K-\theta_0)
 =\sqrt{n}\,\Pn\phi_0
  +\sqrt{n}\,b_K
  +\sqrt{n}\,\Pn\big\{(s-\psi_0)-\E[s-\psi_0]\big\}
  -\sqrt{n}\,\Pn L
  -\sqrt{n}\,\mathbb{U}_n d.
\]
We show that the last four terms are $o_p(1)$. First, $\sqrt{n}\,b_K=o_p(1)$ directly by assumption. Second, conditional on~$\calT$, the centered empirical average $R_1:=\sqrt{n}\,\Pn\{(s-\psi_0)-\E[s-\psi_0]\}$ has mean zero and conditional variance
\[
 \Var(R_1\mid\calT)
 =\Var(s-\psi_0\mid\calT)
 \le \E[(s-\psi_0)^2\mid\calT]
 =\norm{s-\psi_0}_{L^2(P)}^2
 =o_p(1).
\]
By conditional Markov's inequality, for every $\varepsilon>0$ we have $\Pr(|R_1|>\varepsilon\mid\calT)\le\min\{1,\Var(R_1\mid\calT)/\varepsilon^2\}\to_p 0$, and the bounded convergence theorem implies $\Pr(|R_1|>\varepsilon)\to 0$, so $R_1=o_p(1)$. Third, the exact same calculation for $R_2:=\sqrt{n}\,\Pn L$ gives $\Var(R_2\mid\calT)=\norm{L}_{L^2(P)}^2=o_p(1)$, whence $R_2=o_p(1)$. Fourth, Proposition~\ref{prop:variance} gives
\[
 \Var(\sqrt{n}\,\mathbb{U}_n d\mid\calT)
 =\frac{2n\E[d^2]}{n(n-1)}
 =\frac{2\E[d^2]}{n-1}
 =o_p(1)
\]
under $\E[d^2]=o_p(n)$, so $\sqrt{n}\,\mathbb{U}_n d=o_p(1)$ by the same conditional Markov argument.

Collecting these bounds yields $\sqrt{n}(\thh_K-\theta_0)=n^{-1/2}\sum_{i=1}^n\phi_0(W_i)+o_p(1)$. Because $\phi_0(W_1),\ldots,\phi_0(W_n)$ are iid with mean zero and finite positive variance $\E[\phi_0^2]$, the Lindeberg--L\'evy central limit theorem gives~\eqref{eq:ordinary-limit}. The two sufficient conditions for $b_K=o_p(n^{-1/2})$ follow immediately from~\eqref{eq:bias-small} and the operator-norm fallback~\eqref{eq:fallback-bound}.
\end{proof}

\begin{proof}[Proof of Proposition~\ref{prop:vhat}]
We first relate the three population moments $(A,B,C)$ in~\eqref{eq:ABC} to the Hoeffding variances $(\sigma_{1,n}^2,\sigma_{2,n}^2)$ of $\thh_K=\mathbb{U}_n q$. Writing $\mu_q:=\E[q_{12}]$, the Hoeffding decomposition of the symmetric kernel~$q$ in~\eqref{eq:q} is
\[
 q_{ij}=\mu_q+\tfrac12\{\phi_K(W_i)+\phi_K(W_j)\}-d(W_i,W_j),
\]
where $\phi_K(W_i)$ has mean zero and variance $\sigma_{1,n}^2$, $d(W_i,W_j)$ is degenerate with variance $\sigma_{2,n}^2$, and all three components are mutually orthogonal. For independent evaluation draws $W_1,W_2,W_3,W_4$ conditional on~$\calT$, direct expansion using orthogonality gives
\begin{align*}
 A&=\E[q_{12}^2]
   =\mu_q^2+\tfrac14(\sigma_{1,n}^2+\sigma_{1,n}^2)+\sigma_{2,n}^2
   =\mu_q^2+\tfrac12\sigma_{1,n}^2+\sigma_{2,n}^2,\\
 B&=\E[q_{12}q_{13}]
   =\mu_q^2+\tfrac14\E[\phi_K(W_1)^2]
   =\mu_q^2+\tfrac14\sigma_{1,n}^2,\\
 C&=\E[q_{12}q_{34}]
   =\mu_q^2.
\end{align*}
Solving this linear system for $\sigma_{1,n}^2$ and $\sigma_{2,n}^2$ yields
\[
 \sigma_{1,n}^2=4(B-C),\qquad
 \sigma_{2,n}^2=A-2B+C.
\]
Next, because $W_1,\ldots,W_n$ are iid conditional on~$\calT$, every ordered distinct pair $(i,j)$ satisfies $\E[q_{ij}^2]=A$, every ordered distinct triple $(i,j,k)$ satisfies $\E[q_{ij}q_{ik}]=B$, and every ordered distinct quadruple $(i,j,k,l)$ satisfies $\E[q_{ij}q_{kl}]=C$. Since there are $(n)_2$, $(n)_3$, and $(n)_4$ such ordered tuples respectively, the estimators in~\eqref{eq:ABC-est} satisfy $\E[\widehat A\mid\calT]=A$, $\E[\widehat B\mid\calT]=B$, and $\E[\widehat C\mid\calT]=C$. Substituting these expectations into~\eqref{eq:vhat} and applying Proposition~\ref{prop:variance} gives
\[
 \E[\widehat V\mid\calT]
 =\frac{4}{n}(B-C)+\frac{2}{n(n-1)}(A-2B+C)
 =\frac{\sigma_{1,n}^2}{n}+\frac{2\sigma_{2,n}^2}{n(n-1)}
 =\tau_n^2.
\]
\end{proof}

\begin{proof}[Proof of Proposition~\ref{prop:general-clt}]
First check that the martingale differences $\Delta_i$ in~\eqref{eq:martingale} sum to the centered estimator. Summing~\eqref{eq:martingale} over $i=1,\ldots,n$ and using symmetry of~$d$ gives
\[
 \sum_{i=1}^n\Delta_i
 =\frac{1}{n}\sum_{i=1}^n\phi_K(W_i)
  -\frac{2}{n(n-1)}\sum_{1\le j<i\le n}d(W_i,W_j)
 =\Pn\phi_K-\mathbb{U}_n d
 =\thh_K-\E[\thh_K\mid\calT].
\]
For each $i\ge 1$, the observation~$W_i$ is independent of $\mathcal F_{i-1}=\sigma(\calT,W_1,\ldots,W_{i-1})$. Accordingly, $\E[\phi_K(W_i)\mid\mathcal F_{i-1}]=\E[\phi_K(W_i)\mid\calT]=0$, and for every $j<i$, degeneracy of~$d$ gives $\E[d(W_i,W_j)\mid\mathcal F_{i-1}]=\E[d(W_i,W_j)\mid\calT,W_j]=0$. Hence $\E[\Delta_i\mid\mathcal F_{i-1}]=0$, so $\{\Delta_i/\tau_n\}_{i=1}^n$ is a martingale difference array with respect to $\{\mathcal F_i\}$ because $\tau_n$ is $\calT$-measurable (and thus $\mathcal F_0$-measurable).

Under the conditional variance convergence and Lindeberg conditions~\eqref{eq:martingale-conditions}, the martingale central limit theorem (cf.\ Hall~\cite{hall} for varying degenerate U-statistic kernels) implies
\[
 \frac{\thh_K-\E[\thh_K\mid\calT]}{\tau_n}
 =\sum_{i=1}^n\frac{\Delta_i}{\tau_n}
 \ad N(0,1).
\]
Since $\widehat V/\tau_n^2\to_p 1$ implies $\sqrt{\tau_n^2/\widehat V}\to_p 1$, Slutsky's theorem immediately gives the centered studentized limit~\eqref{eq:studentized-centered}. When $b_K=o_p(\tau_n)$ as well, rewriting
\[
 \frac{\thh_K-\theta_0}{\sqrt{\widehat V}}
 =\sqrt{\frac{\tau_n^2}{\widehat V}}
  \left(\frac{\thh_K-\E[\thh_K\mid\calT]}{\tau_n}+\frac{b_K}{\tau_n}\right)
\]
and applying Slutsky's theorem again yields~\eqref{eq:studentized}. The width of a Wald confidence interval $\thh_K\pm z_{1-\gamma/2}\sqrt{\widehat V}$ is $2z_{1-\gamma/2}\sqrt{\widehat V}=2z_{1-\gamma/2}\tau_n\{1+o_p(1)\}$, which scales as $n^{-1/2}$ when $n\tau_n^2$ is bounded above and away from zero in probability and as $\tau_n$ otherwise.
\end{proof}

\begin{proof}[Proof of Corollary~\ref{cor:crossfit-ordinary}]
We first establish the pooled linear expansion~\eqref{eq:crossfit-ordinary} and then verify consistency of the pooled out-of-fold score variance~$\widehat\sigma^2$.

\emph{Part 1: Pooled asymptotic linear representation.}
For each fold $k\in\{1,\ldots,K\}$, let $\mathcal T_k$ denote the sigma-field generated by the fold partition, all observations outside~$I_k$, and any independent randomness used during training. Because the partition is independent of the data, the $n_k$ observations in~$I_k$ are iid from~$P$ conditional on~$\mathcal T_k$. Applying Proposition~\ref{prop:ordinary} to fold~$k$ yields
\[
 \widehat\theta_k-\theta_0=\mathbb P_k\phi_0+R_k,
 \qquad \sqrt{n_k}\,R_k=o_p(1),
\]
where $\mathbb P_k\phi_0:=n_k^{-1}\sum_{i\in I_k}\phi_0(W_i)$. Multiplying by $n_k/N$ and summing over $k=1,\ldots,K$ gives every observation weight $1/N$ in the leading oracle average,
\[
 \sum_{k=1}^K\frac{n_k}{N}\mathbb P_k\phi_0
 =\frac1N\sum_{i=1}^N\phi_0(W_i),
\]
so scaling by $\sqrt N$ leaves
\[
 \sqrt N(\widehat\theta_{\rm CF}-\theta_0)
 -\frac1{\sqrt N}\sum_{i=1}^N\phi_0(W_i)
 =\sum_{k=1}^K\sqrt{\frac{n_k}{N}}\,(\sqrt{n_k}\,R_k).
\]
Because $n_k/N\le 1$, each term in the finite sum is $o_p(1)$, and a finite sum of $K$ terms that are each $o_p(1)$ is $o_p(1)$ regardless of dependence across folds. This proves~\eqref{eq:crossfit-ordinary}.

\emph{Part 2: Variance consistency and studentized limit.}
For each fold~$k$, define the held-out mean squared score error $D_k:=\mathbb P_k[(s_{-k}-\psi_0)^2]\ge 0$. Conditional on~$\mathcal T_k$, the observations in~$I_k$ are iid from~$P$, so $\E[D_k\mid\mathcal T_k]=\norm{s_{-k}-\psi_0}_{L^2(P)}^2=o_p(1)$ by the score-convergence assumption in~\eqref{eq:ordinary-conditions}. For any $t>0$, conditional Markov's inequality gives
\[
 \Pr(D_k>t\mid\mathcal T_k)
 \le \min\!\left\{1,\,\frac{\norm{s_{-k}-\psi_0}_{L^2(P)}^2}{t}\right\}
 \to_p 0.
\]
Since the right-hand side is bounded by~$1$, the bounded convergence theorem implies $\Pr(D_k>t)=\E[\Pr(D_k>t\mid\mathcal T_k)]\to 0$, so $D_k=o_p(1)$ for every~$k$. Because $K$ is fixed, the pooled mean squared score error vanishes as well:
\[
 \frac1N\sum_{i=1}^N\{\widehat s_i-\psi_0(W_i)\}^2
 =\sum_{k=1}^K\frac{n_k}{N}D_k
 =o_p(1).
\]
Now let $\widehat{\mathbf s}:=(\widehat s_1,\ldots,\widehat s_N)'$ and $\boldsymbol\psi_0:=(\psi_0(W_1),\ldots,\psi_0(W_N))'$, and let $M_N:=I_N-N^{-1}\mathbf 1\mathbf 1'$ be the sample-centering projection matrix. Writing $\widetilde\sigma:=(N-1)^{-1/2}\|M_N\boldsymbol\psi_0\|$ for the sample standard deviation of the oracle scores and $\widehat\sigma=(N-1)^{-1/2}\|M_N\widehat{\mathbf s}\|$, the reverse triangle inequality and the contraction property $\|M_N\|_{\op}=1$ give
\[
 |\widehat\sigma-\widetilde\sigma|
 \le \frac{\|M_N(\widehat{\mathbf s}-\boldsymbol\psi_0)\|}{\sqrt{N-1}}
 \le \left(\frac{1}{N-1}\sum_{i=1}^N\{\widehat s_i-\psi_0(W_i)\}^2\right)^{1/2}
 =o_p(1).
\]
By the weak law of large numbers for the iid square-integrable scores $\psi_0(W_i)$, we have $\widetilde\sigma^2\to_p\Var(\psi_0)=\E[\phi_0^2]>0$, and hence $\widehat\sigma^2\to_p\E[\phi_0^2]$. Applying the Lindeberg--L\'evy central limit theorem to~\eqref{eq:crossfit-ordinary} together with Slutsky's theorem yields $\sqrt N(\widehat\theta_{\rm CF}-\theta_0)/\widehat\sigma\ad N(0,1)$.
\end{proof}

\begin{proof}[Proof of Proposition~\ref{prop:short-target}]
By linearity of the functional~$m$ and the Riesz representation~\eqref{eq:riesz},
\[
 \theta_0-\theta_{\mathcal H}
 =\E[m(W,g_0)]-\E[m(W,\Pi g_0)]
 =\E[m(W,g_0-\Pi g_0)]
 =\ip{\alpha_0}{(I-\Pi)g_0}.
\]
Because $I-\Pi$ is self-adjoint and idempotent on $L^2(P_X)$, we can move one copy of $I-\Pi$ onto $\alpha_0$, yielding
\[
 \theta_0-\theta_{\mathcal H}
 =\ip{(I-\Pi)\alpha_0}{(I-\Pi)g_0}
 =\ip{\alpha_0-\alpha_{\mathcal H}}{g_0-g_{\mathcal H}},
\]
which proves~\eqref{eq:short-long}.

Now suppose that both fitted nuisances $\gh$ and $\ah$ belong to~$\mathcal H_k$. Then $(I-\Pi)\gh=0$ and $(I-\Pi)\ah=0$, so their orthogonal error components reduce to
\[
 \delta_g^\perp=(I-\Pi)(\gh-g_0)=-(g_0-g_{\mathcal H}),\qquad
 \delta_\alpha^\perp=(I-\Pi)(\ah-\alpha_0)=-(\alpha_0-\alpha_{\mathcal H}).
\]
Substituting these two identities into the exact projection bias formula~\eqref{eq:projection-tail} (from Proposition~\ref{prop:bias}) gives
\[
 \E[\thh_\Pi\mid\calT]-\theta_0
 =-\ip{\delta_\alpha^\perp}{\delta_g^\perp}
 =-\ip{\alpha_0-\alpha_{\mathcal H}}{g_0-g_{\mathcal H}}
 =-(\theta_0-\theta_{\mathcal H}).
\]
Adding $\theta_0$ to both sides yields $\E[\thh_\Pi\mid\calT]=\theta_{\mathcal H}$, proving~\eqref{eq:short-center}.
\end{proof}

\begin{proof}[Proof of Proposition~\ref{prop:working-orthogonality}]
Recall that $K_M(x,x')=b(x)'Mb(x')$, $\Sigma=\E[b(X)b(X)']$, $u_g=\E[b(X)\delta_g(X)]$, and $u_\alpha=\E[b(X)\delta_\alpha(X)]$. For any square-integrable candidate error $\delta_g$, the integral operator $T_M$ induced by $K_M$ acts as
\[
 (T_M\delta_g)(x)
 =\E[b(x)'Mb(X)\delta_g(X)]
 =b(x)'M\E[b(X)\delta_g(X)]
 =b(x)'Mu_g.
\]
Taking the inner product with $\delta_\alpha$ gives
\[
 \ip{\delta_\alpha}{T_M\delta_g}
 =\E[\delta_\alpha(X)b(X)']Mu_g
 =u_\alpha'Mu_g.
\]
In the special case $M=\Sigma^{-1}$, $T_{\Sigma^{-1}}$ is the exact $L^2(P_X)$ orthogonal projection~$\Pi$ onto~$\mathcal H_k$, so $\Pi\delta_g=b'\Sigma^{-1}u_g$ and $\ip{\Pi\delta_\alpha}{\Pi\delta_g}=u_\alpha'\Sigma^{-1}u_g$. Now combine Proposition~\ref{prop:bias} with the orthogonal split~\eqref{eq:working-bias-split} of $-\ip{\delta_\alpha}{\delta_g}$:
\begin{align*}
 \mathcal M_M(g,\alpha)-\theta_0
 &=-\ip{\delta_\alpha}{\delta_g}+\ip{\delta_\alpha}{T_M\delta_g}\\
 &=-\ip{(I-\Pi)\delta_\alpha}{(I-\Pi)\delta_g}
   -\ip{\Pi\delta_\alpha}{\Pi\delta_g}
   +\ip{\delta_\alpha}{T_M\delta_g}\\
 &=-\ip{(I-\Pi)\delta_\alpha}{(I-\Pi)\delta_g}
   +u_\alpha'(M-\Sigma^{-1})u_g,
\end{align*}
which proves~\eqref{eq:orthogonality-bias-decomposition}.

Next suppose the candidates lie in~$\mathcal H_k$, so $g=b'\beta$ and $\alpha=b'\gamma$. By Proposition~\ref{prop:short-target}, the first term on the right-hand side of~\eqref{eq:orthogonality-bias-decomposition} equals $\theta_{\mathcal H}-\theta_0$. In addition, because $g_0-b'\beta_{\mathcal H}=(I-\Pi)g_0$ is orthogonal to every coordinate of~$b(X)$, we have
\[
 u_g
 =\E[b(X)\{b(X)'\beta-g_0(X)\}]
 =\E[b(X)b(X)'](\beta-\beta_{\mathcal H})
 =\Sigma(\beta-\beta_{\mathcal H}),
\]
and identically $u_\alpha=\Sigma(\gamma-\gamma_{\mathcal H})$. Substituting these expressions into~\eqref{eq:orthogonality-bias-decomposition} and moving $\theta_{\mathcal H}-\theta_0$ to the left-hand side yields
\[
 \mathcal M_M(b'\beta,b'\gamma)-\theta_{\mathcal H}
 =(\gamma-\gamma_{\mathcal H})'\Sigma(M-\Sigma^{-1})\Sigma(\beta-\beta_{\mathcal H}),
\]
which is~\eqref{eq:orthogonality-cubic}. Let $\Delta\beta:=\beta-\beta_{\mathcal H}$, $\Delta\gamma:=\gamma-\gamma_{\mathcal H}$, and $\Delta M:=M-\Sigma^{-1}$. Because the right-hand side is a homogeneous trilinear form $\Delta\gamma'\Sigma(\Delta M)\Sigma\Delta\beta$ of degree~3 in $(\Delta\beta,\Delta\gamma,\Delta M)$, its value, all first-order directional derivatives, and all second-order directional derivatives vanish identically at $(\Delta\beta,\Delta\gamma,\Delta M)=(0,0,0)$. Hence $\mathcal M_M(b'\beta,b'\gamma)-\theta_{\mathcal H}$ is second-order Neyman orthogonal jointly in $(\beta,\gamma,M)$ at $(\beta_{\mathcal H},\gamma_{\mathcal H},\Sigma^{-1})$. If $g_0,\alpha_0\in\mathcal H_k$, then $\theta_{\mathcal H}=\theta_0$, so the same orthogonality holds for~$\theta_0$ itself.
\end{proof}

\section{Additional empirical results on partitions and sensitivity}\label{app:empirical}
\subsection{401(k) partitions and variance}\label{app:401k}
We analyze the public 401(k) eligibility dataset from the 1991 Survey of Income and Program Participation~\cite{dml,doublemldata}, covering 9,915 households. Let $D$ denote 401(k) eligibility, $Z$ the nine standard household covariates, and $X=(D,Z)$. The target is the covariate-adjusted ATE functional,
\[
 m(W,g)=g(1,Z)-g(0,Z),\qquad
 \alpha_0(D,Z)=\frac{D}{e_0(Z)}-\frac{1-D}{1-e_0(Z)},
\]
whose causal interpretation requires consistency, conditional exchangeability, and overlap. We treat observations as iid without survey weights.

\paragraph{Sample design and baseline estimates}
We split the sample randomly (seed 20261109) into 4,957 training observations and 4,958 evaluation observations, comparing two nuisance specifications: (i)~linear outcome regressions by eligibility arm with a logistic propensity score, and (ii)~random forests (500 trees, with minimum leaf sizes of 20 for outcomes and 40 for propensity). Propensities are clipped to $[0.1,0.9]$ to stabilize scores without establishing overlap, and all nuisance fits and kernels are frozen on training data.

Table~\ref{tab:emp-basic} compares original net financial assets with capped assets,
\[
 Y^c:=\min\{250{,}000,\max\{-50{,}000,Y\}\},
\]
with outcome predictions clipped identically. Capping changes the outcome for 105 households and defines a distinct estimand. All monetary figures are in 1991 dollars. For capped assets, ordinary debiasing moves the forest plug-in estimate from \$7,285 to \$7,559 and moves the linear plug-in estimate from \$6,440 to \$5,914. Switching the ordinary DML fits from linear to forest reduces the estimated standard error by 9.5\% for capped assets and 41.3\% for original assets.

\begin{table}[H]
\centering\small
\caption{From regression adjustment to ordinary debiasing (401(k) data)}
\label{tab:emp-basic}
\begin{tabular}{lrrrr}
\toprule
 & \multicolumn{2}{c}{Original assets} & \multicolumn{2}{c}{Capped assets}\\ Method & Estimate & SE & Estimate & SE\\
\midrule
Unadjusted contrast & 20,485 & 1,945 & 18,006 & 1,249\\
Linear regression plug-in & 3,626 & --- & 6,440 & ---\\
Linear AIPW & 5,191 & 2,712 & 5,914 & 1,090\\
Forest regression plug-in & 6,289 & --- & 7,285 & ---\\
Forest AIPW & 8,406 & 1,593 & 7,559 & 987\\
\bottomrule
\end{tabular}
\par\vspace{5pt}\begin{minipage}{\textwidth}\footnotesize Amounts are 1991 dollars. The unadjusted row estimates a descriptive group contrast. AIPW uses the usual debiased ATE score. Its reported standard error describes centered sampling variation, whereas valid inference for the ATE also requires negligible learning bias or a bias allowance. Plug-in standard errors are omitted because treating the fitted regression as fixed does not address its first-order estimation bias.\end{minipage}
\end{table}

\paragraph{The correction and its variance}
Using training data alone, we partition household income into $J\in\{2,4,8,16\}$ quantile cells, cross them with eligibility~$D$ to form $2J$ cell indicators~$b(D,Z)$, and construct the cell-projection kernel
\[
 \widehat K(x,x'):=\sum_{\ell=1}^{2J}
 \frac{b_\ell(x)b_\ell(x')}{\widehat p_\ell},
\]
where $\widehat p_\ell$ denotes the training frequency of cell~$\ell$. Table~\ref{tab:emp-kernel} reports the added correction $-\widehat B_K$, the updated estimate, and its full estimated standard error. With four income cells, the linear AIPW estimate rises by \$979 (from \$5,914 to \$6,893), whereas the forest estimate rises by only \$83. The full standard error falls by 1.3\% after linear DML and by 0.5\% after forest DML. The two models then give more similar estimates of the gain in assets at essentially the same precision. Quantifying the bias removed would require knowing the true effect.

\begin{table}[H]
\centering\small
\caption{Add one correction, then enlarge its working space (401(k) data)}
\label{tab:emp-kernel}
\begin{tabular}{lrrrrr}
\toprule
Fit & Income cells & Added correction & Estimate & Full SE & SE red. (\%)\\
\midrule
Linear & 2 & 284 & 6,198 & 1,071 & 1.7\\
Linear & 4 & 979 & 6,893 & 1,075 & 1.3\\
Linear & 8 & 883 & 6,797 & 1,065 & 2.2\\
Linear & 16 & 956 & 6,870 & 1,064 & 2.3\\
Forest & 2 & -3 & 7,556 & 974 & 1.2\\
Forest & 4 & 83 & 7,642 & 982 & 0.5\\
Forest & 8 & 38 & 7,597 & 989 & -0.2\\
Forest & 16 & 149 & 7,708 & 1,018 & -3.1\\
\bottomrule
\end{tabular}
\par\vspace{5pt}\begin{minipage}{\textwidth}\footnotesize Capped assets. The added correction is $-\widehat B_K$, so a positive entry raises the ordinary estimate. Income cutoffs and cell normalizations are computed on training data. The working space has rank $2J$, with a separate indicator for each income cell and eligibility status. The last column gives the percentage reduction in the full standard error relative to ordinary DML with the same initial fit.\end{minipage}
\end{table}

Across the four forest specifications in Table~\ref{tab:emp-var}, the pairwise term accounts for just 0.08\% to 0.55\% of total estimated variance. Updating the first-order influence function shifts the standard error more than adding pairs, which explains why the two components appear separately in Table~\ref{tab:decision-guide}.

\begin{table}[H]
\centering\small
\caption{Separate the change in influence function from pairwise variance (401(k) forests)}
\label{tab:emp-var}
\begin{tabular}{rrrrr}
\toprule
Income cells & Original SE & Corrected first-order SE & Full SE & Pairwise share\\
\midrule
2 & 987 & 974 & 974 & 0.08\%\\
4 & 987 & 981 & 982 & 0.17\%\\
8 & 987 & 988 & 989 & 0.31\%\\
16 & 987 & 1,015 & 1,018 & 0.55\%\\
\bottomrule
\end{tabular}
\par\vspace{5pt}\begin{minipage}{\textwidth}\footnotesize Forest fits, capped assets. The middle standard error is $\sqrt{\widehat V_1}$, and the full standard error is $\sqrt{\widehat V_1+\widehat V_2}$. The last column is $100\widehat V_2/(\widehat V_1+\widehat V_2)$. Both components come from the unbiased distinct-index variance formula and are positive in these calculations without truncation.\end{minipage}
\end{table}

\subsection{ACIC partitions, coverage, and sensitivity}\label{app:acic}
The 2016 ACIC competition generator~\cite{dorie-acic} supplies the true treatment probability and both conditional outcome means for each of 4,802 covariate records, with the average treatment effect over those records as the target. On each replication, we draw records uniformly with replacement, assign treatment according to the true probability, and add independent standard normal noise to the assigned mean. Because the population law is known, we can evaluate the conditional bias of every training fit directly.

\paragraph{Simulation design and choice of working space}
We use settings 27 and 29, which share polynomial treatment assignment, step-function outcome means, full overlap, and medium effect heterogeneity, while differing in how strongly treatment and outcome models share covariate terms (alignment parameter $0.25$ in setting 27 versus $0.75$ in setting 29). Fixing the first four response surfaces in each setting, we draw 50 independent pairs of training and evaluation samples per surface, giving 200 draws per setting with $n=2{,}400$ per split.

Ordinary DML fits ridge outcome regressions by treatment arm and a penalized logistic propensity on main effects, clipping propensities to $[0.05,0.95]$. To build a data-adaptive working space~$\mathcal H_k$ on training data, we fit a single regression tree to the standardized treatment and outcome residuals, cross its leaves with treatment arm, require at least 12 training observations per leaf, and add one pseudo-count per arm-by-leaf cell when estimating~$\widehat p_c$. We report all three pre-specified leaf caps (4, 32, and 128 leaves) alongside ordinary DML.

\begin{table}[H]
\centering\small
\caption{Bias, root mean squared error, and nominal 95\% Wald coverage on ACIC response surfaces}
\label{tab:acic-performance}
\begin{tabular}{@{}lrrrrrr@{}}
\toprule
& \multicolumn{3}{c}{Setting 27, alignment 0.25} & \multicolumn{3}{c}{Setting 29, alignment 0.75}\\
\cmidrule(lr){2-4}\cmidrule(l){5-7}
Procedure & Bias & RMSE & Coverage & Bias & RMSE & Coverage\\
\midrule
Ordinary DML       & 0.0372 & 0.0668 & 74.0\% & 0.0719 & 0.0976 & 52.5\%\\
DDML, up to 4 leaves   & 0.0190 & 0.0401 & 85.5\% & 0.0409 & 0.0582 & 67.5\%\\
DDML, up to 32 leaves  & 0.0131 & 0.0333 & 94.0\% & 0.0258 & 0.0498 & 83.5\%\\
DDML, up to 128 leaves & 0.0166 & 0.0496 & 92.5\% & 0.0222 & 0.0669 & 89.0\%\\
\bottomrule
\end{tabular}
\par\vspace{4pt}\begin{minipage}{\textwidth}\footnotesize
Bias is the mean over training fits of the absolute conditional bias, computed by integrating the fitted score and correction under the known law. Bias and RMSE use the marginal outcome standard deviation on each surface as the unit. Coverage is the fraction of 200 draws per setting in which the nominal 95\% Wald interval contains the target. A nonpositive variance estimate counts as a miss. This occurred once at each of the 4- and 32-leaf caps in setting 29.
\end{minipage}
\end{table}

At 32 leaves, the correction lowers RMSE on every one of the eight response surfaces and roughly halves it in both settings (Table~\ref{tab:acic-performance}), lifting Wald coverage from 74.0\% to 94.0\% in setting 27 and from 52.5\% to 83.5\% in setting 29. The 32-leaf correction reduces the estimated standard error by 16.4\% and 15.5\% on average over draws with valid standard errors for both methods. Even so, the absolute conditional bias increases in 33.5\% of individual fits in setting 27, reflecting the cancellation mechanism in Example~\ref{ex:cancellation}.

A finer partition estimates the correction within smaller groups. Raising the cap from 32 to 128 leaves increases RMSE in both settings. In setting 29, the finer partition reduces mean absolute conditional bias from 0.0258 to 0.0222, but the mean estimated standard error rises by 40.4\%, from 0.0326 to 0.0458. We report all three pre-specified caps so that the comparison at 32 leaves remains purely descriptive.

\paragraph{Sensitivity to the errors left by a partition}
For a cell with true probability $p_c$ and training mass $\widehat p_c$, the operator $T$ acts within the cell-indicator space with eigenvalue $p_c/\widehat p_c$. The definition in~\eqref{eq:q-errors} weights the squared error along that indicator by $|1-p_c/\widehat p_c|$. The error outside the cell space receives weight one. Table~\ref{tab:acic-sensitivity} applies~\eqref{eq:q-budget} to the resulting 32-leaf correction. For comparison, widening by the true absolute conditional bias, observable under this generator, gives coverage of 98.5\% in setting~27 and 95.0\% in setting~29.

\begin{table}[H]
\centering\small
\caption{Coverage and width under assumed residual error budgets, 32-leaf cap (ACIC)}
\label{tab:acic-sensitivity}
\begin{tabular}{@{}lrrrrrr@{}}
\toprule
& \multicolumn{3}{c}{Setting 27, alignment 0.25} & \multicolumn{3}{c}{Setting 29, alignment 0.75}\\
\cmidrule(lr){2-4}\cmidrule(l){5-7}
Budget $q$ & Coverage & Width increase & $|b_K|\le B_q$ & Coverage & Width increase & $|b_K|\le B_q$\\
\midrule
0\% & 94.0\% & 0.0\% & 0.0\% & 83.5\% & 0.0\% & 0.0\%\\
1\% & 97.0\% & 19.1\% & 52.0\% & 89.0\% & 17.9\% & 22.0\%\\
2\% & 98.5\% & 38.3\% & 83.0\% & 95.5\% & 36.1\% & 53.5\%\\
\bottomrule
\end{tabular}
\par\vspace{4pt}\begin{minipage}{\textwidth}\footnotesize
Width increase is the percentage increase in mean interval width relative to the Wald interval among draws with positive estimated variance. The columns headed $|b_K|\le B_q$ give the fraction of fits for which the allowance covers the true absolute conditional bias. Only this diagnostic uses the known bias.
\end{minipage}
\end{table}

A 1\% budget raises coverage by 3.0 and 5.5 percentage points, even though none of the 400 fits satisfies both bounds on weighted mean squared error at that level. In setting 29, the 2\% budget reaches 95.5\% coverage with intervals 36.1\% wider than the corrected Wald intervals, yet $B_q$ bounds the true absolute conditional bias in only 53.5\% of fits. Sampling width helps bridge that gap, showing that a formal coverage guarantee requires both a justified bias allowance and valid centered inference.

\paragraph{Sensitivity to the errors left by a learned direction}
Table~\ref{tab:acic-q} accompanies the coverage results in Table~\ref{tab:rankacic-ci}. The common Riesz scale makes differences in the allowance on a given draw attributable to the outcome residual scale. In setting~29, a 2\% budget yields an average allowance of 0.0231 with the linear outcome fit and 0.0103 with the boosted fit. These allowances cover the absolute conditional bias of the forest-direction estimator in 31.0\% and 40.5\% of fits, respectively.

\begin{table}[H]
\centering\small
\caption{Allowances, adjusted widths, and absolute conditional bias for the ACIC rank-one corrections}\label{tab:acic-q}
\begin{tabular}{@{}l*{7}{@{\hspace{.72em}}r}@{}}
\toprule
 & & \multicolumn{3}{c}{$q=1\%$} & \multicolumn{3}{c}{$q=2\%$}\\
\cmidrule(lr){3-5}\cmidrule(lr){6-8}
Initial fits / direction & Draws & Mean $B_q$ & Width & Bias covered & Mean $B_q$ & Width & Bias covered \\
\midrule
\multicolumn{8}{l}{\textit{Setting 27}}\\
Linear, logistic / forest & 200 & 0.0113 & 0.133 & 48.0 & 0.0227 & 0.156 & 72.5 \\
Boosted, boosted / forest & 200 & 0.0037 & 0.087 & 43.0 & 0.0074 & 0.095 & 74.0 \\
Linear, logistic / PFN & 20 & 0.0113 & 0.111 & 90.0 & 0.0227 & 0.134 & 95.0 \\
PFN, logistic / forest & 20 & 0.0037 & 0.091 & 10.0 & 0.0075 & 0.099 & 40.0 \\
PFN, boosted / forest & 20 & 0.0037 & 0.089 & 30.0 & 0.0075 & 0.097 & 65.0 \\
\addlinespace
\multicolumn{8}{l}{\textit{Setting 29}}\\
Linear, logistic / forest & 200 & 0.0115 & 0.141 & 14.5 & 0.0231 & 0.164 & 31.0 \\
Boosted, boosted / forest & 200 & 0.0051 & 0.092 & 17.5 & 0.0103 & 0.103 & 40.5 \\
Linear, logistic / PFN & 20 & 0.0110 & 0.117 & 55.0 & 0.0220 & 0.139 & 85.0 \\
PFN, logistic / forest & 20 & 0.0036 & 0.090 & 20.0 & 0.0073 & 0.097 & 25.0 \\
PFN, boosted / forest & 20 & 0.0036 & 0.087 & 30.0 & 0.0073 & 0.094 & 35.0 \\
\bottomrule
\end{tabular}
\par\vspace{4pt}\begin{minipage}{\textwidth}\footnotesize
Allowances use one debiased Riesz scale per draw and the residual scale of each outcome fit. Allowances and adjusted widths are in marginal outcome SDs. Bias covered is the percentage of draws for which $|b_K|\le B_q$, computed using the known conditional bias. Coverage of the target for the same draws and budgets appears in Table~\ref{tab:rankacic-ci}.
\end{minipage}
\end{table}
\end{appendix}

\end{document}